\documentclass[12pt]{article}
\usepackage{mathrsfs}
\usepackage{amsmath}
\usepackage{mathrsfs}
\usepackage{amssymb}
\usepackage{color}
\pdfoutput=1
\usepackage[unicode]{hyperref}
\usepackage{graphicx}
\usepackage{xcolor}
\usepackage{hyperref}
\usepackage{color}
\usepackage[utf8]{inputenc}
\usepackage{setspace}
\usepackage{amsmath, amssymb, amsthm}
\numberwithin{equation}{section}
\hypersetup{colorlinks=false, linkcolor=blue, citecolor=red}
\newcommand{\be}{\begin{equation}}
\newcommand{\ee}{\end{equation}}
\def\g{\gamma}
\def\G{\Gamma}

\def\s{\sigma}

\def\a{\alpha}

\def\e{\epsilon}
\def\l{\lambda}
\def\k{\kappa}
\def\b{\beta}
\def\d{\delta}

\def\f{\phi}

\def\D{\Delta}

\newcommand{\bea}{\begin{eqnarray}}
\newcommand{\eea}{\end{eqnarray}}

\renewcommand{\a}{\alpha}
\renewcommand{\b}{\beta}

\renewcommand{\d}{\delta}
\newcommand{\dsl}{\pa \kern-0.5em /}
\newcommand{\la}{\lambda}
\newcommand{\half}{\frac{1}{2}}
\newcommand{\pa}{\partial}

\newcommand{\nn}{\nonumber\\}

\renewcommand{\eqref}[1]{eq.(\ref{#1})}
\def\intl{\displaystyle\int_{-i\infty}^{i\infty}}
\begin{document}
	\title{\bf  Simplifying large spin bootstrap in Mellin space}
	\date{}
	
\author{Parijat Dey\footnote{parijat@chep.iisc.ernet.in},~ Kausik Ghosh\footnote{kau.rock91@gmail.com} ~and Aninda Sinha\footnote{asinha@chep.iisc.ernet.in} \\ ~~~~\\
	\it Centre for High Energy Physics,
	\it Indian Institute of Science,\\ \it C.V. Raman Avenue, Bangalore 560012, India. \\}	
\maketitle
\abstract{We set up the conventional conformal bootstrap equations in Mellin space and analyse the anomalous dimensions and OPE coefficients of large spin double trace operators. By decomposing the equations in terms of continuous Hahn polynomials, we derive explicit expressions as an asymptotic expansion in inverse conformal spin to any order, reproducing the contribution of any primary operator and its descendants in the crossed channel. The expressions are in terms of known mathematical functions and involve generalized Bernoulli (N{\o}rlund) polynomials and the Mack polynomials and enable us to derive certain universal properties. Comparing with the recently introduced reformulated equations in terms of crossing symmetric tree level exchange Witten diagrams, we show that to leading order in anomalous dimension but to all orders in inverse conformal spin, the equations are the same as in the conventional formulation. At the next order, the polynomial ambiguity in the Witten diagram basis is needed for the equivalence and we derive the necessary constraints for the same.}
\tableofcontents
%\tableofcontents

\onehalfspacing
\section{Introduction and summary of results }

The resurgence of interest in the conformal bootstrap since the work of \cite{Rattazzi:2008pe} has seen several nontrivial numerical results in the last few years. The most accurate estimates for the 3d Ising model critical exponents \cite{3dising,mostprecise} arises from this formalism (see \cite{reviews, susy, others, spins, hogerv, Li2, gliozzi2} for a sampling of recent work discussing many applications of bootstrap).

In the last few years, there has been progress in extracting analytic results using conformal field theory techniques.
There have been two different developments. First, certain analytic results are possible for the so-called double trace operators with large spin. In a series of nice papers \cite{Alday:2015eya, Alday:2015ota, aldayzhiboedov, Alday:2016mxe, Alday:2016njk, Alday:2016jfr, Aharony:2016dwx, holorecon}, building on the work of \cite{Fitzpatrick:2012yx, Komargodski:2012ek}, a systematic approach in position space has been advocated which enables one to get the asymptotic expansion in inverse spin, at least in principle--for a related approach see \cite{dsd}. In certain situations, the inverse spin series can be resummed and a finite support piece can be added yielding results for all spins, not just large spins. For instance, the leading anomalous dimensions in the epsilon expansion ($d=4-\e$) can be obtained, using this approach and a set of resummed blocks called ``twist conformal blocks'', in terms of an undetermined parameter\footnote{Certain leading anomalous dimensions can be nicely obtained by exploiting three point functions \cite{rychkovtan}.}. While in principle, this method can be pushed to obtain results at higher orders in epsilon, this has not been achieved as yet. In \cite{KSAS, RGAKKSAS, PDAKAS}, an alternative formulation of the bootstrap was given following a 1974 work by Polyakov \cite{polya}--the success of this approach is not tied to large spin. Specifically, in \cite{RGAKKSAS, PDAKAS}, the power of this formalism was demonstrated in Mellin space. This approach uses manifestly crossing symmetric blocks which turn out to be tree level exchange Witten diagrams. Thus, while crossing symmetry is guaranteed, consistency with the operator product expansion (OPE) needs to be checked which gives an infinite set of consistency conditions. Using this, it was shown how to systematically extract information upto $\epsilon^3$ order for anomalous dimensions (which are in agreement with Feynman diagram techniques) and OPE coefficients (which are typically unknown from the Feynman diagram approach but indirect arguments \cite{PDAKAS, osbornrecent} exist to show consistency with the bootstrap calculations). 

A question naturally emerges: What is the relation between the two different approaches--are they the same? It is our goal in this paper to shed some light on this. Of course, in order to answer this question we need an analytic handle on both sides--thus enters large spin. We will develop Mellin space techniques which turn out to simplify many of the existing calculations that follow from and build on \cite{aldayzhiboedov}. In particular, the algebraic bootstrap approach of \cite{aldayzhiboedov} gives the anomalous dimension of ``double trace'' large spin operators as an asymptotic expansion in inverse conformal spin (to be defined below). The reason inverse conformal spin is a more suitable expansion parameter rather than inverse spin can be explained transparently in Mellin space. Furthermore, for arbitrary spin and twist, the approach of \cite{aldayzhiboedov} enabled one to obtain the coefficients in the large spin series through a recursion relation. While in principle, this could be automated, the question arises if these coefficients can be given an all order expression in terms of known mathematical functions. This exercise will turn out to be quite simple in Mellin space. We will find that the answers for the anomalous dimensions and OPE coefficients can be written in terms of generalized Bernoulli polynomials and the Mack polynomials.

To distinguish the usual position space approach from the one advocated in \cite{RGAKKSAS, PDAKAS}, we will refer to the former as the ``usual approach'' and the latter\footnote{Since it is based on Polyakov's 1974 idea \cite{polya}, more appropriately it is a new take on an old idea but for economy of nomenclature, we will continue referring to it as the ``new approach''! Another suggested terminology is to refer to the usual formalism as the ``Associativity-bootstrap'' or A-bootstrap and the one in \cite{polya,RGAKKSAS} as the ``crossing symmetry-bootstrap'' or C-bootstrap.} as the ``new approach.'' Let us begin introducing some notation (unfortunately there will be a lot of them!). We will focus on the four point function of identical scalars having conformal dimensions $\D_\phi$. We will write the conformal dimension $\Delta$ of the double trace operator with spin $\ell$ as $\Delta=2\D_\phi+\ell+\gamma_\ell$ where $\gamma_\ell$ is the anomalous dimension.  The conformal spin $J_{\g_\ell}^2$ is defined through $J_{\g_\ell}^2=(\D_\phi+\ell+\gamma_\ell/2)(\D_\phi+\ell+\gamma_\ell/2-1)$. For large spin $\gamma_\ell$ is small.  In studying the equations carefully in Mellin space, we will be able to show that the usual approach and the new approach are in fact equivalent in the leading order in $\gamma_\ell$. In fact, when there is an explicit small parameter--let us generically call{\footnote{To be clear, this can also be the $\epsilon$ in the epsilon expansion. A further comment is that since the OPE corrections and the polynomial ambiguity contribution in the new approach set in at $O(\gamma_\ell^2)$, if the series $\sum c_n/J^{2\tau_m+n}$ and $\sum d_p/J^{\tau_m+p}$ did not have any terms in common, even then the calculations would hold to all orders in $1/J$ without such an explicit small factor.}} this $1/N$ --that sits in front of $\gamma_\ell$, one can get all the terms in the asymptotic expansion in inverse $J_0$ to leading order in $1/N$. The equivalence between the two approaches are at the level of the equations to all orders in $1/J_0$ to leading order in $1/N$. This is transparent in Mellin space. Thus we can claim that the large spin sector of the bootstrap equations effectively use the tree level exchange Witten diagrams as the basis at leading order in $1/N$. At the next order in $\gamma_\ell$, namely $\gamma_\ell^2$, there is a mismatch between the two sets of equations. This mismatch is not unexpected. There are polynomial ambiguities in Mellin amplitudes. This mismatch is reflecting this fact. Turning this around, we can say that the polynomial pieces of the Mellin amplitude basis must be constrained in order to agree with the usual conformal block expansion. We will work out this constraint.

There are several key differences and features in the starting point of the two approaches. Let us begin by highlighting the main ones. 
\begin{itemize}
	\item In both approaches we will be focusing on $u^{\D_\phi+k}$ and $u^{\D_\phi+k}\log u$ terms in the bootstrap equation. In the usual approach these arise due to expanding $u^{(\Delta-\ell)/2}=u^{\D_\phi+\gamma_\ell/2}$ in small $\gamma_\ell$. Thus one generates not only $\log u$ but all powers of $\log u$. In the new approach, the Witten diagram basis has double poles through $\G^2(\D_\phi-s)$. Thus one gets only $\log u$'s not powers of $\log u$--to emphasise, we do not expand in small powers to generate logs in this approach. The consistency with OPE demands that these are spurious and hence must cancel. This gives rise to consistency conditions. When we say that we compare at the level of the equations, this is what we mean--we compare the crossing symmetry condition from the usual approach with the OPE consistency condition in the new approach. 
	\item The usual approach crossing symmetry condition looks like an equality between the direct channel (sometimes referred to as the $s$-channel) and the crossed channel (sometimes referred to as the $t$-channel). Schematically, we have
	$$
	s_{usual}=t_{usual}\,.
	$$
	The OPE consistency condition in the new approach involves writing the correlator as the sum of all 3 channels and isolating the spurious poles. At the level of the equations for a four point function involving identical scalars, the $t$ and $u$ channels give the same contribution. Schematically we have here
	$$
	(s_{new}+2 t_{new})|_{spurious}=0\,.
	$$
	These two schematic forms of the equations appear quite different! There is not only a sign but also a factor of 2 that are different. In the large spin case that is of interest to us, we have to explain how the usual form metamorphoses into the new form. Both these issues will explained in due course.
	\item In both formulations we sum over all physical operators. This is just to emphasise that this is unlike what happens in AdS/CFT where the leading order effect of the double trace operators is captured by a measure factor and we only sum over single trace operators.
	\item In the usual approach in the small $u,v$ limit, the direct channel has $\log v$ as the leading singularity while the crossed channel has a power law singularity. After summing over large spin operators of the form discussed above, the $\log v$ gets converted into the power law singularity. In the new approach, in principle, there is {\it no resummation} needed in writing down the OPE consistency conditions. The large spin approximation is needed in order to compare the two approaches. The fact that the new approach gives the correct epsilon expansion (up to $O(\e^3)$) provides some a posteriori justification for this.
\end{itemize}

Since the calculations are somewhat technical in nature, let us summarize the key ingredients here. The four point function for four identical scalar operators $\mathcal{O}$ having conformal dimension $\D_\phi$ is written as 
\begin{equation}
\begin{split}
\mathcal{A}(x_1,x_2,x_3,x_4)&=\left\langle \mathcal{O}(x_1)\mathcal{O}(x_2)\mathcal{O}(x_3)\mathcal{O}(x_4)\right\rangle\\
& =\frac{1}{(x_{12}^2x_{34}^2)^{\Delta_{\phi}}} \mathcal{A}(u,v).
\end{split}
\end{equation}
Here we have pulled out the factors appropriate for an $s$-channel decomposition and defined $x_{ij}=x_i-x_j$.
The cross ratios $(u,v)$ are defined in the conventional way
\begin{equation}
u=\frac{x_{12}^2 x_{34}^2}{x_{13}^2 x_{24}^2},\,\, v=\frac{x_{14}^2x_{23}^2}{x_{13}^2 x_{24}^2}.
\end{equation}
The $s,t$ variables in Mellin space are introduced via %{\bf AS: we should change $\G()^2$ to $\G^2()$},
\begin{equation}\label{Mdef}
\mathcal{A}(u,v)=\int^{i \infty}_{-i \infty} \frac{ds}{2\pi i}\frac{dt}{2\pi i} u^s v^t \Gamma^2(\Delta_{\phi}-s)\, \Gamma^2(s+t) \,\Gamma^2(-t) \mathcal{M}(s,t).
\end{equation}
$\mathcal{M}$ is frequently referred to as the ``Mellin amplitude'' \cite{Mack:2009mi, Mellin, fitz, pau, costa}. This can be expanded in terms of the Mellin transform of the usual conformal blocks \cite{dolanosborn,do2,Dolan:2011dv} or in terms of the tree level exchange Witten diagram basis. The Mellin transform of the direct channel conformal block has an explicit inverse factor $\Gamma^2(\D_\phi-s)$ which gets rid of the potential double poles at $s=\D_\phi+n$. After using a projection operator \cite{fkap} to eliminate the shadow poles $s=d-\D-\ell+q\equiv 2h-\D-\ell+q$, only the physical $s$ poles contribute to the amplitude.  The only way to generate $\log u$'s in the direct channel therefore, is to expand powers of $u$. In this sense, $\log u$ terms in the usual approach are not unphysical. In the Witten diagram basis, the diagrams are typically defined through the $s$ physical poles upto some polynomials in $t$. There is some ambiguity in this definition since one could multiply the physical pole answer by a suitable polynomial in $s$ which leaves the residue at the physical poles unchanged \footnote{cf.  Mittag-Leffler theorem.}. Furthermore, for a spin-$\ell$ exchange there is also an apparent freedom to add a spin-($\ell-1)$ polynomial in $s,t$. It is this that we will refer to as the polynomial ambiguity. An important role of this polynomial ambiguity will become clear in this paper. In the new approach, the double poles from $\Gamma^2(\D_\phi-s)$  are there. These are in conflict with the OPE as in general (barring isolated examples involving protected operators) there are no physical operators having dimensions precisely $\D=2\D_\phi+\ell+2n$. This is distinct from how we handle these double poles in the AdS/CFT literature---there, the double poles are relate to double trace operators and we do not include them in the sum over the spectrum; however, here we include all primary operators in the spectrum and the $s=\D_\phi+n$ poles are spurious. 

Now in Mellin space there are a set of ``natural'' orthonormal polynomials using which we can expand the $t$ dependence. These are called continuous Hahn polynomials (to make life simpler we will refer to these as the ``Q''-polynomials in the rest of the paper) and involve a ${}_3F_2$ hypergeometric function. The measure with respect to which these are orthonormal is proportional to $\G^2(-t)$. Now let us consider what happens to the $t$ channel. In the usual approach the inverse factor $\Gamma^2(\D_\phi-s)$ in the $s$ channel becomes $\G^2(-t)$. Hence there are no contributions from the double poles arising from the $\G^2(-t)$ in the measure factor. However, in the new approach these poles contribute since such an inverse factor is absent. 
%Because of the absence of these poles we won't get the correct expression for finite spin after doing the $t$-integral in usual approach. The large spin story enters into the game to remove this disparity. %
 Here is where the large spin story enters.
  We will derive the necessary large spin asymptotics in this paper.
The large spin approximation of the relevant $_3F_2$ \eqref{Qasym}
\begin{align}\label{3f2a}
{}_3F_2\bigg[\begin{matrix} -\ell,\, 2\s+\ell-1,\,\s+t\\
\ \ \s \ \ , \  \  \s
\end{matrix};1\bigg]  & \sim   \sum_{n, k_1, k_2=0}^{\infty}\frac{(-1)^n}{n!}\frac{\G^2(\s)\,(\s+t)_n}{\G^2(-t-n)}\,\mathfrak{b}_{k_1}(\s)\,\mathfrak{b}_{k_2, n}(t)\,  J^{-2k_1-2k_2-2n-2\s-2t}\nn
&+  \sum_{n, k_1, k_2=0}^{\infty}\frac{(-1)^n}{n!}\frac{\G^2(\s)\,(-t)_n}{\G^2(\s+t-n)}\,\mathfrak{b}_{k_1}(\s)\,\mathfrak{b}_{k_2, n}(-\s-t)\, J^{-2k_1-2k_2-2n+2t}
\end{align}
where the $\mathfrak{b}$'s are defined via eqs. \ref{dd1},\ref{dd2},\ref{dd3} in terms of the generalized Bernoulli polynomials.
Here $\ell$ is the spin of the exchange operator. $J$ on the RHS is defined through $J^2=(\s+\ell)(\s+\ell-1)$. We will frequently choose $\s=s$ in our calculations. When $s=\D_\phi+\g_\ell/2$, $J$ is the conformal spin $J_{\g_\ell}$. Since we will be interested in leading order results in $\g_\ell$, we will frequently not differentiate $J_{\g_\ell}$ from $J_0$ and refer to both as $J$. To distinguish what we are using we will use the notation $Q^{2\s+\ell}_{\ell,0}(t)$ for explicit reference. 
In the $t$ channel we pick up contributions from the first line. As we see there is an inverse $\G^2(-t-n)$ factor which will cancel the contribution from the $\G^2(-t)$ factor in the measure alluded to above. This is the main reason why in the large spin approximation, in terms of residues, the two approaches have the same contributions in the crossed channel. This large spin form of the ${}_3F_2$ is the key player in our story. 

The direct channel in the two approaches have the following structures. In the usual approach, if we focus on the $u^{\D_\phi}, u^{\D_\phi}\log u$ terms, then the $v$ dependence is such that in Mellin space ($t$ dependent part) we have the schematic expansion
$$
\sum_{\D,\ell} u^{(\D-\ell)/2}  \tilde C_{\D,\ell} Q^{\D}_{\ell,0}(t)\,,
$$
in other words $\s=(\D-\ell)/2$. We have absorbed some $\Delta,\ell,\D_\phi$ dependence into the $\tilde C_{\D,\ell}$ which is related to the OPE coefficient (squared) to avoid cluttering the expressions. We can explicitly separate out the contribution from the double trace operators having dimensions $\D=2\D_\phi+\ell+2n+\g_\ell$ from the other operators in the sum and assume that the contribution from the other operators is suppressed through their OPE coefficients--{\it i.e.,} their contributions are at least $O(\gamma_\ell^2)$ in the above equation\footnote{When we claim a difference at this order, a logical possibility is that the contributions from these suppressed operators could in fact make the equations equivalent. However, it seems highly unlikely that these contributions would miraculously cancel the difference to all orders in $1/J$ . A more reasonable possibility is that fixing the polynomial ambiguity is what would resolve the difference. }. Then this piece in the above sum is an expansion in terms of $Q^{2\D_\phi+\ell+2n+\g_\ell}_{\ell,0}$. Now the explicit $\gamma_\ell$ dependence does not make this a suitable basis to expand in as $\gamma_\ell$ depends on which operator we are considering. We can easily find the connection coefficients relating $Q^{2\D_\phi+\ell+2n+\g_\ell}$ to $Q^{2\D_\phi+\ell+2n}$ and use the latter as the basis. The extra contribution due to this is proportional to $\g_\ell$ as an explicit calculation shows. Thus we have schematically for the log term
$$
\sum_\ell \tilde C_{n,\ell} \frac{\g_\ell}{2}u^{\D_\phi+n} \log u ~Q^{2\D_\phi+\ell+2n}_{\ell,0}(t)+O(\g_\ell^2)\,.
$$
For $u^{\D_\phi}\log u$, we set $n=0$.
Now in the new approach, we already know from \cite{RGAKKSAS} that the direct channel $u^{\D_\phi}\log u$ in the consistency equation is in fact in terms of $Q^{2\D_\phi+\ell}_{\ell,0}(t)$! Thus it should not come as too much of a surprise that at the level of the equations the usual and new approaches are in fact equivalent at least to $O(\g_\ell^2)$---of course there was apriori no guarantee that the explicit outside factors would agree to all orders in inverse spin. An explicit calculation confirms this fact. As we mentioned above, the equivalence is to all orders in inverse spin. Quite strikingly, the difference at $O(\g_\ell^2)$ can be attributed only to the direct channel since the crossed channels are the same in the large spin limit to all orders in $O(\g_\ell)$. This difference at $O(\g_\ell^2)$ can be attributed to the polynomial ambiguity in the Witten diagram basis and demanding the equivalence between the two approaches at this order will serve as a constraint for this polynomial ambiguity. In other words, to have a Witten diagram {\it basis} we cannot have an arbitrary polynomial ambiguity. Once we add such a polynomial to the Witten diagram basis in a crossing symmetric manner, the crossed channel large spin answer will not change since there only the physical poles are picked up.  In the epsilon expansion the ambiguity is at $O(\gamma_\ell^2)=O(\e^4)$. In \cite{RGAKKSAS, PDAKAS} we got the correct $O(\e^3)$ results not only because other unknown operators started contributing at the next order but also because the polynomial ambiguity kicks in at the next order \footnote{Further evidence for this exists from explicit calculations using known expressions \cite{charlotte,spinningads} for conserved higher spin currents \cite{ASnew}.}. Except for these peripheral observations, we will not have any further  insights to offer in this paper for the ambiguity.

The expansion in terms of the $Q$-polynomials leads to explicit all order expressions for the anomalous dimensions and OPE coefficients of the large spin double trace operators. Schematically these operators are 
$$
O_{n,\ell}\approx \phi \partial_{\mu_1}\cdots \partial_{\mu_\ell} \partial^{2n} \phi\,.
$$
The asymptotic expansion for the anomalous dimension of a large spin double trace operator $O_{0, \ell}$ due to an exchange of twist $\tau_m$ and spin $\ell_m$ operator in the crossed channel works out to be 
\be
\g_{0, \ell} \sim \sum_{i=0}^{\infty}\frac{\g^{(i)}_{0, \ell} }{J^{2i}}
\ee
with the $\g^{(i)}_{0, \ell} $ given explicitly in \eqref{anmdim2}, 
while the asymptotic expansion for the correction to the OPE coefficient can be expressed as,
\be
\delta C_{0,\ell} \sim \sum_{i=0}^{\infty}\frac{\delta C^{(i)}_{0, \ell} }{J^{2i}}
\ee
where $\delta C^{(i)}_{0, \ell}$ is given in \eqref{ope2}. Our expressions are in exact agreement with \cite{aldayzhiboedov} where comparison is possible \footnote{We can also compare with \cite{dsd} where explicit expressions are given by retaining contributions from primary operators.}. 
The explicit expressions have the generalized Bernoulli polynomials that enter the ${}_3F_2$ asymptotics as well as the Mack polynomials which contain the information of the exchanged operator. We will investigate some applications of these explicit expressions.

As we mentioned above, Witten diagrams as defined in Mellin space have polynomial ambiguities related to contact terms \cite{costa}. The equations for the anomalous dimension and OPE coefficients for the double trace operators differ at $O(\g_\ell^2)$ between the usual and new approaches. Thus, the inequivalence between the equations at the $O(\gamma_\ell^2)$ order can potentially be cured by introducing suitable contact terms. 
The differences between the usual and new bootstraps lead to the constraints in \eqref{diff} and \eqref{derdiff}. It remains to be seen if the systematic approach of \cite{spinningads,sleight} can be made use of to solve these constraints.

This paper is organized as follows. In section 2 we begin by recasting the conventional bootstrap equations in Mellin space and rederive some of the results in \cite{aldayzhiboedov}. In section 3 we derive an all order  expression for the anomalous dimension and OPE coefficient of the double trace operators at large spin. In section 4, we review the approach introduced in \cite{RGAKKSAS} based on tree level exchange Witten diagrams. We show that the leading order equations in the usual approach and the new approach are the same. We derive the difference at the next order and the constraint on the polynomial ambiguity in the Witten diagram basis. We discuss some applications of the all order expressions in section 5 and conclude with a discussion of future directions in section 6. There are a number of appendices which include useful calculational details needed in the paper. In particular, in appendix A we derive the asymptotic expansion of the ${}_3F_2$ entering the definition of the continuous Hahn polynomial in terms of the inverse conformal spin. As is the convention in many papers, the dimensionality $d$ of the CFT is frequently written as $2h$.

\section{Usual bootstrap in Mellin space}
%{\bf AS: uniformize notation--change $c_q$ to $\gamma_{(i)}$}
We begin by re-doing the analysis of  \cite{aldayzhiboedov} using Mellin space techniques. As will be clear this already leads to some simplifications. In the following section, we will use the $Q$-basis to simplify the equations which will enable us to get an all order  expression for the asymptotic expansion of the anomalous dimension. For simplicity and to make the comparison with \cite{aldayzhiboedov} explicit, in this section we will focus on extracting the anomalous dimension of $O_{0,\ell}$--in the next section we will work out the all order  expressions for both the anomalous dimensions and OPE coefficients.
The bootstrap equation, for identical scalar external operators of conformal dimension $\D_\phi$, in position space reads,
\be\label{boot}
G^{(s)}(u, v) \equiv\sum_{\D, \ell} C_{\D, \ell} \, G_{\D, \ell}(u, v) = \left(\frac{u}{v}\right)^{\D_{\f}}\, \sum_{\D, \ell} C_{\D, \ell} \, G_{\D, \ell}(v, u)\equiv G^{(t)}(u, v) 
\ee
where $C_{\D, \ell}$ is the square of the OPE coefficient of the exchange  operator with spin $\ell$, dimension $\D$ and $G_{\D, \ell}$ is the conformal block.

We will work  in the limit $v \ll u \ll 1$ following \cite{aldayzhiboedov}. In this limit the bootstrap equation reduces to,
\be\label{lhs2}
\sum_{\D, \ell} C_{\D, \ell} \,u^{\frac{\D-\ell}{2}} (f_{\D, \ell}( v) +O(u))= \left(\frac{u}{v}\right)^{\D_{\f}}\, \sum_{\D, \ell} C_{\D, \ell} \,v^{\frac{\D-\ell}{2}}\, (f_{\D, \ell}(v, u)+O(v))\,.
\ee

 We denote the left and right side of \eqref{lhs2} by $lhs$ and $rhs$
respectively in what follows. In the $lhs$ the $O(u)$ terms stand for descendant contributions which get mapped to the $O(v)$ terms in the crossed channel.

Following the analysis of \cite{Fitzpatrick:2012yx,Komargodski:2012ek} one can show that there must exist large spin ``double trace'' operators $O_{n, \ell} \sim \phi \partial_{\mu_1}\partial_{\mu_2}\cdots \partial_{\mu_{\ell}} \Box^{n}\phi$ ($\ell \gg 1$) on the $lhs$ to reproduce the leading behaviour on the $rhs$. Let us denote the conformal dimensions and the OPE coefficients of these large spin operators by,
\begin{align}\label{defss}
&\D= 2\D_{\f} +2n+\ell+ \g_{n, \ell} \nn
& C_{\D, \ell} = C_{n, \ell}(1+ \delta C_{n, \ell})
\end{align}
where $C_{n, \ell}$ is the leading order and $\delta C_{n, \ell}$ is the correction to the OPE coefficient. In this section, for simplicity, we will focus on the operators with $n=0$. However, this method can be easily generalised to non-zero $n$. Plugging (\ref{defss}) in \eqref{lhs2} the $lhs$ reduces to the following,
\be\label{sch2}
 \sum_{\ell} C_{0, \ell}(1+\delta C_{0,\ell})\, u^{\D_\f+\frac{\g_{0, \ell}}{2}}\,(f_{\ell}(v)+O(u))\,,
\ee
where,
\be\label{fdef}
f_{\ell}(v)= (1-v)^{\ell}\, _2F_1(\D_\f+\ell+\g_{0, \ell}/2, \D_\f+\ell+\g_{0, \ell}/2, 2\D_\f+2 \ell+\g_{0, \ell}; 1-v)\,.
\ee
 To compute the anomalous dimension $\g_{0, \ell}$ we need to focus on the coefficient of $u^{\D_\phi}\log u$  on the $lhs$. To leading order in $\g_{0, \ell}$ we have,
\begin{align}\label{lhs3}
lhs|_{\log u }%&=  \sum_{\ell} \, \half\,C_{0, \ell}\,{\g_{0, \ell}}\,f_{\ell}(v)\nn
& =\sum_{\ell} C_{0, \ell} \,\frac{\g_{0, \ell}}{2}  \,f^{0}_{\ell}(v)
\end{align}
where,
\be
f^{0}_{\ell}(v)= (1-v)^{\ell}\, _2F_1(\D_\f+\ell, \D_\f+\ell, 2\D_\f+2 \ell; 1-v)\,.
\ee
%we have set $\g_{0, \ell}=0$ in \eqref{fdef}.
We take $C_{0, \ell}$ to be the mean field OPE coefficients  (which can be derived from demanding that the disconnected piece in the crossed channel is reproduced by the direct channel in the bootstrap equation)
\be\label{mft}
C^{MFT}_{0, \ell} =\frac{2\,\G^2(\D_\f+\ell)\,\G(2\D_\f+\ell-1)}{\ell!\, \G^2(\D_\f)\,\G(2\D_\f+2\ell-1)}
\ee
%are the mean field OPE coefficients (which can be derived from demanding that the disconnected piece in the crossed channel is reproduced by the direct channel in the bootstrap equation).
%Now we will
and migrate to Mellin space. The Mellin transform of  \eqref{lhs3} with respect to the Mellin variable $t$ is given by, 
\begin{align}\label{lhslog}
lhs|_{\log u} = &\sum_{\ell}  \g_{0, \ell} \frac{\,\G(2\D_\f+\ell-1)}{\ell!\,\G^4(\D_{\f})}\,(2\D_\f+2\ell-1)\intl \frac{dt}{2\pi i}\,  v^t\,\G^2(\D_{\f}+t)\,\G^2(-t) \,\nn
&\times {}_3F_2\bigg[\begin{matrix} -\ell,\, 2\D_{\f}+\ell-1,\,\D_\f+t\\
	\ \ \D_{\f} \ \ , \   \D_{\f}
\end{matrix};1\bigg]\,.
\end{align}
Note that we have explicitly pulled out the Gamma functions for convenience.
Now we will take the large spin $\ell \gg 1$ limit of the above expression. We will use the approximation \eqref{Qasym} for the $_3F_2$ hypergeometric function in the variable $J_0$ where $J^2_0= (\ell+\D_\f)(\ell+\D_\f-1)$. At this stage we make a change of  variable from the usual spin $\ell$ to the conformal spin $J_0$ %\footnote{Strictly this is $J_0$ but since we are working to leading order in $\gamma_{0,\ell}$ we will keep calling it $J$.} 
. The $\ell$ dependent piece other than the $_3F_2$ can be approximated as,
\be\label{lhsgg}
(2\ell+2\D_\f-1)\,\frac{\G(2\D_\f+\ell-1)}{\G(\ell+1)} \sim {J_0}^{2\D_\f-1} \sum_{r, k_0=0}^{\infty}\binom{\half}{r} {4^{-r}} \mathfrak{b}_{k_0}(2-\D_\f) \,{J_0}^{-2k_0-2r}\,,
\ee
where we have used \eqref{gamma2} and $\mathfrak{b} $ is defined in \eqref{dd3} in terms of the generalized Bernoulli polynomials. %with $\a=\D_\f-1=-\b$.  
This shows that the choice of $C_{0, \ell}$ in \eqref{mft} naturally gives an expansion in terms of even powers of $1/J_0$. 

We can in fact do something better. Let us write the Mellin transform of the coefficient of $\log u$ term in \eqref{sch2} exactly as,
\begin{align}\label{lhslogg}
lhs|_{\log u} = &\sum_{\ell}  \half\,\g_{0, \ell}\, C_{0, \ell}\,\intl \frac{dt}{2\pi i}\,  v^t\,\G^2(\frac{\D-\ell}{2}+t)\,\G^2(-t) \, \frac{\G(\D+\ell)}{\G^2(\frac{\D+\ell}{2})\,\G^2(\frac{\D-\ell}{2})}\nn & \times {}_3F_2\bigg[\begin{matrix} -\ell,\, \D-1,\,\frac{\D-\ell}{2}+t\\
	\ \ \frac{\D-\ell}{2} \ \ , \   \frac{\D-\ell}{2}
\end{matrix};1\bigg]\,.
\end{align}
If we use $C_{0, \ell}$ to be 
\be\label{pmftnew}
C_{0, \ell}=\frac{2\,\G^2(\D_\f+\ell+\frac{\g_{0, \ell}}{2})\,\G(2\D_\f+\ell+\frac{\g_{0, \ell}}{2}-1)}{\G(\ell+1+\frac{\g_{0, \ell}}{2})\, \G^2(\D_\f)\,\G(2\D_\f+2\ell+\g_{0, \ell}-1)} c^{(j)}(\ell)
\ee
where $c^{(j)}(\ell)$ is an $\ell$ dependent factor we will determine shortly,
 %\left(1+\half\, \frac{\partial}{\partial \ell}\g_{0, \ell}\right)
%This reason for this redefinition will become clear in a bit. 
and substitute $\D=2\D_\f+\ell+\g_{0, \ell}$ in \eqref{lhslogg} we obtain the following,
%begin{align}
%C_{0, \ell}\,\frac{\G(\D+\ell)}{\G^2(\frac{\D+\ell}{2})\,\G^2(\frac{\D-\ell}{2})}&=\frac{2\,\G(2\D_\f+\ell+\frac{\g_{0, \ell}}{2}-1)}{\G(\ell+1+\frac{\g_{0, \ell}}{2})\, \G^2(\D_\f)\,\G(2\D_\f+2\ell+\g_{0, \ell}-1)}\times \frac{\G(2\D_\f+2\ell+\g_{0, \ell})}{\G^2(\D_\f+\frac{\g_{0,\ell}}{2})}
%\end{align}
\begin{align}\label{lhs4}
lhs|_{\log u} = &\sum_{\ell} c^{(j)}(\ell) \g_{0, \ell} \intl \frac{dt}{2\pi i}\,  v^t\,\frac{\G^2(\D_\f+\frac{\g_{0,\ell}}{2}+t)}{\G^2(\D_\f+\frac{\g_{0,\ell}}{2})}\,\G^2(-t) \,\nn
&\times \frac{\G(2\D_\f+\ell+\frac{\g_{0, \ell}}{2}-1)\,(2\D_\f+2\ell+\g_{0, \ell}-1)}{\G(\ell+1+\frac{\g_{0, \ell}}{2})\, \G^2(\D_\f)} {}_3F_2\bigg[\begin{matrix} -\ell,\, \D-1,\,\frac{\D-\ell}{2}+t\\
	\ \ \frac{\D-\ell}{2} \ \ , \   \frac{\D-\ell}{2}
\end{matrix};1\bigg]\,.
\end{align}
Quite nicely, if we take the large spin limit of the factors outside ${}_3F_2$ in \eqref{lhs4} using  \eqref{gamma2}, we get,
\be\label{lhsg2}
(2\D_\f+2\ell+\g_{0, \ell}-1)\,\frac{\G(2\D_\f+\ell+\frac{\g_{0, \ell}}{2}-1)}{\G(\ell+1+\frac{\g_{0, \ell}}{2})} \sim {J}^{2\D_\f-1} \sum_{r, k_0=0}^{\infty}\binom{\half}{r} {4^{-r}}\,\mathfrak{b}_{k_0}(2-\D_\f) \,{J}^{-2k_0-2r}
\ee
where $J^2= (\D_\f+\ell+\g_{0, \ell}/2)(\D_\f+\ell+\g_{0, \ell}/2-1)$ and $\mathfrak{b} $ is defined in \eqref{dd3}. %$\a=\D_\f-1=-\b$. %Comparing \eqref{lhsgg} and \eqref{lhsg2} 
We see that the  redefinition of $C_{0, \ell}$ in \eqref{pmftnew} allows us to write the $lhs$ in terms of $J$. 

At this point, we assume that $\g_{0, \ell}$ has the following expansion in $J$,
\be
\g_{0, \ell} \sim \sum_{i=0}^{\infty} \frac{\g^{(i)}_{0, \ell}}{J^{\tau_m+2i}}\,.
\ee
Here $\tau_m$ is the twist $\tau=\D-\ell$ of an operator in the crossed channel. While usually this is the minimal twist operator, %if there are no twist degeneracies, i.e., $\tau_1+k_1\neq \tau_2+k_2$ for any two operators with twists $\tau_1$ and $\tau_2$ with $k_1,k_2\in \mathbb{Z}$, then 
we can consider any operator in the crossed channel and ask how its contribution is reproduced by the double trace operators in the direct channel. If there are twist degeneracies, i.e., $\tau_1+k_1\neq \tau_2+k_2$ for any two (or more) operators with twists $\tau_1$ and $\tau_2$ with $k_1,k_2\in \mathbb{Z}$, then their contributions to the relevant power of $1/J$ have to be added. For a finite number of such contributions, we can simply add up the effects but if infinite number of operators contribute, for instance the large spin double trace operators themselves have $\tau\approx 2\D_\phi$, then summing up their effects may require more work.

Keeping the above issues in mind, we will take the large $J$ limit of \eqref{lhs4}. The $J$ sum can be done using the steps below.
\begin{align}
& \sum_{\ell=\ell_0}^{\infty} \frac{c^{(j)}(\ell)}{{((\ell+a+\frac{\g_{0, \ell}}{2})(\ell+a+\frac{\g_{0, \ell}}{2}-1))}^{\a}}
%&=\sum_{\ell=\ell_0}^{\infty} \frac{1}{{(\ell+a)}^{2\a}}\left(1-\frac{1}{\ell+a}\right)^{-\a}\nn
%&= \sum_{\ell=\ell_0}^{\infty} \frac{1}{{(\ell+a)}^{2\a}} \sum_{k=0}^{\infty} (-1)^k \binom{-\a}{k}\frac{1}{{(\ell+a)}^{k}} \nn
=\sum_{k=0}^{\infty} (-1)^k \binom{-\a}{k}\sum_{\ell=\ell_0}^{\infty}\frac{c^{(j)}(\ell)}{{(\ell+a+\frac{\g_{0, \ell}}{2})}^{2\a+k}}\,.
\end{align}
This sum can be converted to an integral using the standard Euler-Maclaurin summation formula. Since we are summing over only even spins $\ell$ we have to include a factor of $\half$ when we replace the sum by an integral,
\be\label{sum2}
%\sum_{k=0}^{\infty} (-1)^k \binom{-\a}{k}
\sum_{\ell=\ell_0}^{\infty}\frac{c^{(j)}(\ell)}{{(\ell+a+\frac{\g_{0, \ell}}{2})}^{2\a+k}}= \half \int_{\ell_0}^{\infty} d\ell \frac{ c^{(j)}(\ell)}{{(\ell+a+\frac{\g_{0, \ell}}{2})}^{2\a+k}}+\cdots
\ee
where $\cdots$ represent remainder terms which will not produce a $t$ pole.
Now we make a change of variable from $\ell$ to $\tilde{\ell}= \ell+\frac{\g_{0, \ell}}{2}$ and choose 
\be\label{cj}
c^{(j)}(\ell)\bigg(1+\half\,\frac{\partial}{\partial \ell}\g_{0, \ell}\bigg)^{-1}=1\,,
\ee
then  \eqref{sum2} reads,
\be\label{gg2}
\half \int_{\ell_0+\frac{\g_{0, \ell}}{2}}^{\infty} \frac{d{\tilde{\ell}}}{{(\tilde{\ell}+a)}^{2\a+k}}= \frac{1}{2\,(2\a-1+k)} + {\rm regular~in~}t\,.
\ee
%where we have used the fact that there is a pole at $$
%\begin{align}\label{Jsum2}
%\sum_{\ell=\ell_0}^{\infty} \frac{1}{{((\ell+a)(\ell+a-1))}^{\a}}
%&= \sum_{k=0}^{\infty} (-1)^k \binom{-\a}{k} \frac{1}{2\,(2\a-1+k)} +\, \rm{regular}\nn
%&= 2^{2\a-3}\,\frac{\G(1-\a)\,\G(\a-\half)}{\sqrt{\pi}} + \, \rm{regular}
%\end{align}
Hence we are left with \footnote{It is easy to see that to generate higher order poles at $2\alpha-1+k=0$ we would need to introduce powers of $\log (\tilde \ell+a)$ in the integrand in eq.\ref{gg2}.},
\be\label{Jsum}
\sum_{\ell=\ell_0}^{\infty} \frac{1}{{((\ell+a+\frac{\g_{0, \ell}}{2})(\ell+a+\frac{\g_{0, \ell}}{2}-1))}^{\a}}= 2^{2\a-3}\,\frac{\G(1-\a)\,\G(\a-\half)}{\sqrt{\pi}} + \, {\rm regular}\,.
\ee
This justifies the normalisation in \eqref{pmftnew} with \eqref{cj} and is exactly the same as in \cite{aldayzhiboedov}. 
This will introduce poles at specific values of $t$. The $t$ integral in \eqref{lhs4} can be evaluated using the residue theorem. Finally, 
%Using the steps outlined in Appendix \ref{Mellinconf}{\bf which appendix? what steps? explain more in words} 
for the large $J$ approximation  we get the following,

\begin{align}\label{lhs1}
&lhs|_{\log u}  = -{\widehat{\sum}} v^{-\Delta_\phi +\hat{k}+\frac{{\tau_m}}{2}}\,\g^{(q)}_{0, \ell}\, {\G_s}
\end{align}
where ${\widehat{\sum}}$ and $\G_s$ are defined in \eqref{hatsum}.

Now let us focus on the piece proportional  to $\log u$ on the $rhs$ of the bootstrap equation \eqref{lhs2} in the small $u$ limit. We assume that there exists a minimal twist operator in the spectrum \cite{Fitzpatrick:2012yx, Komargodski:2012ek} and consider (for explicitness)  the exchange of minimal twist scalar of dimension $\D_m$ in the $t$ channel. The coefficient of $\log u$  on the $rhs$ is given by, %For simplicity let us first consider the scalar exchange.
\begin{align}\label{rhs}
rhs &= -C_m v^{\frac{\tau_m}{2}-\D_\f} \log u \, \frac{\G({\D_m})}{\G^2(\frac{\D_m}{2})}\, _2F_1\left(\frac{\D_m}{2}, \frac{\D_m}{2}, 1-\frac{d}{2}+\D_m,; v \right)\nn
&=-\,C_m \, \frac{\G({\D_m})}{\G^2(\frac{\D_m}{2})}\, \sum_{\a=0}^{\infty} \frac{1}{\a!} \frac{\left(\frac{\D_m}{2}\right)^2_{\a}}{(1-\frac{d}{2}+\D_m)_{\a}}\, v^{\frac{\tau_m}{2}-\D_\f+\a}\log u
\end{align}
where $C_m$ is the OPE coefficient of the minimal twist scalar.

Now we are in a position to compare the powers of $v$ from both side of the bootstrap equation. Matching the powers of $v^{{\frac{\tau_m}{2}-\D_\f+\hat{k}}}$ from \eqref{lhs1} and \eqref{rhs} we get the following  relation, %{\bf AS: where $\hat d$'s are....., Should the $\sum$ but $\widehat\sum$?.},
%\begin{align}
%&{{\sum_{new}}}\g^{(q)}_{0, \ell} \frac{(-1)^{1+n+p}\,4^{-p-r}}{2\,\sqrt{\pi }\, n!\, p!\, \G^2(\D_\f)\,\G^2(-n+\hat{k}+\frac{\tau_m}{2})}\binom{\half}{r}\,\G(\half+p)\nn & \mathfrak{b}_{k_1}(\D_\f)\,\mathfrak{b}_{k_0}(2-\D_\f) \mathfrak{b}_{k_2, n}(-k_0-r-q-n-k_1-k_2-p-\frac{\tau_m}{2})\nn & \times \G(-\hat{k}+\D_\f-\frac{\tau_m}{2})\,\G(n-\hat{k}+\D_\f-\frac{\tau_m}{2})\,
%\G^2(\hat{k}+\frac{\tau_m}{2})\nn
%&=  -\,C_m  \, \frac{\G({\D_m})}{\G^2(\frac{\D_m}{2})}\,  \frac{1}{\hat{k}!} \frac{\left(\frac{\D_m}{2}\right)^2_{\hat{k}}}{(1-\frac{d}{2}+\D_m)_{\hat{k}}}
%\end{align}
\begin{align}
&{\sum_{new}}\g^{(q)}_{0, \ell}\,\frac{(-1)^{n+p}\,4^{-p-r}}{2\,\sqrt{\pi }\, n!\, p!\, \G^2(\D_\f)\,\G^2(-n+\hat{k}+\frac{\tau_m}{2})}\binom{\half}{r}\G(\half+p)
 \G(-\hat{k}+\D_\f-\frac{\tau_m}{2})\,\G(n-\hat{k}+\D_\f-\frac{\tau_m}{2})\nn & \times \mathfrak{b}_{k_1}(\D_\f)\,\mathfrak{b}_{k_0}(2-\D_\f)\,\mathfrak{b}_{\hat{k}-k_0-k_1-n-p-q-r, n}(-\hat{k}-\frac{\tau_m}{2})\G^2(\hat{k}+\frac{\tau_m}{2})\nn
%\nn & \times \G(\half+p)
% \G(-\hat{k}+\D_\f-\frac{\tau_m}{2})\,\G(n-\hat{k}+\D_\f-\frac{\tau_m}{2})\,
%\G(\hat{k}+\frac{\tau_m}{2})^2\nn
&=  -\,C_m  \, \frac{\G({\D_m})}{\G^2(\frac{\D_m}{2})}\,  \frac{1}{\hat{k}!} \frac{\left(\frac{\D_m}{2}\right)^2_{\hat{k}}}{(1-\frac{d}{2}+\D_m)_{\hat{k}}}\,.
\end{align}
where the sum is defined in \eqref{newsum}.
This gives a recursion relation for  $\g^{(q)}_{0, \ell}$. This recursion relation can be solved iteratively for  any $q$. Let us list the first few values  of $\g^{(q)}_{0, \ell}$, 
\begin{align}\label{gamma}
 \g^{(0)}_{0, \ell} &=- \frac{2\,\G(\D_m)\,\G^2(\D_\f)}{\G^2(\frac{\D_m}{2})\,\G^2(\D_\f-\frac{\D_m}{2})} \,C_m, \nn
 \g^{(1)}_{0, \ell}
 % &=- \bigg(-4 h \left(\Delta _m \left(-3 \Delta _{\phi }+\Delta _m+3\right)+2\right)+\Delta _m \left(-12 \left(\Delta _{\phi }-1\right) \Delta _{\phi }+\Delta _m \left(\Delta _m+4\right)+8\right)+8\bigg)\,C_m\nn
&=  \frac{\Gamma^2\left(\Delta _{\phi }\right)\, \Gamma \left(\Delta _m+1\right)}{24 \left(-h+\Delta _m+1\right) \Gamma^2\left(\frac{\Delta _m}{2}\right){} \Gamma^2\left(\Delta _{\phi }-\frac{\Delta _m}{2}\right)}\nn
& \times \bigg(-4 h \left(\Delta _m \left(-3 \Delta _{\phi }+\Delta _m+3\right)+2\right)+\Delta _m \left(-12 \left(\Delta _{\phi }-1\right) \Delta _{\phi }+\Delta _m \left(\Delta _m+4\right)+8\right)+8\bigg)\,C_m,\nn
%\end{align}
%\begin{align}\label{gammaa2}
\g^{(2)}_{0, \ell} &= -\frac{ 2^{\Delta _m-9} \Delta _m \left(\Delta _m+2\right) \Gamma^2\left(\Delta _{\phi }\right)\, \Gamma \left(\frac{1}{2} \left(\Delta _m+1\right)\right)}{45 \sqrt{\pi } \left(-h+\Delta _m+1\right) \left(-h+\Delta _m+2\right) \Gamma \left(\frac{\Delta _m}{2}\right) \Gamma^2\left(\Delta _{\phi }-\frac{\Delta _m}{2}\right)}\nn
& \times \bigg((\Delta _m+4)\left(16 h^2 \left(\Delta _m \left(\Delta _m \left(5 \Delta _m+27\right)+43\right)+6\right)-8 h \left(\Delta _m+1\right) \left(\Delta _m \left(\Delta _m \left(5 \Delta _m+28\right)+66\right)+36\right)\right. \nn & \left.+\left(\Delta _m+2\right) \left(\Delta _m \left(\Delta _m \left(\Delta _m \left(5 \Delta _m+32\right)+100\right)+64\right)+96\right)\right)\nn &- 120 (h+1) \Delta _{\phi } \Delta _m \left(h \left(4 \Delta _m^2+22 \Delta _m+36\right)-\left(\Delta _m+2\right) \left(\Delta _m \left(\Delta _m+4\right)+12\right)\right)\nn
& -120 \Delta _{\phi }^2 \Delta _m \left(-6 h^2 \left(\Delta _m+2\right)-2 h \left(\Delta _m+6\right) \left(2 \Delta _m+5\right)+\left(\Delta _m+2\right) \left(\Delta _m \left(\Delta _m+4\right)+6\right)\right)\nn
& + 720 \Delta _{\phi }^4 \Delta _m \left(\Delta _m+2\right)-1440 (h+1) \Delta _{\phi }^3 \Delta _m \left(\Delta _m+2\right)\bigg)C_m\,.
\end{align}
These results exactly match with \cite{aldayzhiboedov}. The generalisation of this recursion relation for general spin exchange is straightforward. In the next section we will present the explicit formulae. The Mellin space translation (in the direct manner as in this section) yields at least one algebraic simplification compared to  \cite{aldayzhiboedov}.
In \cite{aldayzhiboedov} there are two main steps to compute the anomalous dimension in the large spin limit. There are two recursion relations that one needs to solve in order to get the coefficients $\g^{(q)}_{0, \ell}$. In the first recursion relation, an all order  solution is not known, as a result of which an all order expression for $\g^{(q)}_{0, \ell}$ cannot be obtained this way ( recently in \cite{Aharony:2016dwx}  progress has been made for integer  external operator dimension ).  This makes the computation a bit more involved although it can be automated on a computer. In $d=4$ the twist conformal block approach leads to some simplification \cite{holorecon} for certain exchanges. However, in Mellin space the calculation is simpler and doable in any dimensions for any exchange. In the next section, we will expand the equations in the $Q$-basis and get an all order  expression for the anomalous dimensions and OPE coefficients straightaway.

\section{Explicit expressions}
In this section, we will expand the $t$-dependence in the bootstrap condition in the previous section in terms of the $Q$-basis. This will lead to an all order  expression for the anomalous dimension and OPE coefficient in terms of known mathematical functions which include the generalized Bernoulli polynomials and the Mack polynomials.
\subsection{Anomalous dimensions}
We begin  with the derivation of the anomalous dimension for $O_{0,\ell}$.
 % {\bf AS: This section has to be completed--for $n=0$ we need to give the closed form expression for the anomalous dimensions.}
  
% {\bf AS: factors of $2\pi i$ need to re-instated in the appropriate places}
\subsubsection{$s$ channel}
 Let us start with  the Mellin transform of the $s$ channel conformal block  w.r.t. the Mellin variable $t$ which reads,
\begin{align}\label{qs4}
G^{(s)}(u, v)&=\sum_{\D, \ell} C_{0, \ell} \,u^{\frac{\D-\ell}{2}}\,\intl \frac{dt}{2\pi i} v^t\,\G^2(\frac{\D-\ell}{2}+t)\,\G^2(-t)\,\frac{\G(\D+\ell)}{\G^2(\frac{\D+\ell}{2})\G^2(\frac{\D-\ell}{2})}\nn & \times {}_3F_2\bigg[\begin{matrix} -\ell,\, \D-1,\,\frac{\D-\ell}{2}+t\\
	\ \ \frac{\D-\ell}{2} \ \ , \   \frac{\D-\ell}{2}
\end{matrix};1\bigg]\,.
\end{align}
We introduce the Mellin variable $s$ conjugate to $u$ and write \eqref{qs4} as follows,
%\eqref{lhslogg}
\begin{align}\label{qsm}
G^{(s)}(u, v)&=-\sum_{\D, \ell} C_{0, \ell} \intl \frac{ds}{2\pi i}\frac{dt}{2\pi i} u^s\,v^t\,\G^2(s+t)\,\G^2(-t)\,\frac{1}{s-\frac{\D-\ell}{2}}\frac{\G(\D+\ell)}{\G^2(\frac{\D+\ell}{2})\G^2(\frac{\D-\ell}{2})}\nn & \times \frac{(2s+\ell-1)_{\ell}}{2^{\ell}\,((s)_{\ell})^2}\,Q^{2s+\ell}_{\ell,0}(t)
\end{align}
where we have replaced the $_3F_2$ in \eqref{qs4} by the continuous Hahn polynomial  using \eqref{Qdefn}.
Note that \eqref{qsm} evaluated at the pole $s=\frac{\D-\ell}{2}$ (which comes from $\G(\frac{\D-\ell}{2}-s)$)exactly reproduces \eqref{qs4}.
% In \eqref{qsm} when we consider the large spin sum, we only sum over even spins. As a result $\sum_\ell \rightarrow \frac{1}{2}\int d\ell$ gives a factor of $\half$. %when moved to the {\it rhs}
Now we write the  pole piece in \eqref{qsm} as follows,
\begin{align}\label{qsnew}
\frac{1}{s-\frac{\D-\ell}{2}}\rightarrow \G^2(\D_\f-s)\bigg(\half\g_{0,\ell}+(\D_\f-s)\,(\g_{0, \ell}\,\g_E-1)\bigg)+O(\g^2_{0, \ell})
\end{align}
where we have replaced $\D$ by $2\D_\f+\ell+\g_{0, \ell}$ and expanded in small $\g_{0, \ell}$.  Using \eqref{qsnew} exactly reproduces $u^{\D_\f}$ and $u^{\D_\f}\log u$ term which arise from \eqref{qsm}. The form of the $rhs$ of \eqref{qsnew}  will facilitate a comparison with the new approach in the next section--the additonal poles that we introduce through the form of the {\it rhs} will not play any role as we will only be interested in $s=\D_\phi$. We obtain the following expession for the $s$ channel,
\begin{align}\label{schlog}
G^{(s)}(u, v)&=-\sum_{ \ell}  \intl \frac{ds}{2\pi i}\frac{dt}{2\pi i} u^s\,v^t\,\G^2(s+t)\,\G^2(-t)\,\G^2(\D_\f-s)\,\mathfrak{q}^{(s)}_{0, \ell}(s)\,Q^{2s+\ell}_{\ell,0}(t)
\end{align}
%Now we replace the sum by an integral
where,
\be\label{qsusual2}
\mathfrak{q}^{(s)}_{0, \ell}(s)=C_{0, \ell} \frac{2^{-\ell } \Gamma^2 (s) \Gamma (2 s+2 \ell -1) \Gamma (\ell +\Delta )}{\Gamma^2 (s+\ell ) \Gamma (2 s+\ell -1) \Gamma^2 \left(\frac{\Delta -\ell }{2}\right) \Gamma^2 \left(\frac{\ell +\Delta }{2}\right)} \bigg(\frac{\gamma_{0, \ell} }{2}+(\gamma_{0, \ell}  \gamma_E -1) \left(\Delta _{\phi }-s\right)\bigg)\,.
\ee
Since we are interested in the coefficient of $\log u$ from \eqref{schlog} we have to evaluate the residue at $s= \D_\f$ from the double pole of $\G^2(\D_\f-s)$. The residue at $s=\D_\f$ is given by \footnote{We throw away derivatives acting on $\G^2(s+t) Q_{\ell,0}^{2s+\ell}(t)$ in calculating the residue as in the final equation after including the crossed channel, the coefficient in front will be $\mathfrak{q}^{(s)}_{0, \ell}+2\,\mathfrak{q}^{(t)}_{0, \ell}$ which is zero. This is identical to what happens in the new approach \cite{RGAKKSAS}.} ,
\begin{align}\label{schlog2}
G^{(s)}(u, v)&=-\sum_{ \ell}  u^{\D_\f}\,\log u \int \frac{dt}{2\pi i} \,v^t\,\G^2(\D_\f+t)\,\G^2(-t)\,\mathfrak{q}^{(s)}_{0, \ell}(\D_\f)\,Q^{2\D_\f+\ell}_{\ell,0}(t)\nn &- \sum_{ \ell}  u^{\D_\f}\,\int \frac{dt}{2\pi i} \,v^t\,\G^2(\D_\f+t)\,\G^2(-t)\,\mathfrak{q'}^{(s)}_{0, \ell}(\D_\f)\,Q^{2\D_\f+\ell}_{\ell,0}(t)\,.
\end{align}
%Now the quadratic term {\bf AS:????}
%In the large $\ell' \gg 1$ limit this reduces to {\bf AS What is the necessity of the following? We should present the J dependent expression straightaway},
%\be
%\mathfrak{q}^{(s)}_{0, \ell'}|_{O(\g)} \sim \frac{e^{\ell' } \left(\frac{1}{\ell' }\right)^{\ell' } 2^{2 \Delta _{\phi }+\ell' -\frac{3}{2}}}{\pi  \Gamma \left(\Delta _{\phi }\right){}^2}\,\g_{0, \ell'} + O(1/\ell')
%\ee
\subsubsection{$t$ channel }
Let us now expand  the $t$ channel  in $Q$-basis. We begin with the Mellin transform of the $\Delta_m,\ell_m$ operator contribution to the $t$ channel,
\begin{align}\label{tch2}
G_{\D_m,\ell_m}^{(t)}(u, v)&= c_{\D_m, \ell_m} \, \intl \frac{ds}{2\pi i}\, \frac{dt}{2\pi i}\,u^s\,v^t\,  \G^2(s+t)\,\G^2(\D_\f-s)\,\G^2(-t)\, B_{\D_m, \ell_m}(t+\D_\f, s-\D_\f)
\end{align}
where,
\be
B_{\D, \ell}(t+\D_\f, s-\D_\f)= \frac{\G(\frac{\D-\ell}{2}-\D_\f-t)\,\G(\frac{2h-\D-\ell}{2}-\D_\f-t)}{\G^2(-t)}\,P_{\D-h, \ell}(t+\D_\f, s-\D_\f)\,,
\ee
and where $c_{\D_m, \ell_m}= C_{\D_m, \ell_m} \mathcal{N}_{\D_m, \ell_m}$ with $\mathcal{N}_{\D,\ell}$ given in \eqref{normu}. We will use the shorthand $C_m=C_{\D_m, \ell_m}$.
As in the direct channel, we suppress the projection factor.
Since we are interested in the coefficient of $\log u$ we have to consider the double pole at $s=\D_\f$. The residue at this pole is given by,
\begin{align}
G_{\D_m,\ell_m}^{(t)}(u, v) &= c_{\D_m, \ell_m}\,u^{\D_\f}\,\log u\, \intl \frac{dt}{2\pi i}\,v^t\,  \G^2(\D_\f+t)\,
 {\G(\frac{\D_m-\ell_m}{2}-\D_\f-t)}\nn & \times \G(\frac{2h-\D_m-\ell_m}{2}-\D_\f-t)\, P_{\D_m-h, \ell_m}(t+\D_\f, 0)\,.
\end{align}
In terms of the $Q$-basis, this becomes
\be \label{texp}
\sum_{\ell=0}^{\infty}u^{\D_\f}\,\log u\,\intl\frac{dt}{2\pi i}\, v^t\,\G^2(\D_\f+t)\,\G^2(-t)\, \mathfrak{q}^{(t)}_{0, \ell|\ell_m}\,Q^{2\D_\f+\ell}_{\ell, 0}(t)
\ee
such that,
\begin{align}\label{qt3}
\mathfrak{q}^{(t)}_{0, \ell|\ell_m} &=\intl \frac{dt}{2\pi i}\, \G^2(\D_\f+t)\,\kappa_{\ell}(\D_\f)^{-1}\,Q^{2\D_\f+\ell}_{\ell, 0}(t)\nn & \times c_{\D_m, \ell_m}\,{\G(\frac{\D_m-\ell_m}{2}-\D_\f-t)\,\G(\frac{2h-\D_m-\ell_m}{2}-\D_\f-t)}\,P_{\D_m-h, \ell_m}(t+\D_\f, 0)\,.
\end{align}
%Notice the extra factor of 2 in this expression. 
%the following must be true that a solution to the crossing symmetry condition must satisfy
%\be
%\sum_{\ell'= odd} a_{\ell'}(s)\,Q^{2s+\ell'}_{\ell',0}(t) = \sum_{\ell'= even} a_{\ell'}(s)\,Q^{2s+\ell'}_{\ell',0}(t)
%\ee
%which gives
%\be
%f(s, t) =\sum_{\ell'=0}^{\infty} 2\,a_{\ell'}(s)\,Q^{2s+\ell'}_{\ell',0}(t)\,.
%\ee
%The reason for this extra factor is that in the direct channel, when we consider the large spin sum, we only summed over even spins. As a result $\sum_\ell \rightarrow \frac{1}{2}\int d\ell$ gave this factor of 2 when moved to the {\it rhs}. 
An important point to note at this stage is that each $t$-channel block, as written above, on its own does not have the symmetry $t\rightarrow -s-t$ which exists in the $s$-channel. Thus, for a specific exchange in the $t$-channel, before summing over the spectrum in \eqref{qt3}, we could have odd spin elements in the basis (which naively get equated to zero as there are no odd spin elements in the direct channel\footnote{This would definitely be inconsistent!}) but this conclusion is incorrect as the full $t$-channel does have the $t\rightarrow -s-t$ symmetry which precludes odd spin elements. Put differently, the sum over spectrum in the $s$-channel should know about the crossed channel physical poles; this involves both $t$ and $u$ channel poles as is evident from \eqref{3f2a}. In position space we pick up one or the other depending on how we are closing the contour (see e.g. appendix C). However, in Mellin space the existence of both sets of poles gives a factor of 2 when decomposing in the $Q$-basis as in the new approach.  
Hence, in the final step there will be an extra factor of 2 in the crossed channel --- that this conclusion must be correct can also be verified by comparing the expressions for anomalous dimensions obtained from position space in section\footnote{There the origin of this factor of 2 was due to $\sum_\ell\rightarrow 1/2\int d\ell$.} 2--- giving
\be\label{bootus1}
\mathfrak{q}^{(s)}_{0, \ell}+2\,\mathfrak{q}^{(t)}_{0, \ell|\ell_m}+\cdots=0\,,
\ee
where $\cdots$ denote contribution from other operators in the crossed channel and
where $\mathfrak{q}^{(t)}_{0, \ell|\ell_m}$ is obtained from \eqref{qt3} by calculating the residue at $t= \frac{\D_m-\ell_m}{2}-\D_\f+q$  for  $q=0, 1, 2, \cdots$:

\begin{align}\label{rhsq2}
\mathfrak{q}^{(t)}_{0, \ell|\ell_m}
% &=\sum_{q=0}^{\infty}\bigg( \frac{(-1)^{q}}{q!}\, \G^2(\D_\f+t)\,\kappa^{-1}_{\ell}(\D_\f)\,Q^{2\D_\f+\ell}_{\ell, 0}(t)\nn & \times c_{\D_m, \ell_m}\,{\G(\frac{2h-\D_m-\ell_m}{2}-\D_\f-t)}{P}_{\D_m-h, \ell_m}(t+\D_\f, 0)\bigg)_{t= \frac{\D_m-\ell_m}{2}-\D_\f+q}\nn
& = \sum_{q=0}^{\infty}\bigg( \beta_{\ell}\,\frac{(-1)^{q}}{q!}\, \G^2(\D_\f+t)\, {}_3F_2\bigg[\begin{matrix} -\ell,\, 2\D_{\f}+\ell-1,\,\D_\f+t\\
	\ \ \D_{\f} \ \ , \   \D_{\f}
\end{matrix};1\bigg]\nn & \times c_{\D_m, \ell_m}\,{\G(\frac{2h-\D_m-\ell_m}{2}-\D_\f-t)}{P}_{\D_m-h, \ell_m}(t+\D_\f, 0)\bigg)_{t= \frac{\D_m-\ell_m}{2}-\D_\f+q}
\end{align}
where,
\be\label{betaell}
\beta_{\ell} = \bigg(\kappa^{-1}_{\ell}(\D_\f)\frac{2^{\ell}\,((\D_\f)_{\ell})^2}{(2\D_\f+\ell-1)_{\ell}}\bigg)
\ee
and we have used \eqref{Qdefn}.
As in \cite{RGAKKSAS}, we will use the notation $ \mathfrak{q}^{(t)}_{0, \ell|\ell_m}$ to denote a particular $\Delta_m,\ell_m$ exchange in the crossed channel which contributes to the $Q^{2\D_\phi+\ell}_{\ell,0}(t)$ basis element. From here on we will focus on this single contribution of spin $\ell_m$ and twist $\tau_m\equiv \D_m-\ell_m$ in the crossed channel so that the effective equation we are solving is 
\be\label{bootus}
\mathfrak{q}^{(s)}_{0, \ell}+2\,\mathfrak{q}^{(t)}_{0, \ell|\ell_m}=0\,.
\ee
We now multiply both sides of \eqref{bootus} with $\beta^{-1}_{\ell}$ so that
%Note that that after taking the $\b_{\ell}$ to the $s$ channel 
we are left with only $\g_{0, \ell}$ in the $s$ channel whereas the $\ell$ dependence in the $t$ channel can only come from the $_3F_2$ hypergeometric function in \eqref{rhsq2}. This can be expanded in the large $J$ limit using \eqref{Qasym} which involves only even powers of $J$ through the $J^{-2s-2t}$ series whose contribution is picked up as we are closing the contour on the right. This makes it evident that $\g_{0, \ell}$ in the large spin limit can be expanded in a series involving only even powers of conformal spin \cite{Alday:2015eya}. 
%{\textbf{checked}}
We assume that $\g_{0, \ell}$ admits the following expansion in $J$,
\be
\g_{0, \ell} \sim \sum_{i=0}^{\infty} \frac{\g^{(i)}_{0,\ell}}{J^{2i}}\,,
\ee
where $\g^{(i)}_{0,\ell}$-s are the unknowns we want to solve for.
In order to extract the coefficient of a particular power $J^{-2i}$ in the $t$ channel, we will denote $k_1+k_2+q+n= i$  and replace $k_2$ by $i-k_1-n-q$. We are now in a position to compare the coefficient of $J^{-2i}$ from both sides of the bootstrap equation \eqref{bootus}. 
This  results in the following expression for $\g^{(i)}_{0, \ell}$,
%We assume the following expansion,
%\be
%\g_{0, \ell} =\sum_{i=0}^{\infty} \frac{\g^{(i)}_{0,\ell}}{J^{2i}}\,.
%\ee
%We take the large spin limit of the $_3F_2$ hypergeometric function using \eqref{Qasym}. and multiply both sides of \eqref{bootus} with $\beta^{-1}_{\ell'}$ which directly leads to the following expression for the anomalous dimension,
\begin{align}\label{anmdim2} 
\g^{(i)}_{0, \ell} &=-\frac{1}{J^{\tau_m }} \,C_m\sum_{q=0}^{i}\sum_{n=0}^{i-q}\sum_{k_1=0}^{i-q-n}(-1)^{n}\,2^{1+\ell_m}\,\mathfrak{b}_{k_1}(\D_\f)\,\mathfrak{b}_{i-k_1-n-q, n}(\frac{\tau_m}{2}-\D_\f+q)\,\, \hat{P}_{\tau_m+\ell_m-h, \ell_m}(q+\tau_m/2,0)\nn
& \times \frac{\left(\tau _m+2 \ell _m-1\right)\,\Gamma \left(-h+\ell _m+\tau _m+1\right)\,\Gamma^2\left(2 \ell _m+\tau _m-1\right)\,\Gamma^2\left(\Delta _{\phi }\right)\, \Gamma \left(q+\frac{\tau _m}{2}\right) \Gamma \left(n+q+\frac{\tau _m}{2}\right) }{ n!\, q! \,\G(1-h+q+\ell_m+\tau_m)\,\Gamma^4\left(\ell _m+\frac{\tau _m}{2}\right)\, \Gamma \left(\ell _m+\tau _m-1\right) \Gamma^2\left(-n-q+\Delta _{\phi }-\frac{\tau _m}{2}\right)}
%& \times \left(\tau _m+2 \ell _m-1\right) \Gamma \left(2 \ell _m+\tau _m-1\right){}^2 \, \Gamma \left(-h+\ell _m+\tau _m+1\right)
\end{align}
%where we have used the asymptotics of the $_3F_2$ \eqref{Qasym} and 
where $\hat{P}$ is defined in \eqref{Phat}. The $\mathfrak{b}$'s are defined in \eqref{dd3}. %with $k_2$ replaced by $i-k_1-n-r$
%\be
%$\a_1 =1-\D_\f=-\b_1 $ and  $\a_2= -1-n-\frac{\tau_m}{2}+\D_\f-q= -\b_2$.%,{\rm{and}} %\qquad k_2= i-k_1-n-q
%\,.
%\ee
This completes the derivation of the all order  expression for the anomalous dimension to all orders in inverse conformal spin. The first few values are quoted in \eqref{gamma} and are in agreement with \cite{aldayzhiboedov}.

\subsection{OPE coefficients}
In this section, we will derive an all order expression for the (corrections to) OPE coefficients of the double trace operators \eqref{defss} as an asymptotic expansion. In order to compute the OPE coefficient we have to focus on the coefficient of the power law term $u^{\D_\phi}$ from both sides of the bootstrap equation \eqref{boot}. We will expand the $t$-dependence in the $Q$-basis following the analysis of the previous section.  %Let us denote the OPE coefficient by $C_{0, \ell} \,(1+\delta C_{0, \ell})$.
%{\bf AS: the following equation needs more explanation!! Little c, Capital c's need fixing below carefully}

%{\bf AS: Check the following}
We start with the $s$ channel expression \eqref{sch2}. Expanding this in $\g_{0, \ell}$ we extract the coefficient of $u^{\D_\f}$ to leading order in $\g_{0, \ell}$ and $\delta C_{0,\ell}$,
\be\label{qt22}
G^{(s)}(u, v) = \sum_{\ell}\bigg(-\frac{\g_{0, \ell}}{2}\,C_{0, \ell}\,f_{\ell}(v)\,\log(1-v)+ C_{0, \ell}\,\delta C_{0, \ell}\,f_{\ell}(v)\bigg)
\ee
where $C_{0, \ell}$ is defined in \eqref{pmftnew}.
%where $C_{0, \ell}$ is defined in \eqref{pmftnew} and
The $\log(1-v)$ term arises from the factor $(1-v)^\ell$ on re-expressing $\ell$ in terms of $J$ and expanding in $\gamma_{0,\ell}$ \cite{aldayzhiboedov}.
Now \eqref{qt22} can be written as,
\be\label{qtt}
G^{(s)}(u, v)|_{u^{\D_\f}} = \sum_{\ell}\intl \frac{dt}{2\pi i}\,v^t\, \G^2(\D_\f+t)\, \G^2(-t)\, \mathfrak{\tilde{q}}^{(s)}_{1, \ell}\,Q^{2\D_\f+\ell}_{\ell, 0}(t)\bigg(\delta C_{0, \ell}-\frac{\g_{0, \ell}}{2}\,\log(1-v)\bigg)
\ee
where, %$\mathfrak{q}^{(s)}_{0, \ell}(s)$ is defined in \eqref{qsusual2}.
%$\mathfrak{q}^{(s)}_{0, \ell}(s)$ is defined in
\be
\mathfrak{\tilde{q}}^{(s)}_{1, \ell}= C_{0, \ell}\,\mathcal{N}_{2\D_\f+\ell, \ell}\,4^{-\ell}\,\G(h-2\D_\f-\ell)\,(2\D_\f+\ell-1)_{\ell}\,(2h-2\D_\f-\ell-1)_{\ell}\,.
\ee
Note that we have already expanded the first term in $Q$-basis following \eqref{schlog2}. For the second term we will use the expression which comes from matching the coefficients of the $u^{\D_\f}\, \log u$ term \eqref{bootus},%{\bf AS: which equation?},
\begin{align}
&\sum_{\ell}\intl \frac{dt}{2\pi i}\,v^t\, \G^2(\D_\f+t)\, \G^2(-t)\, \mathfrak{\tilde{q}}^{(s)}_{1, \ell}\,Q^{2\D_\f+\ell}_{\ell, 0}(t)\frac{\g_{0, \ell}}{2}\nn &= \, \intl \frac{dt}{2\pi i}\,v^t\,  \G^2(\D_\f+t)\,\G^2(-t)
c_{\D_m, \ell_m}\,\frac{\G(\frac{\D_m-\ell_m}{2}-\D_\f-t)\,\G(\frac{2h-\D_m-\ell_m}{2}-\D_\f-t)}{\G^2(-t)}\,P_{\D_m-h, \ell_m}(t+\D_\f,0)\,.
\end{align}
For simplicity let us first assume the case of scalar (dimension $\D_m$) exchange in the $t$ channel i.e.  $\ell'=0$. %We denote the dimension of the minimal twist scalar by $\D_m$. 
Using,
\be
\log (1-v)= -\sum_{r=0}^{\infty} \frac{v^{r+1}}{r+1}
\ee
 in \eqref{qtt} the second term simplifies to the following:
\begin{align}\label{qtt2}
 -\sum_{r=0}^{\infty} c_{\D_m, 0}\,\intl\frac{dt}{2\pi i}\, \frac{v^{t+r+1}}{r+1}\,\G^2(\D_\f+t)\,{\G(\frac{\D_m}{2}-t-\D_\f)\,\G(\frac{2h-\D_m}{2}-\D_\f-t)}\,.
\end{align}
Making a change of variable $t \rightarrow t-r-1$ makes \eqref{qtt2},
\begin{align}
 &-\sum_{r=0}^{\infty} c_{\D_m, 0}\,\intl \frac{dt}{2\pi i}\,v^t\,\G(\frac{\D_m}{2}-t-\D_\f+1+r)\,\G(\frac{2h-\D_m}{2}-\D_\f-t+1+r)
 \frac{\G^2(\D_\f+t-1-r)}{(r+1)} \nn
& = \sum_\ell \intl \frac{dt}{2\pi i}\,v^t\,\G^2(\D_\f+t)\,\G^2(-t)\,\tilde{\mathfrak{q}}^{(s)}_{2,\ell}\, Q^{2\D_\f+\ell}_{\ell,0}(t)
\end{align}
where,
\begin{align}
\tilde{\mathfrak{q}}^{(s)}_{2,\ell} &= -2\,\beta_{\ell}\sum_{n, k_1, k_2, r=0}^{\infty} c_{\D_m, 0}\intl \frac{dt}{2\pi i}\, \G^2(\D_\f+t-1-r)\,\G(\frac{\D_m}{2}-\D_\f-t+1+r)\,\nn & \times \G(\frac{2h-\D_m}{2}-\D_\f-t+1+r) \frac{(-1)^n\,\G^2(\D_\f)\,(\D_\f+t)_n}{n!\,(r+1)\,\G^2(-t-n)} \,\mathfrak{b}_{k_1}(\D_\f)\,\mathfrak{b}_{k_2, n}(t)\, J^{-2k_1-2k_2-2n-2\D_\f-2t}\,.
\end{align} 
%and
%\be
%\beta_{\ell} = \bigg(\kappa^{-1}_{\ell}(\D_\f)\frac{2^{\ell}\,((\D_\f)_{\ell})^2}{(2\D_\f+\ell-1)_{\ell}}\bigg)\,.
%\ee
The residue at $t=\frac{\D_m}{2}-\D_\f+1+r+q$ ($q=0, 1, \cdots $) reads,
\begin{align}
\tilde{\mathfrak{q}}^{(s)}_{2,\ell} &= -\,\beta_{\ell}\,\bigg(\sum_{n, q, r, k_1, k_2}2\, c_{\D_m, 0}\, \G^2(\D_\f+t-1-r)\,\G(\frac{2h-\D_m}{2}-\D_\f-t+1+r)\nn & \times \frac{(-1)^{n+q+1}\,\G^2(\D_\f)\,(\D_\f+t)_n}{n!\,q!\,(r+1)\,\G^2(-t-n)} \,\mathfrak{b}_{k_1}(\D_\f)\,\mathfrak{b}_{k_2, n}(t)\, J^{-2k_1-2k_2-2n-2\D_\f-2t}\bigg)_{t=\frac{\D_m}{2}-\D_\f+1+r+q}
\end{align} 
where $\b_{\ell}$ is defined in \eqref{betaell}.

Now let us consider the $t$ channel. The non-log term from the residue at $s=\D_\f$ is given by,
\begin{align}
G_{\D_m,0}^{(t)}(u,v)&=u^{\D_\phi}\intl \frac{dt}{2\pi i}\,v^t\,\G^2(\D_\f+t)\,\G^2(-t)\,\tilde{\mathfrak{q}}^{(t)}_{1,\ell|0}\,Q^{2\D_\f+\ell}_{\ell,0}(t)
\end{align}
where,
\begin{align}
\tilde{\mathfrak{q}}^{(t)}_{1,\ell|0} &= \,\beta_{\ell}\,\sum_{n, q, k_1, k_2} \bigg(2\,c_{\D_m,0}\,\frac{(-1)^{n+q+1}}{n!\,q!}\G^2(\D_\f+t)\,\G(\frac{2h-\D_m}{2}-\D_\f-t)\frac{\G^2(\D_\f)\,(\D_\f+t)_n}{\G^2(-t-n)}\nn
& \times (2\,\g_E+ 2\,\psi(t+\D_\f))\,\mathfrak{b}_{k_1}(\D_\f)\,\mathfrak{b}_{k_2, n}(t)\,J^{-2k_1-2k_2-2n-2\D_\f-2t}\bigg)_{t=\frac{\D_m}{2}-\D_\f+q}\,.
\end{align}
Then the bootstrap condition for the $u^{\D_\phi}$ term is given by,
\be\label{opeboot}
\tilde{\mathfrak{q}}^{(s)}_{1,\ell}=-\tilde{\mathfrak{q}}^{(s)}_{2,\ell}+ \tilde{\mathfrak{q}}^{(t)}_{1,\ell|0}\,.
\ee
If we take the overall  $\beta_{\ell}$ to the left we are left with $2\,\delta C_{0, \ell}$ and the right hand side contains only even powers of $J$ from the asymptotic expansion of $_3F_2$ \eqref{Qasym}. This makes it evident that the correction to the OPE coefficient in the large spin limit is also an expansion containing only even powers of $J$. We assume the following expansion for the OPE coefficient in the large spin limit,
\be
\delta C_{0, \ell}\sim \sum_{i=0}^{\infty}\frac{\delta C^{(i)}_{0, \ell} }{J^{2i}}\,.
\ee
Matching the powers of $J^{-2i}$ from both sides of \eqref{opeboot} we derive the following expression for the OPE coefficient,
\begin{align}
\delta C^{(i)}_{0, \ell} &= \frac{1}{J^{\D_m}}\,C_m\sum_{n+q+k_1=0}^{i}\frac{2(-1)^{n+q+1}}{n!\,q!\,\G^2(-n-q+\D_\f-\D_m/2)}\,\mathcal{N}_{\D_m, 0}\,(q+\D_m/2)_n\,\mathfrak{b}_{k_1}(\D_\f)\nn & \times\mathfrak{b}_{i-k_1-n-q, n}\left(\frac{\D_m}{2}-\D_\f+q\right)
 \G^2(\D_\f)\,\G(h-q-\D_m)\,\G^2(q+\D_m/2)\,H_{q+\D_m/2-1}\nn
& + \frac{1}{J^{\D_m}}\, C_m\sum_{n+q+r+k_1=0}^{i} \frac{(-1)^{n+q}\,\G^2(\D_\f)\,\G(h-q-\D_m)\,\G^2(q+\D_m/2)}{(1+r)\,n!\,q!\,\G^2(-1-r-n-q+\D_\f-\D_m/2)}\nn
& \times \mathcal{N}_{\D_m, 0}\,(1+r+q+\D_m/2)_n\, \mathfrak{b}_{k_1}(\D_\f)\,\mathfrak{b}_{i-1-r-k_1-n-q, n}\left(\frac{\D_m}{2}-\D_\f+q+r+1\right)
\end{align}
where the $\mathfrak{b}$'s are defined in \ref{dd3}.
%$\a_1=1-\D_\f=-\b_1$ in the first sum and  $\a_2= -2-r-n-q+\D_\f-\D_m/2=-\b_2$ in the second sum. 
Let us list the first two terms, %{\bf AS: give also the next term for the heck of it},
\begin{align}
\delta C^{(0)}_{0, \ell}& = -\frac{1}{J^{\D_m}}\,C_m\frac{2\, \Gamma^2 \left(\Delta _{\phi }\right){} \Gamma \left(\Delta _m\right)}{\Gamma^2 \left(\frac{\Delta _m}{2}\right){} \Gamma^2 \left(\Delta _{\phi }-\frac{\Delta _m}{2}\right){}}\,\bigg(\psi \left(\frac{\Delta _m}{2}\right)+\gamma_E\bigg), \nn
\delta C^{(1)}_{0, \ell}& =\frac{1}{J^{\D_m}}\,C_m\,\frac{\Gamma^2 \left(\Delta _{\phi }\right){}\Gamma \left(\Delta _m\right) }{24\, (h-\D_m-1)\,\Gamma^2 \left(\frac{\Delta _m}{2}\right){}  \Gamma^2 \left(\Delta _{\phi }-\frac{\Delta _m}{2}\right){}}\nn
& \times \left( 6 (h-1) \left(-2 \Delta _{\phi }+\Delta _m+2\right){}^2-\left(\psi \left(\frac{\Delta _m}{2}\right)+\gamma_E\right)\right.\nn
& \left.\Delta _m \left(-4 h \left(\Delta _m \left(-3 \Delta _{\phi }+\Delta _m+3\right)+2\right)+\Delta _m \left(-12 \left(\Delta _{\phi }-1\right) \Delta _{\phi }+\Delta _m \left(\Delta _m+4\right)+8\right)+8\right)\right) \,.
%\delta C^{(2)}_{0, \ell}&=
\end{align}
%where $\psi()$
These  results exactly\footnote{There is a small typo in \cite{aldayzhiboedov} in the bracketting for $\delta C^{(1)}_{0, \ell}$.} match with \cite{aldayzhiboedov}. 
%{\bf general spin exchange}
We also give the expression for a general spin $\ell_m$ and twist $\tau_m$ exchange in the $t$ channel,
\be
\delta C_{0, \ell}\sim \sum_{i=0}^{\infty}\frac{\delta C^{(i)}_{0, \ell} }{J^{2i}}
\ee
with
\begin{align}\label{ope2}
\delta C^{(i)}_{0, \ell}= \frac{1}{J^{\tau_m}}\,(\eta^{(i)}_1+ \eta^{(i)}_2+ \eta^{(i)}_3)
\end{align}
where,
\begin{align}\label{f123}
\eta^{(i)}_1 &= C_m\,\sum_{n+q+k_1=0}^{i}\frac{2\,(-1)^{n+q+1}\Gamma^2(\Delta_\phi )\, \Gamma^2 \left(q+\frac{\tau _m}{2}\right) \Gamma \left(h-q-\ell _m-\tau _m\right)}{n! \,q!\, \Gamma^2 \left(-n-q+\Delta_\phi -\frac{\tau _m}{2}\right)}\,(\ell_m+\tau_m-1)_{\ell_m}\,H_{q+\tau_m/2-1} \nn
& \times  \mathcal{N}_{\tau_m+\ell_m, \ell_m}\,(2h-1-\ell_m-\tau_m)_{\ell_m}\,(q+\tau_m/2)_{n}\,\mathfrak{b}_{k_1}(\D_\f)\,\mathfrak{b}_{i-k_1-n-q, n}\left(\frac{\tau_m}{2}-\D_\f+q\right)\nn & \times \hat{P}_{\tau_m+\ell_m-h, \ell_m}(q+\tau_m/2,0),
\end{align}
\begin{align}\label{f123v2}
\eta^{(i)}_2 &= C_m\,\sum_{n+q+k_1=0}^{i} \frac{(-1)^{n+q+1}\Gamma^2(\Delta_\phi )\, \Gamma^2 \left(q+\frac{\tau _m}{2}\right)\,  \Gamma \left(h-q-\ell _m-\tau _m\right)}{n! \,q!\, \Gamma^2 \left(-n-q+\Delta_\phi -\frac{\tau _m}{2}\right)}\left( \ell _m+\tau _m-1\right)_{\ell_m}\nn
& \times \mathcal{N}_{\tau_m+\ell_m, \ell_m}\,(2h-1-\ell_m-\tau_m)_{\ell_m}\,(q+\tau_m/2)_{n}\,\mathfrak{b}_{k_1}(\D_\f)\,\mathfrak{b}_{i-k_1-n-q, n}\left(\frac{\tau_m}{2}-\D_\f+q\right)\nn
& \times\frac{\partial}{\partial s}\hat{P}_{\tau_m+\ell_m-h, \ell_m}(q+\tau_m/2,\D_\f-s)|_{s=\D_\f},\nn
%\end{align}
%\begin{align}
\eta^{(i)}_3 &= C_m\,\sum_{m+n+q+k_1=0}^{i}\frac{(-1)^{n+q}\,\Gamma^2\left(\Delta _{\phi }\right) \Gamma^2\left(q+\frac{\tau _m}{2}\right)  \Gamma \left(h-q-\ell _m-\tau _m\right)}{(m+1)\, n!\, q!\, \Gamma^2\left(-m-n-q+\Delta _{\phi }-\frac{\tau _m}{2}-1\right)}  \left( \ell _m+\tau _m-1\right)_{\ell_m}\nn
& \times \mathcal{N}_{\tau_m+\ell_m, \ell_m}\,(2h-1-\ell_m-\tau_m)_{\ell_m} (1+m+q+\tau_m/2)_n\,\mathfrak{b}_{k_1}(\D_\f)\nn & \times \mathfrak{b}_{i-n-q-m-k_1-1, n}\left(\frac{\tau_m}{2}-\D_\f+q+1+m\right) \hat{P}_{\tau_m+\ell_m-h, \ell_m}(q+\tau_m/2,0)
\end{align}
and the $\mathfrak{b}$'s are defined in \eqref{dd3}.
%where $\a_1= 1-\D_\f= -\b_1$, $\a_2= -1-n-q+\D_\f-\frac{\tau_m}{2}= -\b_2$ and $k_2$ is replaced by $-k_1-n-q+i$ in the first two sums and $\a_1= 1-\D_\f= -\b_1$, $\a_2= -2-m-n-q+\D_\f-\tau_m/2= -\b_2$ and $k_2$ is replaced by $-1-k_1-m-n-q+i$ in the last sum.
%Thus we have an exact expression for the anomalous dimension to all orders in spin in the large spin limit.
\subsection{Anomalous dimension for $n \neq 0$}
For completeness, we give the expression for the anomalous dimension of the double trace operators \eqref{defss} for $n \neq 0$. We focus on the coefficient of $u^{\D_\f+k}\,\log u$ term for $k=0, 1, \cdots$ from both sides of the bootstrap equation \eqref{boot}. The details are given in appendix \ref{sneq}. We simply quote the result here.
We have a recursion relation for the anomalous dimension $\g_{n, \ell}$ for each value of $k$,
\begin{align}
&\sum_{\ell=\ell'}^{\ell'+2k}\sum_{n=0}^{(\ell'+2k-\ell)/2}\sum_{m_2=0}^{k-n}\sum_{n_2=\ell'}^{\ell-m_2}\frac{(-1)^{k-n}}{(k-n)!}\left(\frac{\g_{n, \ell}}{2}\right)\mu^{(\ell)}_{m_2, n_2}\,(n-k)_{m_2}\,\chi^{(n_2)}_{\ell'}(\D_\f+k)\nn & \times\,C_{n, \ell}\,\mathcal{N}_{2\D_\f+2n+\ell, \ell}\,\G(h-2\D_\f-n-\ell-k)\, (2\D_\f+2n+\ell-1)_{\ell}\,(2h-2\D_\f-2n-\ell-1)_{\ell}\nn
&+\sum_{r=0}^{\infty} 2\,\frac{(-1)^{r}}{(k!)^2\,r!}\, \G^2(\D_\f+k+t)\,\kappa_{\ell'}(\D_\f+k)^{-1}\,Q^{2\D_\f+2k+\ell'}_{\ell', 0}(t)\nn & \times c_{\D, \ell}\,{\G(\frac{2h-\D-\ell}{2}-\D_\f-t)}{P}_{\D-h, \ell}(t+\D_\f, k)\bigg]_{t= \frac{\D-\ell}{2}-\D_\f+r}=0
\end{align}
where $\mu^{(\ell)}_{m, n}$ is defined in \eqref{mudef}.
In principle this can be solved for $\g_{n, \ell}$ for any $n$. We believe there is a more efficient way of computing $\g_{n,\ell}$ which would rely on the factorization of the conformal blocks in the large spin limit  for general $n$ (see e.g., \cite{kss}) and which would give these expressions directly as for $\g_{0,\ell}$--we leave this interesting problem for future work which will enable the extension of \cite{heem, holorecon} to complete generality.
%We leave it as a future work.
\section{Mellin space bootstrap in the new approach} 
In this section, we will consider the problem of extracting anomalous dimensions of the double trace operators using the recently introduced reformulation of the conformal bootstrap in terms of tree level exchange Witten diagrams \cite{RGAKKSAS}. This approach efficiently reproduces anomalous dimensions and OPE coefficients of the Wilson-Fisher theory in $d=4-\e$ up to $O(\e^3)$. Up to this order, in the crossed channel, only the $\phi^2$ operator contributes which is the reason for the huge simplification in the equations. At the next order the equations become complicated as new operators start contributing. As will become evident, at the next order, the polynomial ambiguity in the Witten diagram basis will also play a role. We begin with a quick review of \cite{RGAKKSAS} to set the notation.

%{\bf AS: I feel after reviewing the necessary material, we should just demonstrate the equivalence. Once we demonstrate the equivalence, it is obvious that we will get the same answers. There is absolutely no point in presenting the explicit calculation then.}
\subsection{A quick review}
The Mellin transform of four point function for identical scalars defined through eq. (\ref{Mdef}) gives
the Mellin Amplitude $\mathcal{M}(s,t)$ which in the spectral representation has the following form,
\begin{equation}
	\begin{split}
		\mathcal{M}(s,t)&=\sum_{\Delta,\ell} c_{\Delta,\ell} (M^{(s)}_{\Delta,\ell}(s,t)+M^{(t)}_{\Delta,\ell}(s,t)+M^{(u)}_{\Delta,\ell}(s,t))\\
		& = \sum_{\Delta,\ell} c_{\Delta,\ell} \int_{-i \infty}^{i \infty} d\nu \mu_{\Delta,\ell}(\nu)\bigg(\Omega^{(s)}_{\nu,\ell}(s)P^{(s)}_{\nu,\ell}(s,t)\\
		& + \Omega^{(t)}_{\nu,\ell}(t)P^{(t)}_{\nu,\ell}(s-\Delta_{\phi},t+\Delta_{\phi})+\Omega^{(u)}_{\nu,\ell}(s+t)P^{(u)}_{\nu,\ell}(s-\Delta_{\phi},t)\bigg).
	\end{split}
\end{equation}
The spectral function is
\begin{equation}
	\mu_{\Delta,\ell}(\nu)=\frac{\Gamma^2(\frac{2\Delta_{\phi}-h+\ell+\nu}{2}) \Gamma^2(\frac{2\Delta_{\phi}-h+\ell-\nu}{2})}{2\pi i ((\Delta-h)^2-\nu^2)\Gamma(\nu)\Gamma(-\nu)(h+\nu-1)_{\ell}(h-\nu-1)_\ell}
\end{equation}
and
\begin{equation} \label{omega}
	\Omega_{\nu,\ell}^{(s)}(s)=\frac{\Gamma(\frac{h+\nu-\ell}{2}-s)\Gamma(\frac{h-\nu-\ell}{2}-s)}{\Gamma^2(\Delta_{\phi}-s)}\,.
\end{equation}
 $\Omega_{\nu,\ell}^{(t)}(t)$ and $\Omega_{\nu,\ell}^{(u)}(s+t)$ are obtained from \eqref{omega} by the replacements $s\rightarrow t+\Delta_{\phi}$ and $s\rightarrow \Delta_{\phi}-s-t$ respectively. 

%Now the physical poles in the spectral function is the one at $\nu=\Delta-h$. The residue at this pole in $\nu$ has factors of $\Gamma$ functions in the numerator of \eqref{omega} that gives rise to physical poles in the $s$-variable (alongwith shadow ones too),i.e.,
%\begin{equation}
	%s=\frac{\Delta-\ell}{2}+n, \, \, \, s=\frac{2h-\Delta-\ell}{2}+n,\,\,(n=0,1,2,3...).
%\end{equation}
%While carrying out the $s$-integral we close the contour in the right (deforming it appropiately if necessary), so we pick up the contributions only from physical poles. \\
There are poles in $\nu$ which leads to the spurious poles in the $s$-variable \cite{RGAKKSAS}. There are poles at $\nu=\pm (2\Delta_{\phi}+2n+\ell-h)$ and $\nu=\pm (2s-2n+\ell-h)$ which will give rise to a term proportional to $\Gamma(\Delta_{\phi}+n-s)$. Now for $n=0$ we have spurious pole at $s=\Delta_{\phi}$. At this pole,
\begin{equation}
	P_{(2\Delta_{\phi}+\ell-h),\ell}(s,t)=4^{-\ell} \,(2\Delta_{\phi}+\ell-1)_{\ell}\, (2h-2\Delta_{\phi}-\ell-1)_{\ell}\, Q^{2\Delta_{\phi}+\ell}_{\ell,0}(t)
\end{equation}
so
for the $s$ channel we have
\begin{equation}
	M^{(s)}_{\Delta,\ell}(s,t)=\sum_{\ell}q^{(s)}_{\Delta,\ell}(s)Q^{2s+\ell}_{\ell,0}\,,
\end{equation}
with,
\be\label{qs}
q^{(s)}_{\D, \ell}(s) = -\frac{C_{0, \ell} \,{\mathfrak{N}}_{\D, \ell} 4^{1-\ell}\G^2(\D_\f+s+\ell-h)}{(\ell-\D+2s)\,(\ell+\D+2s-2h)\,\G(2s+\ell-h)}\,.
\ee
%So in the $s$-channel there is only one spin.
The $t$-channel is given by,
\begin{equation}
	M^{(t)}_{\Delta,\ell'}(s,t)=\sum_{\ell}q^{(t)}_{\Delta,\ell|\ell'}(s)Q^{2s+\ell}_{\ell,0}
\end{equation}
with,
\begin{equation}
	\begin{split}
		q^{(t)}_{\Delta,\ell|\ell'}(s)&=\kappa_{\ell}(s)^{-1} \intl \frac{dt}{2\pi i} {d\nu} \Gamma^2(s+t) \Gamma(\frac{h+\nu-\ell}{2}-t-\Delta_{\phi})\Gamma(\frac{h-\nu-\ell}{2}-t-\Delta_{\phi})\\
		& \mu^{t}_{\Delta,\ell'}(\nu) P^{(t)}_{\nu,\ell'}(s-\Delta_{\phi},t+\Delta_{\phi})Q^{2s+\ell}_{\ell,0}(t)\,.
	\end{split}
\end{equation}
For identical scalars the $u$-channel is related to the $t$-channel via,
\begin{equation}
	M^{(u)}_{\Delta,\ell'}(s,t)=M^{(t)}_{\Delta,\ell'}(s,-s-t)\,.
\end{equation}
In terms of the $Q$-basis expansion, this translates into,
\begin{equation}
	\sum_{\ell} Q_{\ell}^{2\Delta_{\phi}+\ell}(-\Delta_{\phi}-t)q^{t}_{\Delta,\ell|\ell'}(s)=\sum_{\ell}Q_{\ell}^{2\Delta_{\phi}+\ell}(t)q^{(u)}_{\Delta,\ell|\ell'}(s).
\end{equation}
Since,
\begin{equation}
	Q_{\ell}^{2\Delta_{\phi}+\ell}(-\Delta_{\phi}-t)=(-1)^\ell Q_{\ell}^{2\Delta_{\phi}+\ell}(t)
\end{equation}
we get,
\begin{equation}
	q^{(t)}_{\Delta,\ell|\ell'}(s)=(-1)^\ell q^{(u)}_{\Delta,\ell|\ell'}(s)\,.
\end{equation}
Now demanding that the spurious pole at $s=\D_\phi$ cancels we will get 
the following consistency conditions,
\begin{align}\label{boott}
&q^{(s)}_{\D, \ell} (\D_\f) +2\,q^{(t)}_{\D, \ell|\ell'} (\D_\f)=0\, , \nn
&{q'}^{(s)}_{\D, \ell} (\D_\f) +2\,{q'}^{(t)}_{\D, \ell|\ell'} (\D_\f)+ q^{disc}=0\,.
\end{align}
$q^{disc}$ is the disconnected contribution which is added separately. Further we are focusing on a single operator in the crossed channel.
%In this section, we present a systematic way to compute the anomalous dimension and OPE coefficient of double field operators $\mathcal{O}_{n,\ell}=\phi \Box^{n} \partial_{\mu_1} \partial_{\mu_2}\cdots \partial_{\mu_\ell} \phi$ using the algorithm developed in \cite{Gopakumar:2016wkt,Gopakumar:2016cpb}. 
%We denote the dimension and OPE coefficient of the double trace operators as,
%\begin{align}
%&\D = 2\D_\f+ \ell+2n +\g_{n, \ell} + O(\g^2)\nn
%& C_{2\D_\f+ \ell+2n} = C_{n, \ell} (1+ \delta C_{n, \ell})\,.
%\end{align}
%Our goal will be to compute the anomalous dimension $\g_{n, \ell}$ and the correction to the OPE coefficient $\delta C_{n, \ell}$ in the limit $\ell \gg 1$. We will treat the case  $n=0$ and $n \neq 0$ separately.
%\subsection{$n = 0$}
The $t$ channel at $s=\D_\f$ can be explicitly worked out as follows. We will denote the dimension and spin of the exchange operator in the $t$ channel by $\D$ and $\ell'$ respectively.
\begin{align}
q^{(t)}_{\D, \ell|\ell'} &=\kappa^{-1}_{\ell}(\D_\f)\, c_{\D, \ell'}\intl \frac{dt}{2\pi i}\,d\nu\,\G^2(\D_\f+t)\,\G(\l_2-t-\D_\f)\,\G({\bar{\l}_2}-t-\D_\f)\nn
& \times \mu^{(t)}_{\D,\ell'}(\nu)\,P_{\nu, \ell'}(t+\D_\f, s-\D_\f)\, Q^{2\D_\f+\ell}_{\ell,0}(t)\nn
& =\beta_{\ell}\,C_{\D, \ell'}\,\mathfrak{N}_{\D, \ell'}\,\intl \frac{dt}{2\pi i}\,d\nu\,\G^2(\D_\f+t)\,\G(\l_2-t-\D_\f)\,\G({\bar{\l}_2}-t-\D_\f)\nn
& \times \mu^{(t)}_{\D,\ell'}(\nu)\,P_{\nu, \ell'}( t+\D_\f, 0)\, {}_3F_2\bigg[\begin{matrix} -\ell,\, 2\D_\f+\ell-1,\,\D_\f+t\\
\ \ \D_\f \ \ , \ \ \ \ \ \ \D_\f
\end{matrix};1\bigg]\,,
\end{align}
where $\l_2= \frac{h+\nu-\ell'}{2}$ and $\bar{\la}_2 =\frac{h-\nu-\ell'}{2}$,
\be\label{normt}
{\mathfrak{N}}_{\D, \ell}= \frac{(-2)^{\ell}\,(\ell+\D-1)\,\G(1-h+\D)\,\G^2(\ell+\D-1)}{\G(\D-1)\,\G^4(\frac{\D+\ell}{2})\,\G^2(\frac{\ell+2\D_\f-\D}{2})\,\G^2(\frac{\ell+2 \D_\f+\D-2h}{2})}
\ee
%\be
%\mathfrak{N}_{\D, \ell}= \frac{2^{\ell}\,\G(\D-h+1)\,\G(\D+\ell)\,\G(\D+\ell-1)}{\G(\D-1)\,\G^4(\frac{\D+\ell}{2})\,\G^2(\D_\f+\frac{\D-\ell}{2})\,\G^2(\D_\f+\frac{\D+\ell}{2}-h)}
%\ee
and we have  used  the expression for $Q^{2\D_\f+\ell}_{\ell,0}(t)$ as defined in \eqref{Qdefn}.  Here $C_{\D, \ell'}$ denotes the OPE coefficient of the operator getting exchanged in the $t$ channel. 
Then we have,
\begin{align}\label{qt2}
\frac{1}{\b_{\ell}}\,q^{(t)}_{\D, \ell|\ell'} &=C_{\D, \ell'}\,\mathfrak{N}_{\D, \ell'}\,\intl \frac{dt}{2\pi i}\,d\nu\,\G(\D_\f+t)^2\,\G(\l_2-t-\D_\f)\,\G({\bar{\l}_2}-t-\D_\f)\nn
& \times \mu^{(t)}_{\D,\ell'}(\nu)\,P_{\nu, \ell'}( t+\D_\f, 0)\, {}_3F_2\bigg[\begin{matrix} -\ell,\, 2\D_\f+\ell-1,\,\D_\f+t\\
\ \ \D_\f \ \ , \ \ \ \ \ \ \D_\f
\end{matrix};1\bigg]\,.
\end{align}
 Now we take the large spin ($\ell \gg 1$) limit of the above expression. It is convenient to make a change of variable from the usual spin $\ell$ to the conformal spin $J$. In terms of $J$ it reads,
 \begin{align}
 \frac{1}{\b_{\ell}}\,q^{(t)}_{\D, \ell|\ell'} &=C_{\D, \ell'}\,\mathfrak{N}_{\D, \ell'}\,\intl \frac{dt}{2\pi i}\,d\nu\,\G^2(\D_\f+t)\,\G(\l_2-t-\D_\f)\,\G({\bar{\l}_2}-t-\D_\f)\,\mu^{(t)}_{\D,\ell'}(\nu)\nn
 & \times P_{\nu, \ell'}( t+\D_\f, 0)\, \sum_{n,k_1,k_2=0}^{\infty}\frac{(-1)^n}{n!}\,\frac{\G^2(\D_\f)\,(\D_\f+t)_n}{\G^2(-t-n)}\,\mathfrak{b}_{k_1}(\D_\f)\,\mathfrak{b}_{k_2, n}(t)\,J^{-2k_1-2k_2-2n-2\D_\f-2t}
 \end{align}
 where the $\mathfrak{b}$'s are defined in \eqref{dd3}.
 %%%%%%%%%%%%%%%%%%%%%%%%%%%%%%%%%%%%%%%%%%
There are poles from the Gamma functions at $t= \la_2-\D_\f+q$ and $t= \bar{\la}_2-\D_\f+q$  for $q=0, 1, 2, \cdots$. There is a  $J^{-2t}$ power in the integrand which allows us to close the $t$ contour on the right. The residue at $t= \la_2-\D_\f+q$ is given by,
\begin{align}
& \frac{1}{\b_{\ell}}	q^{(t)}_{\D, \ell|\ell'}\nn &=\intl \frac{d\nu}{2\pi i}\,\sum_{n, k_1, k_2, q=0}^{\infty}C_{\D, \ell'}\,\mathfrak{N}_{\D, \ell'}\,\frac{(-1)^{n+q}}{n!\,q!}\,\mathfrak{b}_{k_1}(\D_\f)\,\mathfrak{b}_{k_2, n}(\frac{h+\nu-\ell'}{2}-\D_\f+q)\,J^{-h-2k_1-2k_2-2q-2n+\ell'-\nu}\nn
	&\times\frac{\Gamma^2 (\Delta \phi )\, \Gamma (h+\nu -1) \Gamma (-q-\nu )}{((h-\Delta )^2-\nu ^2)\,\Gamma (-\nu ) \Gamma (\nu ) (h +\nu -1)_{\ell'} (h -\nu -1)_{\ell'}}\,P_{\nu, \ell'}\left(\frac{h+\nu-\ell'}{2}+q, 0\right)\nn
	& \times \frac{\Gamma^2 \left(\Delta_\phi +\frac{1}{2} (-h+\ell' -\nu )\right)\,\Gamma^2 \left(\Delta_\phi +\frac{1}{2} (-h+\ell' +\nu )\right)\,\Gamma \left(q+\frac{1}{2} (h-\ell' +\nu )\right) \Gamma \left(n+q+\frac{1}{2} (h-\ell' +\nu )\right)}{ \Gamma^2 \left(-n-q+\Delta_\phi +\frac{1}{2} (-h+\ell' -\nu )\right)}\,.
\end{align}
The power $J^{-\nu}$ requires that we close the $\nu$ contour on
 the right side of the complex $\nu$ plane.
 %{\textbf{PD: Comment on the other $\nu$ poles!!}}. The other possible $\nu$ pole lying inside the contour will come from $\Gamma \left(\Delta_\phi +\frac{1}{2} (-h+\ell' -\nu )\right)^2$.
% The ratio of these two residues is given by,
% \be
% \frac{Res_{\nu=\D-h}}{Res_{\nu=2\D_\f-h+\ell'}} \sim J^{2\D_\f-\D+\ell'}\,.
% \ee
%{\textbf{PD: other poles}}
In the numerator we can have a pole from $\Gamma \left(\Delta_\phi +\frac{1}{2} (-h+\ell' -\nu )\right)$ for $\Delta_\phi +\frac{1}{2} (-h+\ell' -\nu )=-m$. But this will in turn introduce a $\G(-n-q-m)$ in the denominator which has a zero. So this pole will get cancelled off by the zero. The other possible pole lying inside the contour can come from the $\G(-q-\nu)$ at $\nu=-q+m$ when $m > q$. But we get rid off this pole by the zero at $\G(-\nu)$ in the denominator. Hence we are only left with the pole at $\nu=\D-h$.

  We evaluate the residue at $\nu=\D-h$ which is given by,
\begin{align}
&	-\sum_{n, k_1, k_2, q=0}^{\infty}\,C_{\D, \ell'}\,\mathfrak{N}_{\D, \ell'}\,\frac{(-1)^{n+q}}{n!\,q!}\,\mathfrak{b}_{k_1}(\D_\f)\,\mathfrak{b}_{k_2, n}(\frac{\D-\ell'}{2}-\D_\f+q)\,J^{-2k_1-2k_2-2q-2n+\ell'-\D}\nn
	&\times \frac{\Gamma (2 h-\Delta -1) \Gamma (h-q-\Delta ) \Gamma \left(q+\frac{\Delta }{2}-\frac{\ell' }{2}\right) \Gamma \left(n+q+\frac{\Delta }{2}-\frac{\ell' }{2}\right)}{2 \pi  \Gamma (\ell' +\Delta -1) \Gamma (2 h+\ell' -\Delta -1) \Gamma \left(-n-q+\frac{\ell' -\Delta }{2}+\Delta_\phi \right)^2}\,P_{\D-h, \ell'}\left(\frac{\D-\ell'}{2}+q, 0\right)\nn
	& \times \Gamma (\Delta -1) \Gamma^2 (\Delta_\phi ) \, \Gamma^2 \left(\frac{\ell' -\Delta }{2}+\Delta_\phi \right) \Gamma^2 \left(-h+\frac{\ell' +\Delta }{2}+\Delta_\phi \right)\,\sin (\pi  (h-\Delta ))\,.
\end{align}
The pole at $t=\bar{\la}_2-\D_\f + q$ and $\nu=\D-h$ will give the same residue as the above and we have to include an overall factor of $2$ in the above expression, 

\begin{align}\label{tchasym}
\frac{1}{\b_{\ell}}\, q^{(t)}_{\D, \ell|\ell'}  &=-\sum_{n, k_1, k_2, q=0}^{\infty}C_{\D, \ell'}\,\mathfrak{N}_{\D, \ell'}\,\frac{(-1)^{n+q}}{n!\,q!}\, \mathfrak{b}_{k_1}(\D_\f)\,\mathfrak{b}_{k_2, n}(\frac{\D-\ell'}{2}-\D_\f+q)\,J^{ -2k_1-2k_2-2n-2q-\D+\ell'}\nn
 &\times \frac{\Gamma (2 h-\Delta -1) \Gamma (h-q-\Delta ) \Gamma \left(q+\frac{\Delta }{2}-\frac{\ell' }{2}\right) \Gamma \left(n+q+\frac{\Delta }{2}-\frac{\ell' }{2}\right)}{\pi  \Gamma (\ell' +\Delta -1) \Gamma (2 h+\ell' -\Delta -1) \Gamma \left(-n-q+\frac{\ell' -\Delta }{2}+\Delta_\phi \right)^2}P_{\D-h, \ell'}\left(\frac{\D-\ell'}{2}+q, 0\right)\nn
 & \times \Gamma (\Delta -1) \Gamma^2 (\Delta_\phi ) \, \Gamma^2 \left(\frac{\ell' -\Delta }{2}+\Delta_\phi \right) \Gamma^2 \left(-h+\frac{\ell' +\Delta }{2}+\Delta_\phi \right)\,\sin (\pi  (h-\Delta ))\,.
\end{align}

\subsection{Equivalence of  the two bootstraps}
In this section we will show that the $s$ channel expressions from both the bootstraps are exactly the same at $O(\g)$. We will also show that the $t$ channel expressions identical in the large spin limit. The latter is to be expected since in the large spin limit only the physical pole contributions are picked up in both approaches.

\subsubsection{$s$ channel}
First let us consider the $s$ channel. We will show that the $s$ channels are identical at $O(\g_{0, \ell})$ and the difference arises only at $O(\g^2_{0,\ell})$. We substitute $\D= 2\D_\f+\ell+\g_{0, \ell}$ in \eqref{qs} and expand in $\g_{0, \ell}$ to obtain the following,
\begin{align}\label{mellinqs}
q^{(s)}_{\D, \ell}(\D_\f)&=C_{0, \ell}\frac{2^{-\ell -1} \Gamma \left(2 \left(\ell +\Delta _{\phi }\right)\right) \Gamma \left(2 \ell +2 \Delta _{\phi }-1\right)}{\Gamma^4 \left(\ell +\Delta _{\phi }\right) \Gamma \left(\ell +2 \Delta _{\phi }-1\right)}\,\g_{0, \ell} \nn &+ C_{0, \ell}\,\frac{2^{-\ell -2} \Gamma^2 \left(2 \ell +2 \Delta _{\phi }-1\right){}}{\left(2 \Delta _{\phi }-h+\ell \right) \Gamma^4 \left(\ell +\Delta _{\phi }\right)\, \Gamma \left(\ell +2 \Delta _{\phi }-1\right)}\nn
&\times\bigg(6 \Delta _{\phi }-2 \left(2 \Delta _{\phi }+2 \ell -1\right) \left(2 \Delta _{\phi }-h+\ell \right) \left(2 H_{\ell +\Delta _{\phi }-1}-2 H_{2 \left(\ell +\Delta _{\phi }-1\right)}+H_{\ell +2 \Delta _{\phi }-2}\right)\nn&-2 h+4 \ell -1\bigg)\g^2_{0, \ell} + O(\g^3_{0, \ell})\,,  
\end{align}
\begin{align}
{q'}^{(s)}_{\D, \ell}(\D_\f)&= C_{0, \ell}\frac{2^{4 \Delta _{\phi }+3 \ell -4} \left(2 \Delta _{\phi }+2 \ell -1\right) \Gamma^2 \left(\ell +\Delta _{\phi }-\frac{1}{2}\right)}{\pi  \Gamma^2 \left(\ell +\Delta _{\phi }\right)\, \Gamma \left(\ell +2 \Delta _{\phi }-1\right)}
+ C_{0, \ell}\frac{2^{4 \Delta _{\phi }+3 \ell -4} \Gamma^2 \left(\ell +\Delta _{\phi }-\frac{1}{2}\right)}{\pi  \Gamma^2 \left(\ell +\Delta _{\phi }\right)\, \Gamma \left(\ell +2 \Delta _{\phi }-1\right)}\nn & \times \bigg(1-\left(2 \Delta _{\phi }+2 \ell -1\right) \left(2 H_{\ell +\Delta _{\phi }-1}-2 H_{2 \left(\ell +\Delta _{\phi }-1\right)}+H_{\ell +2 \Delta _{\phi }-2}\right)\bigg)\,\g_{0, \ell}\nn
& +b_1 \,\g^2_{0, \ell} + O(\g^3_{0, \ell})\,.
\end{align}
The expression for $b_1$ is too big to show here.

Let us now turn to the $s$ channel expression in usual bootstrap in Mellin space. In order to make the comparison easy we will proceed  as follows. We expand \eqref{qsusual2} in $\D=2\D_\f+\ell+\g_{0, \ell}$ ,
\begin{align}\label{usualqs}
\mathfrak{q}^{(s)}_{0, \ell}(\D_\f)&= C_{0, \ell}\frac{2^{-\ell -1} \Gamma \left(2 \left(\ell +\Delta _{\phi }\right)\right) \Gamma \left(2 \ell +2 \Delta _{\phi }-1\right)}{\Gamma^4 \left(\ell +\Delta _{\phi }\right) \Gamma \left(\ell +2 \Delta _{\phi }-1\right)}\,\g_{0, \ell}\nn
&- C_{0, \ell}\frac{2^{-\ell -1} \Gamma \left(2 \left(\ell +\Delta _{\phi }\right)\right) \Gamma \left(2 \ell +2 \Delta _{\phi }-1\right)}{\Gamma^4 \left(\ell +\Delta _{\phi }\right)\, \Gamma \left(\ell +2 \Delta _{\phi }-1\right)}\bigg(H_{\D_\f-1}+H_{\ell+\D_\f-1}-H_{2\ell+2\D_\f-1}-\g_E\bigg)\,\g^2_{0, \ell} \nn &+ O(\g^3_{0, \ell})\,,\nn
\mathfrak{q'}^{(s)}_{0, \ell}(\D_\f)&=C_{0, \ell}\frac{2^{-\ell } \Gamma \left(2 \left(\ell +\Delta _{\phi }\right)\right) \Gamma \left(2 \ell +2 \Delta _{\phi }-1\right)}{\Gamma^4 \left(\ell +\Delta _{\phi }\right)\, \Gamma \left(\ell +2 \Delta _{\phi }-1\right)}
 + C_{0, \ell}\frac{2^{-\ell } \Gamma \left(2 \left(\ell +\Delta _{\phi }\right)\right) \Gamma \left(2 \ell +2 \Delta _{\phi }-1\right)}{\Gamma \left(\ell +\Delta _{\phi }\right){}^4 \Gamma \left(\ell +2 \Delta _{\phi }-1\right)}\nn & \times\bigg(H_{2 \left(\ell +\Delta _{\phi }-1\right)}-H_{\ell +2 \Delta _{\phi }-2}-2H_{\ell+\D_\f-1}+H_{2\ell+2\D_\f-1}\bigg)\,\g_{0, \ell}
 + b_2\,\g^2_{0, \ell} + O(\g^3_{0, \ell})\,.
\end{align}
Comparing \eqref{mellinqs} and \eqref{usualqs} it is easy to verify that,
\begin{align}\label{diff}
{q}^{(s)}_{\D, \ell}(\D_\f)-\mathfrak{q}^{(s)}_{0, \ell}(\D_\f) &= C_{0, \ell}\frac{2^{-\ell -2} \Gamma \left(2 \left(\ell +\Delta _{\phi }\right)\right) \Gamma \left(2 \ell +2 \Delta _{\phi }-1\right)}{\Gamma^4 \left(\ell +\Delta _{\phi }\right) \Gamma \left(\ell +2 \Delta _{\phi }-1\right)}\nn & \!\!\!\!\!\!\!\!\!\!\!\!\times\bigg(2 H_{\D_\f-1}-2\g_E+\frac{1}{2 \Delta _{\phi }-h+\ell }-2 H_{\ell +\Delta _{\phi }-1}+2\, H_{2 \left(\ell +\Delta _{\phi }-1\right)}-2 H_{\ell +2 \Delta _{\phi }-2}\bigg)\g^2_{0, \ell}\nn & + O(\g^3_{0, \ell})
\end{align}
and we find,
%
%\begin{align}\label{derdiff}
%& {q'}^{(s)}_{\D, \ell}(\D_\f)-\mathfrak{q'}^{(s)}_{0, \ell}(\D_\f) =C_{0, \ell} \frac{2^{4 \Delta _{\phi }+3 \ell -6} \Gamma \left(\ell +\Delta _{\phi }-\frac{1}{2}\right) \Gamma \left(\ell +\Delta _{\phi }+\frac{1}{2}\right)}{3\, \pi\,  \Gamma^2 \left(\ell +\Delta _{\phi }\right) \Gamma \left(\ell +2 \Delta _{\phi }-1\right)}\nn &
%\left(60 \gamma ^2-\pi ^2+24 \left(H_{2 \left(\ell +\Delta _{\phi }-1\right)}\right){}^2-24 H_{2 \left(\ell +\Delta _{\phi }-1\right)} \left(-H_{\D_\f-1}+H_{\ell +\Delta _{\phi }-1}+H_{\ell +2 \Delta _{\phi }-2}+2\,\gamma_E \right)\right.\nn & -12(H_{\D_\f-1}-\g_E)^2 +24 (H_{\D_\f-1}-\g_E) \left(H_{\ell +\Delta _{\phi }-1}+H_{\ell +2 \Delta _{\phi }-2}\right)\nn & +12 \left(H_{\ell +2 \Delta _{\phi }-2}+H_{\ell+\D_\f-1}+2 \gamma_E \right)-12 (H_{2\ell+2\D_\f-1}-\g_E)^2\nn
%& H_{\ell+\D_\f-1}+H_{\ell+2\D_\f-2}-2\g_E+6 \psi ^{(1)}\left(\Delta _{\phi }\right)-6 \psi ^{(1)}\left(\ell +\Delta _{\phi }\right)-12 \psi ^{(1)}\left(\ell +2 \Delta _{\phi }-1\right)\nn &+6 \psi ^{(1)}\left(-h+\ell +2 \Delta _{\phi }\right) +12 \psi ^{(1)}\left(2 \ell +2 \Delta _{\phi }-1\right)-\left.\frac{6}{\left(2 \Delta _{\phi }-h+\ell \right){}^2}-\frac{12}{\left(2 \Delta _{\phi }+2 \ell -1\right){}^2}+\frac{24(H_{2\ell+2\D_\f-1}-\g_E) }{2 \Delta _{\phi }+2 \ell -1}\right)\nn & \times \g^2_{0, \ell} + O(\g^3_{0, \ell})\,.
%\end{align}
\begin{align}\label{derdiff}
& {q'}^{(s)}_{\D, \ell}(\D_\f)-\mathfrak{q'}^{(s)}_{0, \ell}(\D_\f) =C_{0, \ell}\frac{2^{4 \Delta _{\phi }+3 \ell -6} \Gamma \left(\ell +\Delta _{\phi }-\frac{1}{2}\right) \Gamma \left(\ell +\Delta _{\phi }+\frac{1}{2}\right)}{3 \pi  \Gamma^2 \left(\ell +\Delta _{\phi }\right)\, \Gamma \left(\ell +2 \Delta _{\phi }-1\right)}\nn
& \times \left(60 \gamma ^2_E-\pi ^2\right.+24 \left(H_{2 \left(\ell +\Delta _{\phi }-1\right)}\right){}^2-24 H_{2 \left(\ell +\Delta _{\phi }-1\right)} \left(-H_{\Delta _{\phi }-1}+H_{\ell +\Delta _{\phi }-1}+H_{\ell +2 \Delta _{\phi }-2}+2\,\gamma_E \right)\nn
&+12 (H_{\D_\f-1}-\g_E)^2-24 (H_{\D_\f-1}-\g_E)\ \left(H_{\ell +\Delta _{\phi }-1}+H_{\ell +2 \Delta _{\phi }-2}\right)-12(H_{2\ell+2\D_\f-1}-\g_E)^2 \nn
&+12 \left(H_{\ell +2 \Delta _{\phi }-2}+H_{\ell+\D_\f-1}+2 \gamma_E \right)\left(H_{\ell+\D_\f-1}+H_{\ell+2\D_\f-2}-2\,\g_E\right)+6 \psi ^{(1)}\left(\Delta _{\phi }\right)-6 \psi ^{(1)}\left(\ell +\Delta _{\phi }\right)\nn
&+6 \psi ^{(1)}\left(-h+\ell +2 \Delta _{\phi }\right)-12 \psi ^{(1)}\left(\ell +2 \Delta _{\phi }-1\right)+12 \psi ^{(1)}\left(2 \ell +2 \Delta _{\phi }-1\right)-\frac{6}{\left(2 \Delta _{\phi }-h+\ell \right){}^2}\nn & \left.+\frac{24\,(H_{2\ell+2\D_\f-1}-\g_E) }{2 \Delta _{\phi }+2 \ell -1}-\frac{12}{\left(2 \Delta _{\phi }+2 \ell -1\right){}^2}\right)\g^2_{0, \ell} + O(\g^3_{0, \ell})
\end{align}
where $\psi^{(1)}(x)$ is the polygamma function.
\subsubsection{$t$ channel}
The $t$ channel expression from the coefficient of $u^{\D_\f}\,\log u$ in the new and usual bootstraps are given in \eqref{tchasym} and \eqref{rhsq2} respectively. %{\textbf{PD: interchanging $\ell, \ell'$}},
In the large spin limit we can use \eqref{Qasym} for the Hahn polynomial. Hence the ratio of the Mellin and usual bootstrap in the $t$ channel is given by,
\begin{align}
{q^{(t)}_{\D, \ell|\ell_m}}&={\mathfrak{q}^{(t)}_{0, \ell|\ell_m}}\,.
\end{align}
This shows that in the large spin limit the coefficient of $u^{\D_\f}\log u$ from $t$ channels are exactly the same.

Now we will show that the coefficient of $u^{\D_\f}$ from both the bootstraps are also related. We have the coefficient of $u^{\D_\f}$ from \eqref{tch2},
\begin{align}
\mathfrak{q'}^{(t)}_{0, \ell|\ell_m}&= \sum_{q=0}^{\infty}\frac{(-1)^{1+q}\,2^{\ell_m+1}\,\G(h-q-\tau_m-\ell_m)\,\G^2(q+\frac{\tau_m}{2})\,\G(2\ell_m+\tau_m-1)\,\G(2\ell_m+\tau_m)}{q!\G(h-\ell_m-\tau_m)\,\G^4(\ell_m+\frac{\tau_m}{2})\,\G(\ell_m+\tau_m-1)}\nn
& \times \bigg(2\,H_{q-1+\tau_m/2}\,\hat{P}_{\tau_m+\ell_m-h, \ell_m}(q+\tau_m/2, 0)+ \frac{\partial}{\partial s}\hat{P}_{\tau_m+\ell_m-h, \ell_m}(q+\tau_m/2, 0)\bigg)\nn & \times \k^{-1}_{\ell}(\D_\f)\,Q^{2\D_\f+\ell}_{\ell} (\frac{\tau}{2}-\D_\f+q)\,.
\end{align}
The expression for the Mellin bootstrap can be read off from \cite{RGAKKSAS},
\begin{align}
{q'}^{(t)}_{\D, \ell|\ell_m} &=-\sum_{q=0}^{\infty} \frac{2^{\ell_m}\left(\tau _m+2 \ell _m-1\right) \Gamma^2 \left(q+\frac{\tau _m}{2}\right){} \Gamma^2 \left(2 \ell _m+\tau _m-1\right){} \Gamma \left(-h+\ell _m+\tau _m+1\right)}{q!\,\G^4(\ell_m+\frac{\tau_m}{2})\,\G(\tau_m+\ell_m-1)\G(\ell_m+1-h+q+\tau_m)}\nn
& \times \bigg(2\,H_{q-1+\tau_m/2}\,\hat{P}_{\tau_m+\ell_m-h, \ell_m}(q+\tau_m/2, 0)+ \frac{\partial}{\partial s}\hat{P}_{\tau_m+\ell_m-h, \ell_m}(q+\tau_m/2, 0)\bigg)\nn & \times\k^{-1}_{\ell}(\D_\f)\,Q^{2\D_\f+\ell}_{\ell} (\frac{\tau}{2}-\D_\f+q)\,.
\end{align}
 Comparing both the terms by using the reflection identity judiciously, we find,
\begin{align}
{q'}^{(t)}_{\D, \ell|\ell_m}&={\mathfrak{q'}^{(t)}_{0, \ell|\ell_m}}\,.
\end{align}
Thus the $O(\gamma_\ell^2)$ difference between the two approaches at large spin is contained only in the $s$-channel expressions. A word of caution: While the above discussion certainly holds for large spin, we cannot blindly set $\ell=0$ e.g., in \eqref{diff} as there are finite support pieces in both formalisms that need to be added before we compare the expressions.
%\begin{align}
%\frac{(-1)^{1+q}\,\G(h-q-\tau_m)\,\G^2(q+\tau_m/2)\G(\tau_m)}{q!\,\G(h-\tau_m)\,\G^4(\frac{\tau}{2})}
%\end{align}
%{\bf PD: Derivative of t}
%n the large spin limit we have,
%\be
%C_{\D_m, \ell_m}\,{q}^{(t)}_{\ell'}(s=\D_\f)+ C_{\D_m, \ell_m}\,{q}^{(u)}_{\ell'}(s=\D_\f) = -\hat{q}^{(t)}_{\ell'}
%\ee
\subsection{Comments on finite support}
%%%%
In \cite{Alday:2016jfr},\cite{holorecon} it was pointed out that there can be contribution to the anomalous dimension which have finite support in spin, i.e., 
$$
\gamma_{\ell}=\gamma_{\ell}^{asymp}+\gamma_{\ell}^{fin}
$$
where the last piece is an extra contribution and contributes only to $\ell<s$ where $s$ is some finite spin. In the new bootstrap, it is easy to see where such an extra contribution comes from. In the large spin limit there is a $\G(-n-t)^2$ inverse factor for the crossed channels that kills the contribution from the $\G(-t)^2$ poles in the measure. For finite spin, e.g., $\ell=0$ clearly these poles will contribute (for instance in the epsilon expansion these poles are crucial to give the right answer \cite{RGAKKSAS, PDAKAS}). Hence there will be an extra contribution from these poles for low spins in the crossed channel. When we set up the difference equation for low lying spins, it will be important to take into account these contributions in both the approaches as these pieces will be different in the two formalisms from $O(\gamma^2)$.

%%%%%%%%%%%%%%%%%%

\subsection{Polynomial ambiguity of Witten diagram}
%Notation: $\mathfrak{q}$ and $q$ stand for the usual bootstrap and Mellin bootstrap respectively.
In this section we will show how the polynomial ambiguity in the Witten diagrams are needed for the equivalence of the two bootstraps at $O(\g^2)$. Since the $t$ channels are the same in both the bootstraps we focus only on the $s$ channel which reads \eqref{qsusual2} \eqref{qs},
\be\label{us2}
s_{usual}=\sum_{\ell=0}^{\infty}%mathfrak{q}^{(s)}_{0, \ell'}\, Q^{2\s+\ell'}_{\ell'}(t)
\mathfrak{q}^{(s)}_{0, \ell}(s)\,Q^{2\D_\f+\ell}_{\ell}(t)\,,\quad
s_{Mellin}=\sum_{\ell=0}^{\infty}{q}^{(s)}_{\D, \ell}(s)\, Q^{2\D_\f+\ell}_{\ell}(t)\,.
\ee
%where, 
%\begin{align}
%Q^{2\s+\ell'}_{\ell', 0}(t)&= \delta_{\ell, \ell'} Q^{2s+\ell}_{\ell, 0}(t)+ \sum_{\ell=0}^{\ell'-1}\bar{c}_{\ell', \ell}\,Q^{2s+\ell'}_{\ell', 0}(t)\nn
%&= \sum_{\ell=0}^{\infty}\delta_{\ell, \ell'} Q^{2s+\ell}_{\ell, 0}(t)+ \sum_{\ell=0}^{\ell'-1}\bar{c
%}_{\ell', \ell}\,Q^{2s+\ell'}_{\ell', 0}(t)
%\end{align}
%with $s=\D_\f+\g_{0, \ell}, \s=\D_\f+k$
%and $\bar{c}_{\ell', \ell}$ is defined in \eqref{c}.  
In the Witten diagram basis, for a spin $\ell$ exchange, we can add a $\ell-1$ polynomial piece in $t$ \cite{costa}. For the double trace operators,  we can write the combination of these terms as,
\be\label{dell}
q^{(s)}_{amb}=\sum_{\ell=0}^{\infty}\sum_{\ell'=0}^{\ell-1} d_{\ell', \ell}(s)\,Q^{2\D_\f+\ell'}_{\ell',0}(t)\,,
\ee
where $d_{\ell', \ell}$-s are the unknowns and need to be fixed. The upper limit of the sum makes it evident that the polynomial ambiguity of Witten diagram  for spin $\ell$ is a polynomial of degree $\ell-1$.

Now the $s$ channel in Mellin bootstrap reads,
\be\label{M2}
s^{now}_{Mellin}=\sum_{\ell=0}^{\infty}\bigg({q}^{(s)}_{\D, \ell}(s)\, Q^{2\D_\f+\ell}_{\ell}(t)+ \sum_{\ell'=0}^{\ell-1} d_{\ell', \ell}(s)\,Q^{2\D_\f+\ell'}_{\ell',0}(t)\bigg)\,.
\ee
 %The $s$ channel expression with this ambiguous term is given by,
%\be\label{M2}
%\sum_{\ell=0}^{\infty} \bigg({q}^{(s)}_{\D, \ell}\,Q^{2\D_\f+\ell}_{\ell,0}(t)+\sum_{\ell'=0}^{\ell-1} d_{\ell', \ell}\,Q^{2\D_\f+\ell'}_{\ell',0}(t)\bigg)
%\ee
%Usual bootstrap
%In the usual bootstrap we have the following expression from the $s$ channel,
%\begin{align}\label{us2}
%s_{usual}&=\sum_{\ell'=0}^{\infty} \mathfrak{q}^{(s)}_{0,\ell'}\,\bigg(\delta_{\ell, \ell'}\,Q^{2\D_\f+\ell}_{\ell,0}(t)+\sum_{\ell=0}^{\ell'-1} \bar{c}_{\ell, \ell'}\,Q^{2\D_\f+\ell}_{\ell,0}(t)\bigg)
%&= \sum_{\ell=0}^{\infty} \bigg({q}^{(s)}_{\ell}\,Q^{2\D_\f+\ell}_{\ell,0}(t)+\sum_{\ell'=0}^{\ell-1} d_{\ell', \ell}\,Q^{2\D_\f+\ell'}_{\ell',0}(t)\bigg)
%\end{align}
Now we demand the equality of \eqref{M2} and \eqref{us2},
\be\label{diff4}
\sum_{\ell=0}^{\infty}{q}^{(s)}_{\D, \ell}(s)\, Q^{2\D_\f+\ell}_{\ell}(t)+ \sum_{\ell=0}^{\infty}\sum_{\ell'=0}^{\ell-1} d_{\ell', \ell}(s)\,Q^{2\D_\f+\ell'}_{\ell',0}(t)= \sum_{\ell=0}^{\infty}%mathfrak{q}^{(s)}_{0, \ell'}\, Q^{2\s+\ell'}_{\ell'}(t)
\mathfrak{q}^{(s)}_{0, \ell}(s)\,Q^{2\D_\f+\ell}_{\ell}(t)\,.
\ee
We multiply both side by $Q^{2\D_\f+\tilde{\ell}}_{\tilde{\ell},0}(t)$ and use the orthonormality of the continuous Hahn polynomials. 
%Note that there is the OPE coefficient $C_{0, \ell}$ sitting outside every term in \eqref{diff4}.
 Comparing the coefficients of $Q^{2\D_\f+\tilde{\ell}}_{\tilde{\ell},0}(t)$ we obtain,
\begin{align}\label{diff1}
&\sum_{\ell=\tilde{\ell}+1}^{\infty} d_{\tilde{\ell}, \ell} (\D_\f)= \mathfrak{q}^{(s)}_{0,\tilde{\ell}}(\D_\f)-q^{(s)}_{\D, \tilde{\ell}}(\D_\f) %+\sum_{\ell'=\tilde{\ell}+1}^{\infty} \mathfrak{q}^{(s)}_{0,\ell'}\,\bar{c}_{\tilde{\ell}, \ell'}
\qquad {\rm{for \quad even}}\quad \tilde{\ell},\nn
%\ee
%\be
&\sum_{\ell=\tilde{\ell}+1}^{\infty} d_{\tilde{\ell}, \ell} (\D_\f)= 0 \qquad {\rm{for \quad odd}}\quad \tilde{\ell}\,.
\end{align}
Similarly we have an analogous constraint from the derivative expression,
\begin{align}\label{diff2}
&\sum_{\ell=\tilde{\ell}+1}^{\infty} d'_{\tilde{\ell}, \ell}(\D_\f) = \mathfrak{q'}^{(s)}_{0,\tilde{\ell}}(\D_\f)-{q'}^{(s)}_{\D, \tilde{\ell}}(\D_\f) %+\sum_{\ell'=\tilde{\ell}+1}^{\infty} \mathfrak{q}^{(s)}_{0,\ell'}\,\bar{c}_{\tilde{\ell}, \ell'}
\qquad {\rm{for \quad even}}\quad \tilde{\ell},\nn
%\ee
%\be
&\sum_{\ell=\tilde{\ell}+1}^{\infty} d'_{\tilde{\ell}, \ell}(\D_\f) = 0 \qquad {\rm{for \quad odd}}\quad \tilde{\ell}\,.
\end{align}
%Here we have used the fact that $\tilde{\ell} \leq \ell-1$  and $\tilde{\ell} \leq \ell'-1$ in order to have a non-zero $d_{\ell', \ell}$ and $\bar{c}_{\ell, \ell'}$. 
%Finally we have,
%\begin{align}\label{ambig}
%\sum_{\ell=\tilde{\ell}+1}^{\infty} d_{\tilde{\ell}, \ell}& = \mathfrak{q}^{(s)}_{0,\tilde{\ell}}-q^{(s)}_{\tilde{\ell}} +\sum_{\ell=\tilde{\ell}+1}^{\infty}\mathfrak{q}^{(s)}_{0,\ell}\, \bar{c}_{\tilde{\ell}, \ell}\nn
%&=b_1-b_2\,.
%\end{align}
%where the right hand side is given in \eqref{g2} and \eqref{g3} with $\ell'$ replaced by $\tilde{\ell}$.
The difference can be read off from \eqref{diff} and \eqref{derdiff} respectively. It will be interesting to see if the results of \cite{spinningads,sleight} can be used to solve these constraints (at least for the large spin operators). 
\section{Universal asymptotics and analyticity in spin}
Let us discuss a very interesting consequence of the formulae in eq. (\ref{anmdim2}) and \eqref{ope2}. First, following \cite{aldayzhiboedov} let us plot $r_\g=\sqrt{|\frac{\gamma^{(k+1)}_{0,\ell}}{\gamma^{(k)}_{0,\ell}}|}$. As is evident from the log-log plot below, for large $k$, $r_\g$ asymptotes to $k/\pi$. This is exactly what was found analytically in \cite{aldayzhiboedov} for certain integer dimension scalar exchange and numerically for the $\epsilon$ exchange in the 3d Ising case. Remarkably, this behaviour seems to be universal for any exchange, not just scalars. Using the explicit equation (\eqref{anmdim2}) we can explain this finding. By explicitly checking the expression\footnote{While this can presumably be established more rigorously, our claim is based on explicit checks and the numerical results in the plots. A further comment is that for twist 2 exchange, the behaviour is completely different and the graph flattens out--this appears to be an exception and is consistent with the claims in \cite{holorecon}.}, the large $k$ limit is dominated by $k_1=k, q=n=k_2=0$ with the ensuing sum over generalized Bernoulli polynomials in eq.(\ref{dd1}) being dominated by the top term. Using the asymptotic behaviour of the generalized Bernoulli polynomial \cite{temme}
\be \label{Basy}
B_n^{(\mu)}(z)=\frac{2 n! n^{\mu-1}}{(2\pi)^n \G(\mu)}\left(\cos \pi(2z+\mu-\frac{1}{2}n)+O(1/n)\right)\,,
\ee 
we find that 
\be\label{rasy}
r_\g\sim \frac{k}{\pi}+\frac{1}{4\pi}+O(1/k)\,.
\ee
Here the $O(1/k)$ correction in eq.(\ref{rasy}) and $\D_\phi$ enter only at the $O(1/k)$ order.
An immediate feature of this is that it is the same ratio for any spin or any dimension exchange for any $\D_\phi$. Furthermore, the straight line behaviour is approached from above according to this formula. $\gamma^{(k)}_{0,\ell}$'s in the large $k$ limit are alternating in sign. The OPE ratio $r_{OPE}=\sqrt{|\frac{\delta C^{(k+1)}_{0,\ell}}{\delta C^{(k)}_{0,\ell}}|}$ asymptotic behaviour is the same (see fig. 2).

\begin{figure}
	\begin{tabular} {c c c}
		\includegraphics[width=0.31\textwidth,height=3.5cm]{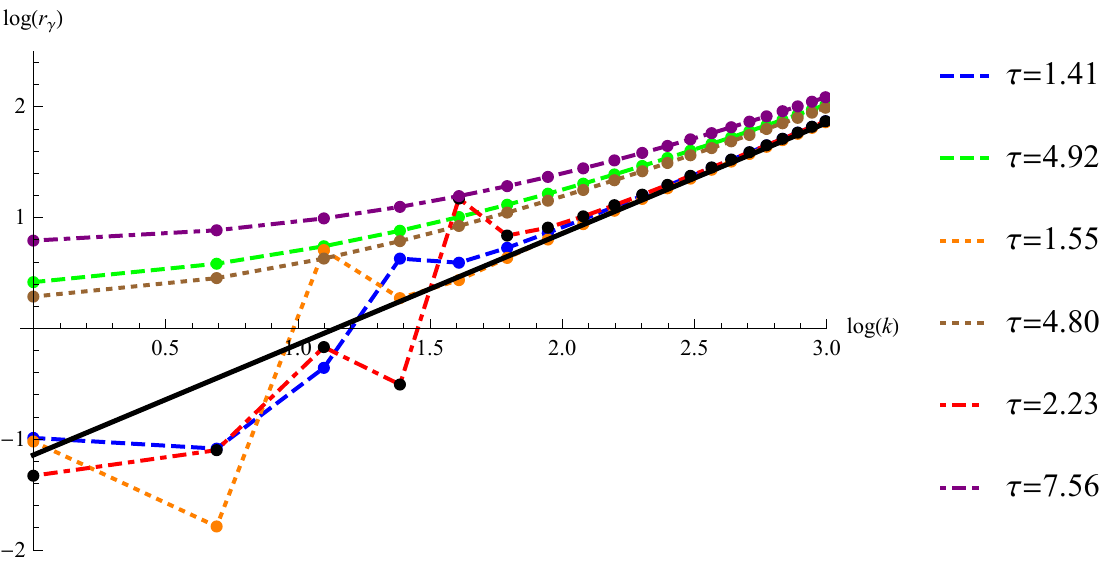} &\includegraphics[width=0.31\textwidth, height=3.5cm]{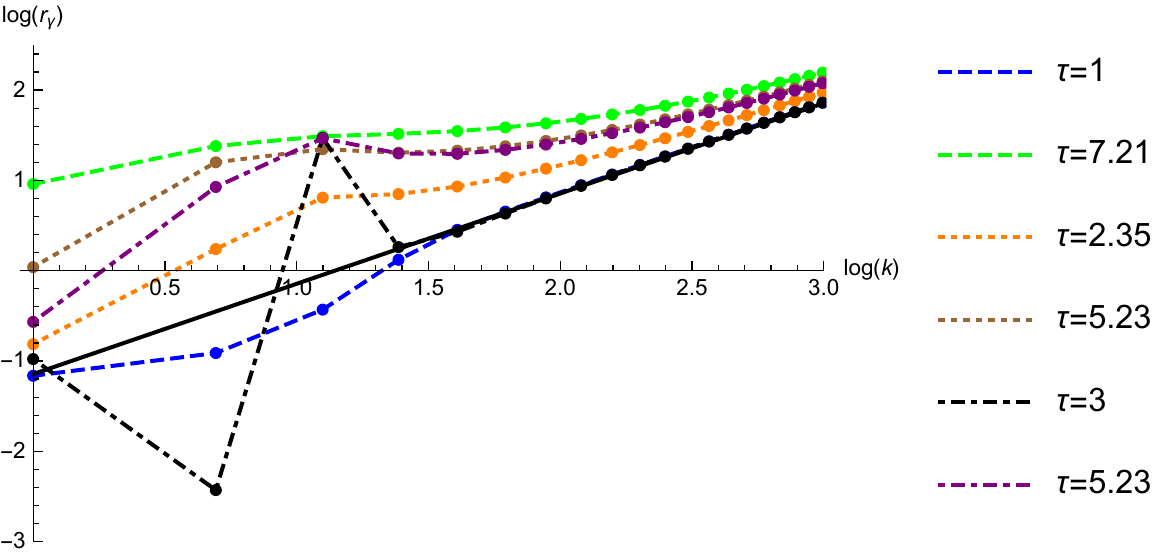} & \includegraphics[width=0.31\textwidth,height=3.5cm]{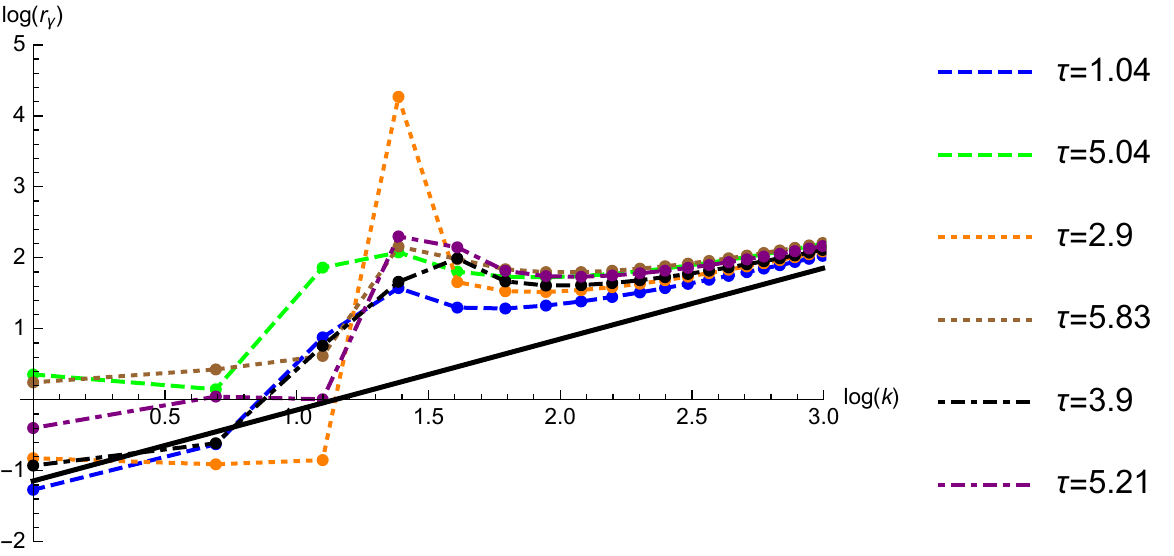}\\
		$ \ell=0$ & $\ell=2$ & $\ell=4$ \\
	\end{tabular} 
	\caption{Plots of $\log {r_\gamma}$ vs $\log k$ for various twists and spins. The solid black line is $r_\g=k/\pi$. The dashed lines are for $d=3,\D_\phi=0.518$, the dotted lines are for $d=4,\D_\phi=1.28$ and the dot-dashed lines for $d=5,\D_\phi=1.88$.}
\end{figure}

So what does one gain by knowing this? In \cite{aldayzhiboedov} it was pointed out that the series can be Borel resummed. We will point out another feature of these asymptotics. A series of the sort 
$$\sum_{n=0}^\infty (-1)^n \frac{a_n}{J^{2n}}\,,
$$
is called a Stieltjes series \cite{bender} if $$ a_n=\int_0^\infty W(y) y^n dy\,, $$ where $W(y)$ is a positive weight function (i.e., $W(y)>0$ for $y>0$). An example where $a_{n+1}/a_n\rightarrow n^2/\pi^2+n/(2\pi^2)$ as in the anomalous dimensions above for large $n$ (for the choice we make this is for any $n$) is obtained by choosing $W(y)=\exp(-2\pi\sqrt{y}-\log y)$.  From here one can construct an analytic function $f(j)$
\begin{figure}
	\begin{tabular} {c c c}
		\includegraphics[width=0.31\textwidth,height=3.5cm]{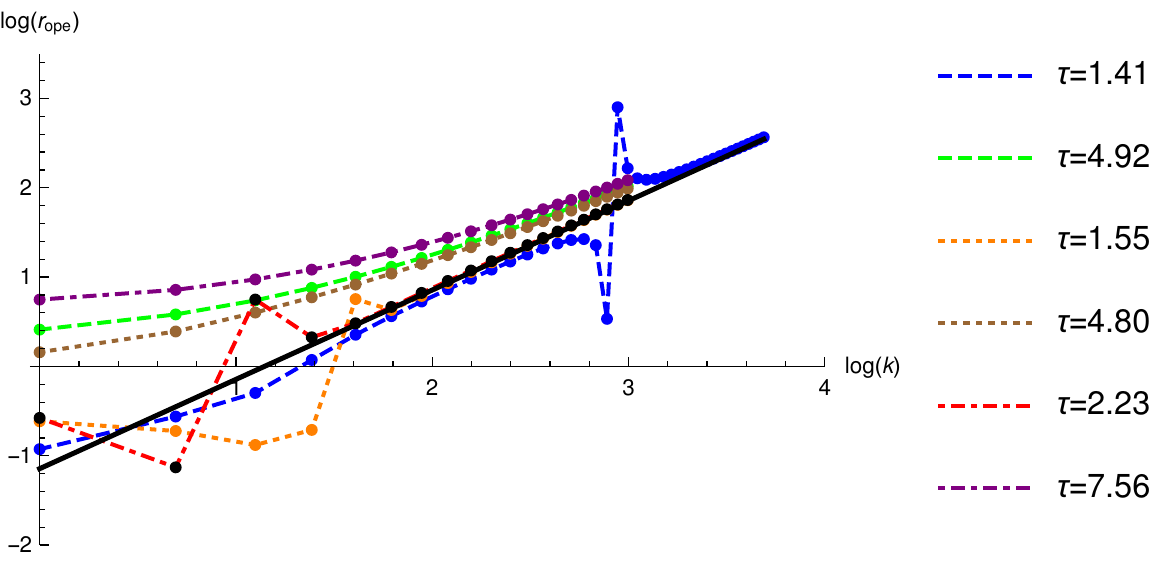} &\includegraphics[width=0.31\textwidth, height=3.5cm]{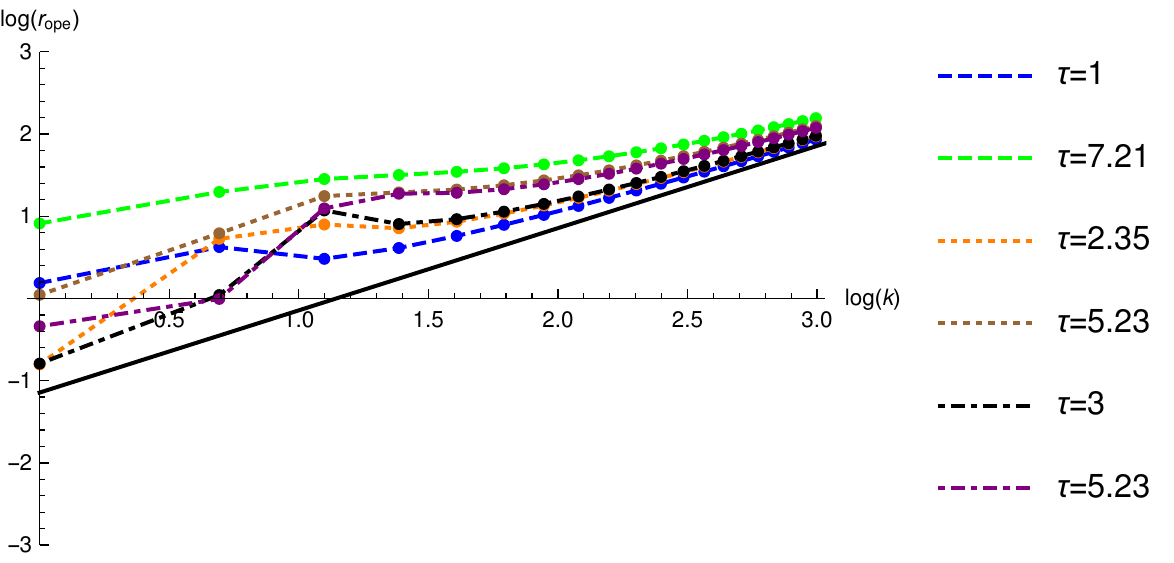} & \includegraphics[width=0.31\textwidth,height=3.5cm]{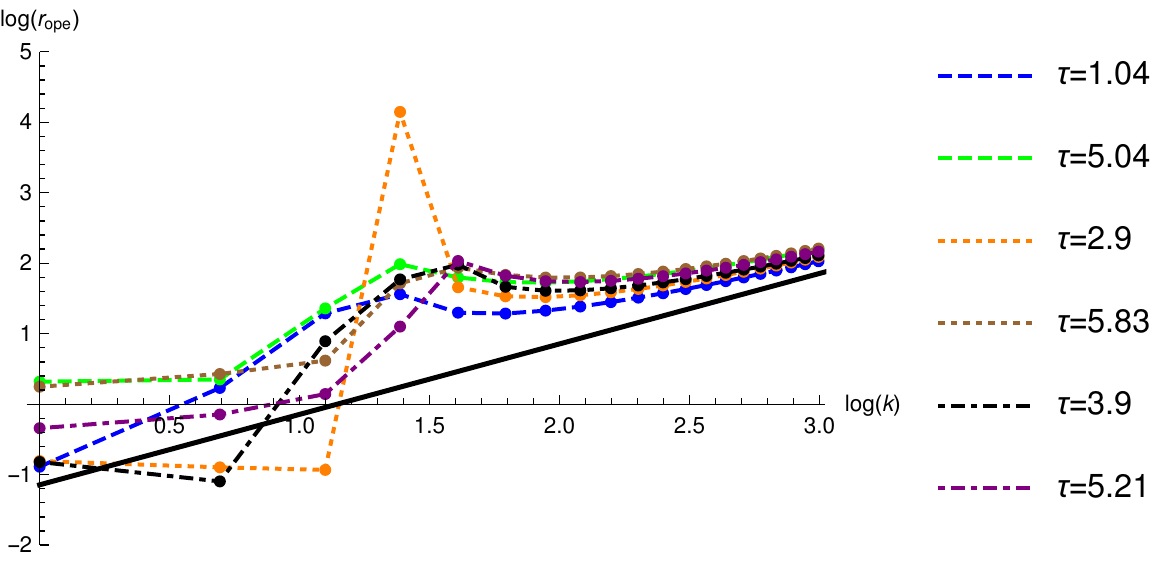}\\
		$ \ell=0$ & $\ell=2$ & $\ell=4$ \\
	\end{tabular} 
	\caption{Plots of $\log {r_{OPE}}$ vs $\log k$ for various twists and spins. The solid black line is $r_{OPE}=k/\pi$. The dashed lines are for $d=3,\D_\phi=0.518$, the dotted lines are for $d=4,\D_\phi=1.28$ and the dot-dashed lines for $d=5,\D_\phi=1.88$. The OPE for the $\tau=1.41, d=3$ exhibits an interesting feature which needed us to go to a higher number of points to see the asymptotic behaviour.}
\end{figure}

$$
f(j)=\int_0^\infty \frac{W(y)}{1+j y } dy\,,
$$
which is analytic except on the negative real axis in the $j$ complex plane, goes to zero as $j\rightarrow 0$, admits the asymptotic expansion $f(j)\sim \sum_{n=0}^\infty (-1)^n a_n j^{n}$ so that we can identify $j=1/J^2$ in our case.  $-f(j)$ also satisfies and important property called the Herglotz property which means that $-{\rm Im}~ f$ has the same sign as $-{\rm Im}~j$ in complex $j$ plane.  This essentially implies that the function $f(j)$ must have singularities in the complex plane. From here, it can be shown that defining
\be
A(j)=\frac{1}{2 i} \lim_{\e\rightarrow 0} \left(f[j+i\e]-f[j-i\e]\right)\,,
\ee
then one can get a dispersion relation (where the cut is on the negative real axis in $j$) for the coefficients $a_n$:
\be
a_n=\frac{1}{\pi}\int_{-\infty}^0 dj \,\frac{A(j)}{j^{n+1}}\,.
\ee
Another nice property of the above Stieltjes series is that since the $a_n$'s satisfy the Carleman condition \footnote{In our example, this condition is saturated whereas for the energy for an anharmonic oscillator with $x^4$ perturbation, the $a_n$'s grow like $n!3^n$ \cite{bender} while for $x^6$ it grows like $(2n)!$ .}, namely $a_n$'s do not grow faster than $(2n)!\a^n$ with $\a=1/(2\pi)^2$ in this case, then the $1/J$ series can be Pad\'e resummed (to a unique value). Further when the Carleman condition is met, the weight function $W(y)$ is unique--one may idly wonder if the uniqueness is telling us of an (unique?) effective theory for the large spin sector. 
These comments are tied in with the analyticity in spin observations of Caron-Huot's \cite{caronhout}. It should be possible to find the weight function for the actual series for the anomalous dimensions and OPE coefficients rather than the related example above and hence obtain $A(j)$. The uniqueness property of $W(y)$ suggests that it should be possible to dress this function to obtain the weight function for the actual series such that the leading asymptotics for $a_n$'s remain the same. This would be a direct way of calculating $a_n$ much like the inversion formula that \cite{caronhout} has for the OPE coefficients. Our all orders formula is of course doing precisely this but we have not used the powerful analyticity properties--it is very likely many of the steps may simplify further if we were to do this and it will be desirable to establish the relevant Mellin space techniques for the same.

%\section{Discussion}
%$(\log u)^2$--what is the equivalent statement in the new approach; connection with AdS loops; need for a more efficient method for the non-zero $n$ case; connection with Caron-Huot; connection with Simmons-Duffin?? applications of asymptotic form for ${}_3F_2$ to spinning correlators [[need a concrete expression]]; finite support story....there is an exact t channel calculation--what does it contain?
\section{Discussion}

In this paper we have worked out explicit expressions for the anomalous dimensions and OPE coefficients of large spin double trace operators $\D=2\D_\phi+\ell+\g_\ell$ to all orders in inverse conformal spin by exploiting Mellin space techniques. We also demonstrated the equivalence of the usual approach and new approach in terms of Witten diagrams to $O(\g_\ell)$ and worked out the constraint on the polynomial ambiguity piece at $O(\g_\ell^2)$. The main formula that we derived in this paper which made a systematic exploration of this possible was the large parameter asymptotics for the continuous Hahn polynomials. There are several future directions to pursue:

\begin{itemize}

\item It should be possible to develop our methods for other twist operators $O_{n,\ell}$ and extend the holographic reconstruction calculations of \cite{heem, holorecon} to arbitrary dimensions (see \cite{kss, kss2} for related earlier work on this). We believe that Mellin space techniques are the best suited to address this issue and our methods will prove useful in this venture.

\item The connection between the usual approach and the new approach for $\ell=0$ operators rather than large spin operators should be studied. The efficiency of the expansion in the Witten diagram basis that was found in our earlier work \cite{RGAKKSAS, PDAKAS} in the context of the epsilon expansion is still not properly understood and it will be gratifying to find a way to incorporate the same efficiency in the standard formulation of bootstrap. This paper shows that the polynomial ambiguity in the Witten diagram basis will not change the results in \cite{RGAKKSAS, PDAKAS} but will be relevant to go to the next order compared to these papers. Hence, developing the systematics of the epsilon expansion from bootstrap will need an understanding of these polynomial terms.
Our findings in this paper  will now make it possible \cite{PDAKnew} to extract the results for anomalous dimensions for the double trace operators at order $\frac{\e^5}{\ell^2}$ using the usual approach since this term only needs the $\e^3$ OPE coefficient and anomalous dimension of lower operators which can be calculated using the new approach. This will be a new prediction since Feynman diagram results at five loop order for higher spin operators are not available as of now.

%\item In the usual formalism, there are further constraints to be obtained by looking at $(\log u)^n$ terms. Such terms are absent in the new formalism as there we are not expanding small exponents. It will be interesting to understand how the Witten diagram basis encodes this information. Furthermore, there are pieces in the anomalous dimensions which have finite support in spin. These can be obtained by looking at the $\log u \log v$ terms in the bootstrap condition \cite{holorecon}. In \cite{holorecon}, a crossing symmetry condition was written down directly for the coefficient of the $\log u \log v$ term. From this it was shown that a finite support (in spin) piece is needed for the anomalous dimensions. Understanding these terms is vital to extract epsilon expansion results from the standard formulation. Hence these terms can be expected to play an important role in understanding the equivalence (once the polynomial ambiguities are constrained) with the new approach. We have left a study of these pieces from the Mellin space viewpoint  in the standard formulation for future work.

\item So far as the Witten diagram basis is concerned, it will be very important to understand how to fix the polynomial ambiguity so that it is consistent with the usual formulation. As of now, an independent governing principle for these terms is lacking. Our analysis in this paper suggests that in order to have a Witten diagram basis, we do not have freedom to adjust the polynomial ambiguity. It will be important to understand this issue more clearly \cite{ASnew}. Without understanding this point, it will be futile to attempt to do numerics using the Witten diagram basis.
 It will be interesting and important to derive the explicit expressions in this paper by exploiting the analyticity in spin \cite{caronhout}. We believe there will be further simplifications in the intermediate steps to be had and it is very likely that nontrivial constraints on the polynomial ambiguity will be found.

\item Our asymptotic formula for ${}_3F_2$ will likely find use in the analysis of external operators carrying spin \cite{zhiboedov2, polandnew}. It should be possible to find asymptotic anomalous dimensions and OPE coefficients for the relevant double trace operators appearing, for example, in the bootstrap constraints on $\langle J J \phi \phi\rangle$ where $J$ is an external conserved current.

\item Recently Tauberian theorems were proved in \cite{tauberian} which puts the lightcone limit of the bootstrap equations on firmer footing. These theorems were proved making use of position space. It will be interesting to see if any simplifications are to be had by making use of Mellin space.

\end{itemize}
\section*{Acknowledgments}
We thank B. Ananthanarayan, Apratim Kaviraj, Ahmadullah Zahed for discussions and especially Rajesh Gopakumar for numerous helpful discussions and comments on the draft. We also thank Fernando Alday and Eric Perlmutter for correspondence. A.S. acknowledges support from a DST Swarnajayanti Fellowship Award DST/SJF/PSA-01/2013-14.  

\appendix
\section{Asymptotics of continuous Hahn polynomial}
%{\bf AS: This section needs more references to the maths literature}
In the Mellin space approach considered in this paper, the key ingredient is the large spin asymptotics of the continuous Hahn polynomials. Unfortunately, a suitable form does not exist in the literature. However, we can piece together several existing results in the literature and come up with a very convenient form. This form will use the inverse conformal spin as the expansion parameter which will enable us to derive simple all order  expressions for the asymptotic expansions in the large conformal spin limit considered in the paper. These expressions will be in terms of generalized Bernoulli polynomials which are known and studied in the literature. We will outline the derivation in this appendix--the main formula is eq.(\ref{Qasym}).

The continuous Hahn polynomials can be derived as a limit from Wilson polynomials (see e.g.,\cite{AAR}). Wilson polynomials are defined as
\begin{equation}
W_n(x^2; a,b,c,d)=(a+b)_n (a+c)_n (a+d)_n\ {}_4F_3\bigg[\begin{matrix}-n,n+a+b+c+d-1,a+i x, a-ix\\
	\ \ a+b\ \ ,a+c \ \  a+d
\end{matrix};1\bigg]\,.
\end{equation}

The continuous Hahn polynomials are obtained through a limit\footnote{There is an important typo in \cite{AAR} as it misses the $t^n$ factor in the denominator! The correct expression is what we use from \cite{nist-hbk}.}
\begin{eqnarray}\label{lt1}
&&\lim_{t\rightarrow\infty} \frac{W_n((x+t)^2;a-it,b-it,c+it,d+it)}{(-2 t)^n n!}\nonumber \\
&=& i^n \frac{(a+c)_n(a+d)_n}{n!} {}_3F_2\bigg[\begin{matrix} -n,\, n+a+b+c+d-1,\,a+i x\\
	\ \ a+c \ \ , \ \ \ \ \ \  a+d
\end{matrix};1\bigg]\,.
\end{eqnarray}

Now thankfully, Wilson in 1991 \cite{Wilson} worked out the large argument ($n\rightarrow \infty$) asymptotics of the Wilson polynomials. 
\be \label{ltw}
W_n(x^2)=n!\sum_{k=0}^{r-1}u_k(i x)u_{n-k}(-ix)\frac{2ix-n+2k}{2ix}+\pi_r(ix)A(-ix)C_n O(n^{-2ix-2r})+c.c. \,,
\ee
where
\be
\pi_k(x)=(a+x)_k(b+x)_k(c+x)_k(d+x)_k\,,
\ee
and
\be
u_k(x)=\frac{\pi_k(x)}{k!(1+2x)_k}\,,\quad A(x)=\frac{\Gamma(2x)}{\G(a+x)\G(b+x)\G(c+x)\G(d+x)}\,.
\ee

We are interested in finding the asymptotics of the ${}_3F_2$ that appears in the definition of the continuous Hahn polynomials:

\be\label{Qdefn}
{Q}^{2s+\ell}_{\ell,0}(t)= \frac{2^{\ell}\,((s)_{\ell})^2}{(2s+\ell-1)_{\ell}}\,{}_3F_2\bigg[\begin{matrix} -\ell,\, 2s+\ell-1,\,s+t\\
	\ \ s \ \ , \ \ \ \ \ \  s
\end{matrix};1\bigg]\,.
\ee
These polynomials satisfy the orthogonality property \cite{AAR},
\be\label{ortho}
\frac{1}{2\pi i}\int_{-i\infty}^{i\infty} dt \ \G^2(s+t)\G^2(-t) {Q}^{2s+\ell}_{\ell,0}(t) {Q}^{2s+\ell '}_{\ell',0}(t)=(-1)^\ell {\kappa}_{\ell}(s)\d_{\ell,\ell'}\,,
\ee
where,
\be\label{kappadef}
{\kappa}_{\ell}(s)= \frac{4^\ell \ell!}{(2s+\ell-1)_\ell^2}\frac{\G^4(\ell+s)}{(2s+2\ell-1)\G(2s+\ell-1)}\,.
\ee
Comparing with eq. (\ref{lt1}) we find $a=b=s, c=d=0, n=\ell$. Then using eq.(\ref{ltw}) we find 
\be
{}_3F_2\bigg[\begin{matrix} -\ell,\, 2s+\ell-1,\,s+t\\
	\ \ s \ \ , \ \ \ \ \ \  s
\end{matrix};1\bigg]\sim \sum_k {}^\ell C_k  (-1)^k\frac{(-t)_{\ell-k}^2 (s+t)_k^2}{(s)_\ell^2}+(t\rightarrow -s-t)
\ee
Now it can be verified that the asymptotic expansion for $\ell\rightarrow \infty$ satisfies\footnote{This can be verified on Mathematica. A word of warning: There appears to be a bug in Mathematica when it comes to expanding the inbuilt Pochhammer symbol of the sort $(a+n)_n$ for large $n$ as it gives identically 1! It is safer to rewrite the Pochhammers in terms of gamma functions and then do the expansion.}
\be \label{compform}
\sum_k {}^\ell C_k  (-1)^k\frac{(-t)_{\ell-k}^2 (s+t)_k^2}{(s)_\ell^2}\sim \sum_n \frac{(-1)^n \ell! \Gamma^2(s) \Gamma(\ell-n-1+s-t)\Gamma(n+s+t)}{n!\Gamma(2s+\ell-1)\Gamma^2(-n-t)\Gamma(s+t)\Gamma(1+\ell+n+s+t)}\,.
\ee
The reason we prefer this second, seemingly more complicated, form will become clear in a bit. The way we reached this second form was to begin with the inverse Mellin transform formula 
\begin{align}\label{Qint}
{}_3F_2\bigg[\begin{matrix} -\ell,a+\ell,b\\
\ \ c\ \ , \  \  d
\end{matrix};z\bigg]= \frac{1}{2\pi i} \int_{-i \infty}^{i \infty}d\xi\ z^{-\xi} \,F(\xi)
\end{align}
where,
\begin{align}
F(\xi)& = \frac{ \Gamma(c)\,\G(d)\,\G(b-\xi)\,\G(-\ell-\xi)\,\G(a+\ell-\xi)\,\G(\xi)}{\G(b)\,\G(-\ell)\,\G(a+\ell)\,\G(c-\xi)\,\G(d-\xi)}\nn
&= (-1)^{\xi}\,\ell!\,\frac{ \Gamma(c)\,\G(d)\,\G(b-\xi)\,\G(a+\ell-\xi)\,\G(\xi)}{\G(b)\,\G(1+\ell+\xi)\,\G(a+\ell)\,\G(c-\xi)\,\G(d-\xi)}
\end{align}

%Let us keep the arguments of the hypergeometric function general,
%\be
%{}_3F_2\bigg[\begin{matrix} -\ell,a+\ell,b\\
%	\ \ c\ \ , \  \  d
%\end{matrix};z\bigg]\,. 
%\ee
This $_3F_2$ can be mapped to the continuous Hahn polynomial \eqref{Qdefn} with the following mapping: $ a= 2s-1, b = s+t,  c=s, d=s$ and $z=1$. 
%The Mellin transform $F(\xi)$ of this hypergeometric function in terms of Mellin variable $\xi$ is given by,
%\begin{align}\label{Qint}
%{}_3F_2\bigg[\begin{matrix} -\ell,a+\ell,b\\
%\ \ c\ \ , \  \  d
%\end{matrix};z\bigg]= \frac{1}{2\pi i} \int_{-i \infty}^{i \infty} z^{-\xi} \,F(\xi)
%\end{align}
%where,
%\begin{align}
%F(\xi)& = \frac{ \Gamma(c)\,\G(d)\,\G(b-\xi)\,\G(-\ell-\xi)\,\G(a+\ell-\xi)\,\G(\xi)}{\G(b)\,\G(-\ell)\,\G(a+\ell)\,\G(c-\xi)\,\G(d-\xi)}\nn
%&= (-1)^{\xi}\,\ell!\,\frac{ \Gamma(c)\,\G(d)\,\G(b-\xi)\,\G(a+\ell-\xi)\,\G(\xi)}{\G(b)\,\G(1+\ell+\xi)\,\G(a+\ell)\,\G(c-\xi)\,\G(d-\xi)}
%\end{align}
%and we have used the reflection formula in the last line. Now let us consider the poles of $F(\xi)$. 
The RHS of eq.(\ref{compform}) are the residues at $\xi= b+n$ given by

\be\label{r1}
R^{n}_1= \frac{(-1)^n \,\G(\ell+1)\,\G(c)\,\G(d)\,\G(a-b+\ell-n)\,\G(b+n)}{n!\,\G(a+\ell)\,\G(c-b-n)\,\G(d-b-n)\,\G(b)\,\G(b+1+\ell+n)}\,.
\ee
% The residue at the second set gives,

Now let us focus on the $\ell$- dependent terms in \eqref{r1},
\be\label{r11}
\frac{\G(\ell+1)}{\G(a+\ell)}\times\frac{\G(a-b+\ell-n)}{\,\G(b+1+\ell+n)}\,.
\ee
The trick we will use now is to use the following identity \cite{Fields1964},
\be\label{idd}
\frac{\G(\la+\a)}{\G(\la+\b)} \sim \sum_{j=0}^{\infty}  \frac{\G(\b-\a+2j)}{\G(\b-\a)\,(2j)!}\, \mathcal{B}^{1+\a-\b}_{2j} (\frac{1+\a-\b}{2})\,\bigg(\lambda+ \frac{\a+\b-1}{2}\bigg)^{-2j}
\ee
where $\mathcal{B}$-s are the generalised Bernoulli polynomial. These are also known as the N{\o}rlund polynomials and are inbuilt in Mathematica. Explicit expressions in terms of ${}_2F_1$ are known through the work of \cite{srivastava}. 
By shifting the variable, $\lambda \rightarrow J$  where $J =\sqrt{\la\,(\la+\a+\b-1)}$ we can write,
\be\label{gamma2}
\frac{\G(\la+\a)}{\G(\la+\b)} \sim \sum_{k=0}^{\infty} {d}_{\a, \b, k} \, J^{\a-\b-2k}
\ee
where,
\be \label{dd1}
d_{\a, \b, k}= \sum_{j=0}^{k}\, c_j \, \binom{\frac{\a-\b-2j}{2}}{k-j}  \bigg(\frac{-1+\a+\b}{2}\bigg)^{2k-2j}
\ee
and
\be \label{dd2}
c_j = \frac{\G(\b-\a+2j)}{\G(\b-\a)\,(2j)!}\, \mathcal{B}^{1+\a-\b}_{2j} (\frac{1+\a-\b}{2})\,.
\ee
To shorten the notation, we will henceforth denote,
\be \label{dd3}
\mathfrak{b}_{k_1}(s) = d_{\a_1, \b_1, k_1}, \qquad
 \mathfrak{b}_{k_2, n}(t) = d_{\a_2, \b_2, k_2}
\ee
where, 
\be
\a_1=1-s=-\b_1, \quad  \a_2=-t-1-n=-\b_2\,. 
\ee
Now if we use the identity \eqref{idd} we get an asymptotic expansion for the two ratios of the Gamma functions in \eqref{r11},
\begin{align}
\frac{\G(\ell+1)}{\G(a+\ell)}  & \sim  \sum_{k_1=0}^{\infty} \mathfrak{b}_{k_1}(s)\,J^{\a_1-\b_1-2k_1}\nn
\frac{\G(a-b+\ell-n)}{\G(b+1+\ell+n)}  &\sim  \sum_{k_2=0}^{\infty} \mathfrak{b}_{k_2, n}(t)\,J^{\a_2-\b_2-2k_2}
\end{align}
%where $d_i$ stands for $d_{\a_i, \b_i, k_i}$ and 
where $ \la=\ell+s$ and $J^2=(\ell+s)(\ell+s-1)$.
%\be
%\a_1=1-s=-\b_1, \quad  \a_2=-t-1-n=-\b_2, \quad \la=\ell+s, \quad
% {\rm{and}} \,\,\,\,J^2=(\ell+s)(\ell+s-1)\nonumber\,.
% \ee
  Note that this $J$ is nothing but the conformal spin.
Using the above in \eqref{r11} we get  the following from \eqref{r1},
%\be
%\sim J^{\g}_0 \sum_{k_1, k_2=0}^{\infty}d^{(1)}_{k_1}\, d^{(2)}_{k_2} J^{-2k_1-2k_2}_0
%\ee
%with $J^2 = (\ell+s)\,(\ell+s-1)$\,.
%So \eqref{r1} is given by
\be
R^{n}_1 \sim \frac{(-1)^n}{n!}\,\frac{\G^2(s)\,\G(s+t+n)}{\G^2(-t-n)\,\G(s+t)}\,\sum_{k_1, k_2 =0}^{\infty}\mathfrak{b}_{k_1}(s)\, \mathfrak{b}_{k_2, n}(t)\,J^{-2k_1-2k_2-2s-2t-2n}\,.
\ee
 From the above expression it is evident that only even powers of $J$ will appear in the expansion after pulling out $J^{-2s-2t}$. Note that for this to happen it was important for $\alpha_1+\beta_1=\alpha_2+\beta_2=0$ to hold. Plugging it in \eqref{Qdefn} we get the large $\ell$ approximation of the $_3F_2$ hypergeometric function,

\begin{align}\label{Qasym}
{}_3F_2\bigg[\begin{matrix} -\ell,\, 2s+\ell-1,\,s+t\\
\ \ s \ \ , \  \  s
\end{matrix};1\bigg]  & \sim \sum_{n, k_1, k_2=0}^{\infty}\frac{(-1)^n}{n!}\frac{\G^2(s)\,(s+t)_n}{\G(-t-n)^2}\,\mathfrak{b}_{k_1}(s)\,\mathfrak{b}_{k_2, n}(t)\,  J^{-2k_1-2k_2-2n-2s-2t}\nn & 
 +\sum_{n, k_1, k_2=0}^{\infty}\frac{(-1)^n}{n!}\frac{\G^2(s)\,(-t)_n}{\G(s+t-n)^2}\,\mathfrak{b}_{k_1}(s)\,\mathfrak{b}_{k_2, n}(-s-t)\, J^{-2k_1-2k_2-2n+2t}\,.
\end{align}
%where $\hat{d}_i = {d}_i(t \rightarrow -s-t)$. 
The presence of the second term makes it evident that the $_3F_2$ is symmetric under the exchange $t \rightarrow -s-t$. Eq.(\ref{Qasym}) then is our primary formula with $\mathfrak{b}$'s defined via eqs. (\ref{dd1},\ref{dd2},\ref{dd3}). Note that in explicit calculations in position space we pick up either of the series but not both, depending on how we choose to close the contour.

%Note that the second term above can be obtained from the first term by replacing $t \rightarrow -s-t$.
%$\hat{\a}_2=s+t-1-n=-\hat{\b}_2$

%\section{Large spin sum}

\section{Calculation details for section 2}\label{Mellinconf}
We will give some algebraic details of section 2.
The large spin behavior of \eqref{lhslog} in terms of $J$ reads, 
\begin{align}
lhs|_{\log u} & = \frac{2 }{\G(\D_{\f})^2}\sum_{J}{\widetilde{\sum}} \int \frac{dt}{2\pi i} v^t\,\g^{(q)}_{0, \ell} \,\frac{(-1)^n}{4^r n!} \binom{\half}{r} \mathfrak{b}_{k_1}(\D_\f)\,\mathfrak{b}_{k_0}(2-\D_\f) \,\mathfrak{b}_{k_2, n}(-\D_\f-t)\nn & \times \frac{\G^2(\D_\f+t)\,\G^2(-t)\,(-t)_n}{\G^2(\D_\f+t-n)}\,J^{\la}
\end{align}
where $\mathfrak{b}$'s are defined in \eqref{dd3}, 
%\be
%\la= 2\D_\f-2k_0-2r-1-\tau_m-2q+2t-2n-2k_1-2k_2
%\ee
%$\a_1=1-\D_\f=-\b_1, a_2= -1-n+\D_\f-t=-\b_2$,
\be
{\widetilde{\sum}}= \sum_{k_0,k_1,k_2, r,q, n=0}^{\infty}\,.
\ee 
and $\lambda= -\tau_m-2q-2k_1-2k_2-2n+2t+2\D_\f-1-2k_0-2r$.  
%The  $d_i$'s defined in Appendix. 
%We put a lower cut-off  $J_0$ for the large $J$ sum.
Note that there is a $v^t\,J^{2t}$ dependence in the integrand.
Since we will be working in the limit $J^2\,v \gg 1$  we will close the $t$-contour on the left side of the complex $t$ plane. %We put some lower cut-off $\tilde{J}_0$ on the conformal spin $J$. 
 The $J$ sum can be done using the formula given in \eqref{Jsum},
\begin{align}
lhs|_{\log u} & = \frac{2}{\G(\D_{\f})^2} {\widetilde{\sum}}\int \frac{dt}{2\pi i} v^t\,\g^{(q)}_{0, \ell}\, \frac{(-1)^n}{4^r n!} \binom{\half}{r} \, \frac{\G^2(\D_\f+t)\,\G^2(-t)\,(-t)_n}{\G^2(\D_\f+t-n)} \nn
& \times  \mathfrak{b}_{k_1}(\D_\f)\,\mathfrak{b}_{k_0}(2-\D_\f)\, \mathfrak{b}_{k_2, n}(-\D_\f-t)\bigg(2^{2\a-3}\,\frac{\G(1-\a)\,\G(\a-\half)}{\sqrt{\pi}}\bigg)
\end{align}
where $\a=-\frac{\la}{2}$. In the above integral the $\G(\a-\half)$ has poles at $\a=\half-p$ ( for  $p=0,1,2, \cdots$). This will in turn introduce poles at $t=-\D_\f+k_0+r+\frac{\tau_m}{2}+q+n+k_1+k_2+p $\,. We choose the contour such that the poles of $\G(1-\a)$ will always lie outside the contour. Evaluating the residue  at this pole we get,

\begin{align}\label{log2}
 lhs|_{\log u}  =- {\widehat{\sum}} v^{-\Delta_\phi +\hat{k}+\frac{{\tau_m}}{2}}\,\g^{(q)}_{0, \ell}\, {\G_s}
\end{align}
%\begin{align}
%lhs|_{\log u} & = -\frac{2}{\G(\D_{\f})^2} {\widetilde{\sum}}\int \frac{dt}{2\pi i} v^t\,c_q \frac{(-1)^n}{4^r n!} \binom{\half}{r} \hat{d}_{k_1}\,\hat{d}_{k_2}d_{k} \frac{\G(\D_\f+t)^2\G(-t)^2\,(-t)_n}{\G(\D_\f+t-n)^2} \nn
%& \times \bigg(2^{2\a-3}\,\frac{\G(1-\a)\,\G(\a-\half)}{\sqrt{\pi}}\bigg)
%\end{align}
where, 
\begin{align}\label{hatsum}
\hat{k} &=k_0+k_1+k_2+n+p+q+r, \qquad 
 {\widehat{\sum}}  =\sum_{p =0}^{\infty} {\widetilde{\sum}}\nn
{\G_s} &=(-1)^{1+n+p} 4^{-p-r}\binom{\frac{1}{2}}{r} \,\mathfrak{b}_{k_1}(\D_\f)\,\mathfrak{b}_{k_0}(2-\D_\f) \mathfrak{b}_{k_2, n}(-k_0-r-q-n-k_1-k_2-p-\frac{\tau_m}{2}) \nn & \times \G^2(k_0+r+q+n+k_1+k_2+p+\frac{\tau_m}{2})\nn & \times  \frac{\Gamma \left(p+\frac{1}{2}\right)\G(-k_0-k_1-k_2-p-q-r+\D_\f-\frac{\tau_m}{2})\,\G(-k_0-k_1-k_2-n-p-q-r+\D_\f-\frac{\tau_m}{2})}{2\,\sqrt{\pi } \,n!\, p!\,\G^2(\D_\f)\,\G^2(k_0+k_1+k_2+p+q+r+\frac{\tau_m}{2})}\,.
\end{align}
%with
%$\a_1 = 1-\D_\f = -\b_1 =-\a =\b$, 
%$\a_2 = -1+k_0+k_1+k_2+p+q+r+\frac{\tau_m}{2} = -\b_2$.
We want to extract the coefficient of  $v^{-\Delta_\phi +\hat{k}+\frac{{\tau_m}}{2}}$ from \eqref{log2}. In order to do that we will replace the $k_2$ in \eqref{log2} by $\hat{k}-k_0-k_1-n-p-q-r$ which is given by,
\begin{align}
&{{\sum_{new}}}\g^{(q)}_{0, \ell}\, \frac{(-1)^{n+p}\,4^{-p-r}}{2\,\sqrt{\pi }\, n!\, p!\, \G^2(\D_\f)\,\G^2(-n+\hat{k}+\frac{\tau_m}{2})}\binom{\half}{r}\,\G(\half+p) \G(-\hat{k}+\D_\f-\frac{\tau_m}{2})\,\G(n-\hat{k}+\D_\f-\frac{\tau_m}{2})\nn & \times \mathfrak{b}_{k_1}(\D_\f)\,\mathfrak{b}_{k_0}(2-\D_\f)\mathfrak{b}_{\hat{k}-k_0-k_1-n-p-q-r, n}(-k_0-r-q-n-k_1-k_2-p-\frac{\tau_m}{2})\,\G^2(\hat{k}+\frac{\tau_m}{2})%\nn & \times \G(\half+p) \G(-\hat{k}+\D_\f-\frac{\tau_m}{2})\,\G(n-\hat{k}+\D_\f-\frac{\tau_m}{2})\,
%\G(\hat{k}+\frac{\tau_m}{2})^2
\end{align}
where,
\be\label{newsum}
\sum_{new} = \sum_{k_0+k_1+n+p+q+r=0}^{\hat{k}}
\ee
%with,
%\be
%\ee
%\be
%{\sum} = \sum_{k=0}^{\hat{k}}\sum_{k_1=0}^{\hat{k}-k}\sum_{n=0}^{\hat{k}-k-k_1}\sum_{p=0}^{\hat{k}-k-k_1-n}\sum_{q=0}^{\hat{k}-k-k_1-n-p}\sum_{r=0}^{\hat{k}-k-k_1-n-p-q}\,.
%\ee

%Let us consider piece proportional  to $\log u$ on the $rhs$ of the bootstrap equation \eqref{lhs2}. %For simplicity let us first consider the scalar exchange.
%\begin{align}
%&rhs= -C_m v^{\frac{\tau_m}{2}-\D_\f} \log u \, \frac{\G({\D_m})}{\G(\frac{\D_m}{2})^2}\, _2F_1\left(\frac{\D_m}{2}, \frac{\D_m}{2}, 1-\frac{d}{2}+\D_m,; v \right)\,.
%\end{align}
%%where $C_m$ is the OPE coefficient of the minimal twist operator with twist $\tau_m$. 
%This can be expanded in a series in $v$ in the small $v$ limit.
%\begin{align}\label{rhs2}
%& rhs|_{\log u}= -\,C_m v^{\frac{\tau_m}{2}-\D_\f}  \, \frac{\G({\D_m})}{\G(\frac{\D_m}{2})^2}\, \sum_{\hat{k}} \frac{1}{\hat{k}!} \frac{\left(\frac{\D_m}{2}\right)^2_{\hat{k}}}{(1-\frac{d}{2}+\D_m)_{\hat{k}}}\, v^{\hat{k}}
%\end{align}

%Matching the coefficient of $v^{-\Delta_\phi +\hat{k}+\frac{{\tau_m}}{2}}$ from \eqref{lhs2} and \eqref{rhs2} we get the following recursion relation for the $c_q$-s,
\begin{align}
&{\sum_{new}}\g^{(q)}_{0, \ell}\,\frac{(-1)^{n+p}\,4^{-p-r}}{2\,\sqrt{\pi }\, n!\, p!\, \G^2(\D_\f)\,\G^2(-n+\hat{k}+\frac{\tau_m}{2})}\binom{\half}{r}\G(\half+p)
 \G(-\hat{k}+\D_\f-\frac{\tau_m}{2})\,\G(n-\hat{k}+\D_\f-\frac{\tau_m}{2})\nn & \times \mathfrak{b}_{k_1}(\D_\f)\,\mathfrak{b}_{k_0}(2-\D_\f)\,\mathfrak{b}_{\hat{k}-k_0-k_1-n-p-q-r, n}(-\hat{k}-\frac{\tau_m}{2})\,\G^2(\hat{k}+\frac{\tau_m}{2})\nn
%\nn & \times \G(\half+p)
% \G(-\hat{k}+\D_\f-\frac{\tau_m}{2})\,\G(n-\hat{k}+\D_\f-\frac{\tau_m}{2})\,
%\G(\hat{k}+\frac{\tau_m}{2})^2\nn
&=  -\,C_m  \, \frac{\G({\D_m})}{\G(\frac{\D_m}{2})^2}\,  \frac{1}{\hat{k}!} \frac{\left(\frac{\D_m}{2}\right)^2_{\hat{k}}}{(1-\frac{d}{2}+\D_m)_{\hat{k}}}\,.
\end{align}
%where
%\be
%\widetilde{\sum} = \sum_{k=0}^{i}\sum_{k_1=0}^{i-k}\sum_{n=0}^{i-k-k_1}\sum_{p=0}^{i-k-k_1-n}\sum_{q=0}^{i-k-k_1-n-p}\sum_{r=0}^{i-k-k_1-n-p-q}\,.
%\ee
\section{Recovering the $u$ channel}
In this section we will point out how to obtain the $u$-channel expression from the $s$-channel. This will in turn satisfy the bootstrap equation which demands the equality of $s$ and $u$ channel,

\be
\sum_{\D, \ell} C_{\D, \ell} \, G_{\D, \ell}(u, v) = {u}^{\D_{\f}}\, \sum_{\D, \ell} C_{\D, \ell} \, G_{\D, \ell}\left(\frac{1}{u}, \frac{v}{u}\right)
\ee
The Mellin transform of the term associated with the $\log u$ term in the $s$ channel is given by \eqref{lhslog}, 
 \begin{align}
 lhs|_{\log u} &= \sum_{\ell}  \frac{\g_{0, \ell}}{2}\int \frac{dt}{2\pi i}\,  v^t\,\G^2(\D_{\f}+t)\,\G^2(-t) \,\frac{2\,\G(-1+\ell+2\D_\f)}{\ell!\,\G^4(\D_{\f})}(2\ell+2\D_\f-1)\nn & \times
 {}_3F_2\bigg[\begin{matrix} -\ell,\, 2\D_{\f}+\ell-1,\,\D_\f+t\\
 \ \ \D_{\f} \ \ , \   \D_{\f}
 \end{matrix};1\bigg]\,.
 \end{align}
 Now we take the large spin limit of $_3F_2$ given in \eqref{Qasym}. In order to reproduce the $u$-channel we have to consider the first series in $J$ and we will be working in the regime $\frac{J^2}{v} \gg 1$. This will allow us to close the contour of the $t$ integral on the right side. Translated in terms of the conformal spin $J$ this reads,
 \begin{align}
 & lhs|_{\log u}  \sim -\frac{2}{\G(\D_{\f})^2}\sum_{J}{\widetilde{\sum}} \int \frac{dt}{2\pi i} v^t\,\g^{(q)}_{0, \ell} \frac{(-1)^n}{4^r n!} \binom{\half}{r} \mathfrak{b}_{k_1}(\D_\f)\,\mathfrak{b}_{k_0}(2-\D_\f) \,\mathfrak{b}_{k_2, n}(t) \nn & \times \frac{\G^2(\D_\f+t)\,\G^2(-t)\,(\D_\f+t)_n}{\G^2(-t-n)} J^{-2k-2r-1-\tau_m-2q-2t-2n-2k_1-2k_2}\nn
 &= -\frac{2}{\G^2(\D_{\f})}\sum_{J} {\widetilde{\sum}} \int \frac{dt}{2\pi i} v^t\,\g^{(q)}_{0, \ell}\frac{(-1)^n}{4^r n!} \binom{\half}{r}\mathfrak{b}_{k_1}(\D_\f)\,\mathfrak{b}_{k_0}(2-\D_\f) \,\mathfrak{b}_{k_2, n}(t) \nn & \times \frac{\G^2(\D_\f+t)\,\G^2(-t)\,(\D_\f+t)_n}{\G^2(-t-n)}
  2^{2\a-3}\,\frac{\G(1-\a)\,\G(\a-\half)}{\sqrt{\pi}}
 \end{align}
 where $\a=k_0+r+1/2+\tau_m/2+q+t+n+k_1+k_2$. Here the $t$ poles from the $\G(\a-\half)$ function will lie inside the contour which will in turn give rise to poles at $t=-k-r-\tau_m/2-q-n-k_1-k_2-p$. 

%$\hat{k}=k+k_1+k_2+n+p+q+r$, $\sum_{k, k_1,k_2, p, q, r, n =0}^{\infty} = {\widehat{\sum}}$ and 
%\begin{align}
%&\widehat{\G} =(-1)^{n+p} 4^{-p-r}\binom{\frac{1}{2}}{r} \,d_1\, d_2\, d_3 \nn
%&\times  \frac{\Gamma \left(p+\frac{1}{2}\right)\G(-k-k_1-k_2-p-q-r+\D_\f-\frac{\tau_m}{2})\,\G(-k-k_1-k_2-n-p-q-r+\D_\f-\frac{\tau_m}{2})}{\sqrt{\pi } \,n!\, p!\,\G(\D_\f)^2\,\G(k+k_1+k_2+p+q+r+\frac{\tau_m}{2})^2}\,.
%\end{align}

\section{Mack polynomials} %{\bf AS: combine sections on Mack polynomials together. First give conventions of mack polynomials then give the connection coefficients.}
Our conventions for the Mack polynomials\cite{Mack:2009mi, costa, Dolan:2011dv} are,
\be\label{Phat}
{P}_{\nu, \ell}(s, t) = (h+\nu-1)_{\ell}\, (h-\nu-1)_{\ell} \,\hat{P}_{\nu, \ell}(s, t)
\ee
where,
\be
\hat{P}_{\nu, \ell}(s, t)= \sum_{m=0}^{\ell}\sum_{n=0}^{ \ell-m}\mu^{(\ell)}_{m, n}\,\left(\frac{h+\nu-\ell}{2}-s\right)_m\,(-t)_n\,.
\ee
Here 
\begin{eqnarray}\label{mudef}
\mu_{m,n}^{(\ell)}&=&2^{-\ell} \frac{(-1)^{m+n}\ell!}{m! n! (\ell-m-n)!}(\frac{\D+\ell}{2}-m)_m (\frac{\tau}{2}+n)_{\ell-n} (\frac{\tau}{2}+m+n)_{\ell-m-n}(\ell+h-1)_{-m}(\ell+\D-1)_{n-\ell}  \nonumber \\
&\times& {}_4F_3[-m,1-h+\frac{\tau}{2},1-h+\frac{\tau}{2},n-1+\D;2-2h+\tau,\frac{\D+\ell}{2}-m,\frac{\tau}{2}+n;1]
\end{eqnarray}
and $h+\nu=\D$. The last ${}_4F_3$ is a well-balanced one and here $\tau=\D-\ell$ as usual. Further these have the symmetry that under $t\rightarrow -s-t$ they are invariant upto a $(-1)^\ell$ factor.
Hence when we consider the relevant Mack polynomial for the t-channel (which is obtained via $s\rightarrow t+\D_\phi, t\rightarrow s-\D_\phi$ from the s-channel one),
\be
\hat{P}_{\D-h, \ell}(t+\D_\f, s-\D_\f)= \sum_{m=0}^{\ell}\sum_{n=0}^{\ell-m}\mu^{(\ell)}_{m, n}\,(\frac{\D-\ell}{2}-\D_\f-t)_m\,(\D_\f-s)_n\,.
\ee
At $s=\D_\f$, $t=\frac{\D-\ell}{2}-\D_\f+q$ the above sum reduces to the following,
\be\label{phatt}
\hat{P}_{\D, \ell}(\frac{\D-\ell}{2}+q, 0)= \sum_{m=0}^{q}\,\mu^{(\ell)}_{m, n}\,(-q)_m\,.
\ee
We again note
\be
P_{\D-h, \ell}\left(s=\frac{\D-\ell}{2}, t\right) =4^{-\ell} {(\D-1)_{\ell}\,(2h-\D-1)_{\ell}}\,Q^{\D}_{\ell,0}(t)\,.
\ee
For possible future use, we record the relation for ${Q}^{2\s+\ell'}_{\ell',0}(t)$ in terms of ${Q}^{2s+\ell}_{\ell,0}(t)$. 
\begin{align}\label{q1}
{Q}^{2\s+\ell'}_{\ell',0}(t)%= \sum_{\ell=0}^{\ell'} \bar{c}_{\ell, \ell'}\,{Q}^{2s+\ell}_{\ell,0}(t)
=\delta_{\ell, \ell'}\,{Q}^{2s+\ell}_{\ell,0}(t)+\sum_{\ell=0}^{\ell'-1} \bar{c}_{\ell, \ell'}\,{Q}^{2s+\ell}_{\ell,0}(t)
\end{align}
where the connection coefficients are given by,
\begin{align}\label{c}
\bar{c}_{\ell, \ell'}& = \frac{2^{2 s-2 \sigma -\ell '+\ell +1}\,\Gamma \left(s+\ell +\frac{1}{2}\right)  \left( -\ell '\right)_{\ell} \Gamma \left(s+\ell '\right) \Gamma \left(\ell +2 \sigma +\ell '-1\right) \Gamma \left(-s-\ell +\sigma +\ell '\right)}{\Gamma (\ell +1) \Gamma (\sigma -s) \Gamma (\ell +\sigma )  \Gamma \left(2 s+\ell +\ell '\right) \Gamma \left(\sigma +\ell '-\frac{1}{2}\right)}\nn
& \times {}_4F_3\bigg[\begin{matrix} s+\ell,\,\ \ \s-s\ \ ,\,\ell-\ell' \ \ , \, -\sigma -\ell '+1\\
\ \ \sigma +\ell \ \ , \ \  -s-\ell '+1 \ \ , \ \ s-\sigma -\ell '+\ell +1
\end{matrix};1\bigg]\nn
&= -(s-\sigma)\,\frac{2^{-\ell '+\ell +1} \left(-\ell '\right)_{\ell } \Gamma \left(\ell +\sigma +\frac{1}{2}\right) \Gamma \left(\ell '-\ell \right) \Gamma \left(\sigma +\ell '\right) \Gamma \left(\ell +2 \sigma +\ell '-1\right)}{\Gamma (\ell +1) \Gamma (\ell +\sigma ) \Gamma \left(\sigma +\ell '-\frac{1}{2}\right) \Gamma \left(\ell +2 \sigma +\ell '\right)}+ O((s-\s)^2)\,.
\end{align}

\section{Details for $n \neq 0$}\label{sneq}
In this section, we will compute the anomalous dimension of the operators $O_{n, \ell}$ for $n \neq 0$. Let us
consider the Mellin transform of the $s$ channel conformal block,
\begin{align}
G^{(s)}(u, v) %& =\sum_{\D, \ell} c_{\D, \ell}\,G_{\D, \ell}(u, v)\nn
 &=\sum_{\D, \ell} c_{\D, \ell} \, \int \frac{ds}{2\pi i}\, \frac{dt}{2\pi i}\,u^s\,v^t\,  \G(s+t)^2\,\G(\D_\f-s)^2\,\G(-t)^2\, B_{\D, \ell}(s, t)\,.
\end{align}
%where,
%\be
%B_{\D, \ell}(s, t)= \frac{\G(\frac{\D-\ell}{2}-s)\,\G(\frac{2h-\D-\ell}{2}-s)}{\G(\D_\f-s)^2}\,P_{\D-h, \ell}(s, t)\,.
%\ee
%and  ${P}_{\D-h, \ell}(s, t)$ is the Mack polynomial defined in \eqref{P}. 
The above $s$ integral has poles from the Gamma functions in the numerator. We will pick up the pole  at $s=\frac{\D-\ell}{2}+m_1 $ for $m_1=0, 1, 2, \cdots$. Note that the $s$ contour must be closed on the right side of the complex $s$ plane because we will be working in the $u \ll 1$ limit. The residue at this pole is given by,
\begin{align}
G^{(s)}(u, v) &=\sum_{\D, \ell}\sum_{m_1=0}^{\infty}\frac{(-1)^{m_1}}{m_1!}\,u^{\frac{\D-\ell}{2}+m_1}\,c_{\D, \ell} \,\int \frac{dt}{2\pi i}\, v^t\,\G(\frac{\D-\ell}{2}+m_1+t)^2\,\G(-t)^2\,\G(h-\D-m_1)\nn & \times  P_{\D-h, \ell}\left(s=\frac{\D-\ell}{2}+m_1, t\right)\,.
\end{align}
Now we consider the exchange of spin $\ell$ and dimension $\D=2\D_\f+ 2n+ \ell+\g_{n, \ell}$ (where $\ell \gg 1$) operators in the $s$ channel. 
We  focus on the coefficient of the $\log u$ term. To leading order in $\g_{n, \ell}$ we have,
\begin{align}\label{qs1}
G^{(s)}(u, v)|_{\log u}& =\sum_{\ell=0}^{\infty}\sum_{n, m_1=0}^{\infty}\frac{(-1)^{m_1}}{m_1!}\left(\frac{\g_{n, \ell}}{2}\right)\,u^{\D_\f+n+m_1}\,C_{n, \ell}\,\,\int \frac{dt}{2\pi i}\, v^t\,\bigg[\G^2(\frac{\D-\ell}{2}+m_1+t)\,\G^2(-t)\nn & \times  \G(h-\D-m_1)\,\mathcal{N}_{\D, \ell} \, P_{\D-h, \ell}\left(\frac{\D-\ell}{2}+m_1, t\right)\bigg]_{\D=2\D_\f+ 2n+ \ell}
\end{align} 
where $\mathcal{N}_{\D, \ell} $ is defined in \eqref{normu}. 
Now we will use the  relation \eqref{Phat}.
%\be
%P_{\D-h, \ell}(s, t) = (\D-1)_{\ell}\,(2h-\D-1)_{\ell}\, \hat{P}_{\D-h, \ell}(s, t)
%\ee
%where,
%\begin{align}
%\hat{P}_{\D-h, \ell}(s, t) &= \sum_{m_2=0}^{\ell}\sum_{n_2=0}^{\ell-m_2}\,\mu^{\ell}_{m_2, n_2}\,(\frac{\D-\ell}{2}-s)_{m_2}\,(-t)_{n_2}\,,\nn
%\mu^{\ell}_{m_2, n_2} & =
%\end{align}
Note that
\be
\bigg(\frac{\D-\ell}{2}-s\bigg)_{m_2}\vert_{s=\frac{\D-\ell}{2}+m_1}= (-m_1)_{m_2}
\ee
Hence we must have $m_1 \geq m_2$ for the above expression to be non-zero. Then the sum reduces to the following,
\be
\hat{P}_{\D-h, \ell}(s, t) = \sum_{m_2=0}^{m_1}\sum_{n_2=0}^{\ell-m_2}\,\mu^{(\ell)}_{m_2, n_2}\,(\frac{\D-\ell}{2}-s)_{m_2}\,(-t)_{n_2}\,.
\ee
We now use \cite{ASnew} to write 
\be
(-t)_{n_2} = \sum_{\ell'=0}^{\infty} \chi^{(n_2)}_{\ell'}(\s)\,Q^{2\s+\ell'}_{\ell', 0}(t)
\ee
where,
\be
\chi^{(n_2)}_{\ell'}(\s)= 2^{-\ell'}(-1)^{\ell'}\frac{\G(2\s+2\ell')\,\G^2(\s+n_2)}{\ell'!\,\G^2(\ell'+\s)\,\G(2\s+n_2)}\frac{(-n_2)_{\ell'}}{(2\s+n_2)_{\ell'}}
\ee
and $\ell \geq n_2 \geq \ell'$ for $\chi^{(n_2)}_{\ell'}(\s)$ to be non-zero. Using these in \eqref{qs1} , we get,
\begin{align}\label{q2}
G^{(s)}(u, v)|_{\log u}& =\sum_{\ell'=0}^{\infty}\sum_{\ell=\ell'}^{\infty}\sum_{n, m_1=0}^{\infty}\sum_{m_2=0}^{m_1}\sum_{n_2=\ell'}^{\ell-m_2}u^{\D_\f+n+m_1}\int \frac{dt}{2\pi i}\, v^t\,\G^2(\D_\f+m_1+n+t)\,\G^2(-t)\,\frac{(-1)^{m_1}}{m_1!}\nn & \times \left(\frac{\g_{n, \ell}}{2}\right)\,C_{n, \ell}\,\mathcal{N}_{2\D_\f+2n+\ell, \ell}\,\G(h-2\D_\f-2n-\ell-m_1) (2\D_\f+2n+\ell-1)_{\ell}\,\nn & \times (2h-2\D_\f-2n-\ell-1)_{\ell}\,\mu^{(\ell)}_{m_2, n_2}\,(-m_1)_{m_2}\,\chi^{(n_2)}_{\ell'}(\s)\,Q^{2\s+\ell'}_{\ell', 0}(t)
\end{align}
where $\s =\D_\f+ m_1+n$. Let us denote $k=m_1+n$ such that $\s= \D_\f+k$ and we will replace $m_1$ by $k-n$ in what follows.  
We are interested in the coefficient of the continuous Hahn polynomial,
\be
G^{(s)}(u, v)|_{\log u} =\sum_{\ell'=0}^{\infty}\, u^{\D_\f+k}\int \frac{dt}{2\pi i}\, v^t\,\G(\D_\f+m_1+n+t)^2\,\G(-t)^2\, \mathfrak{q}^{(s)}_{k, \ell'}\,Q^{2\D_\f+2k+\ell'}_{\ell', 0}(t)
\ee

we obtain the coefficient of $u^{\D_\f+k}\,\log u$ from \eqref{q2} (to leading order in $\g_{n, \ell}$),

\begin{align}\label{lhsq}
\mathfrak{q}^{(s)}_{k,\ell'}&=\sum_{\ell=\ell'}^{\ell'+2k}\sum_{n=0}^{(\ell'+2k-\ell)/2}\sum_{m_2=0}^{k-n}\sum_{n_2=\ell'}^{\ell-m_2}\frac{(-1)^{k-n}}{(k-n)!}\left(\frac{\g_{n, \ell}}{2}\right)\mu^{(\ell)}_{m_2, n_2}\,(n-k)_{m_2}\,\chi^{(n_2)}_{\ell'}(\D_\f+k)\nn & \times\,C_{n, \ell}\,\mathcal{N}_{2\D_\f+2n+\ell, \ell}\,\G(h-2\D_\f-n-\ell-k)\, (2\D_\f+2n+\ell-1)_{\ell}\,(2h-2\D_\f-2n-\ell-1)_{\ell}\,.
\end{align}
%\section{$t$ channel for $n \neq 0$}
Let us expand  the $t$ channel expression in continuous Hahn polynomial basis. We begin by the Mellin transform of the $t$ channel conformal block.
\begin{align}\label{tch}
G^{(t)}(u, v)%& =\sum_{\D, \ell} c_{\D, \ell}\,G_{\D, \ell}(v, u)\nn
 &=\sum_{\D, \ell} c_{\D, \ell} \, \int \frac{ds}{2\pi i}\, \frac{dt}{2\pi i}\,u^s\,v^t\,  \G^2(s+t)\,\G^2(\D_\f-s)\,\G^2(-t)\, B_{\D, \ell}(t+\D_\f, s-\D_\f)
\end{align}
%where,
%\be
%B_{\D, \ell}(t+\D_\f, s-\D_\f)= \frac{\G(\frac{\D-\ell}{2}-\D_\f-t)\,\G(\frac{2h-\D-\ell}{2}-\D_\f-t)}{\G(-t)^2}\,P_{\D-h, \ell}(t+\D_\f, s-\D_\f)\,.
%\ee
Since we are interested in the coefficient of $\log u$ we have to consider the pole at $s=\D_\f+k$ for $k=0, 1, 2, \cdots$. The residue at this pole is given by,
\begin{align}
G^{(t)}(u, v)|_{\log u} &=\sum_{k=0}^{\infty}\sum_{\D, \ell} \,\frac{1}{(k!)^2}\,u^{\D_\f+k}\, \int \frac{dt}{2\pi i}\,v^t\,  \G^2(\D_\f+k+t)\,\G^2(-t)\nn &
\times c_{\D, \ell}\,\frac{\G(\frac{\D-\ell}{2}-\D_\f-t)\,\G(\frac{2h-\D-\ell}{2}-\D_\f-t)}{\G^2(-t)}\,P_{\D-h, \ell}(t+\D_\f, k)\,.
\end{align}
We will expand the above in continuous Hahn polynomial basis,
\be
G^{(t)}(u, v)|_{\log u} =\sum_{\ell'=0}^{\infty}\sum_{k=0}^{\infty}\,u^{\D_\f+k}\,\int \frac{dt}{2\pi i}\, v^t\,\G^2(\D_\f+k+t)\,\G^2(-t)\, \mathfrak{q}^{(t)}_{k, \ell'|\ell}\,Q^{2\D_\f+2k+\ell'}_{\ell', 0}(t)
\ee
such that,
\begin{align}
\mathfrak{q}^{(t)}_{k, \ell'|\ell} &= \frac{1}{(k!)^2}\,\int \frac{dt}{2\pi i}\, \G^2(\D_\f+k+t)\,\kappa_{\ell'}(\D_\f+k)^{-1}\,Q^{2\D_\f+2k+\ell'}_{\ell', 0}(t)\nn & \times c_{\D, \ell}\,{\G(\frac{\D-\ell}{2}-\D_\f-t)\,\G(\frac{2h-\D-\ell}{2}-\D_\f-t)}\,P_{\D-h, \ell}(t+\D_\f, k)\,.
\end{align}
Note that we are focusing on one particular exchange (spin $\ell$ and dimensin $\D$)  in the $t$ channel.
This $t$ integral can be evaluated using residue theorem. We compute the residue at $t= \frac{\D-\ell}{2}-\D_\f+r$  for  $r=0, 1, 2, \cdots$. The residue is given by,
\begin{align}\label{rhsq}
\mathfrak{q}^{(t)}_{k, \ell'|\ell} &=\sum_{r=0}^{\infty} \frac{(-1)^{r}}{(k!)^2\,r!}\, \G^2(\D_\f+k+t)\,\kappa_{\ell'}(\D_\f+k)^{-1}\,Q^{2\D_\f+2k+\ell'}_{\ell', 0}(t)\nn & \times c_{\D, \ell}\,{\G(\frac{2h-\D-\ell}{2}-\D_\f-t)}{P}_{\D-h, \ell}(t+\D_\f, k)\bigg]_{t= \frac{\D-\ell}{2}-\D_\f+r}\,.
\end{align}
We are now in a position to compare the powers of $u^{\D_\f+k}\, \log u$ from \eqref{lhsq} and \eqref{rhsq}. The bootstrap equation in Mellin space reads,
\be
\mathfrak{q}^{(s)}_{k, \ell'}+2\,\mathfrak{q}^{(t)}_{k, \ell'|\ell}=0\,.
\ee
Thus we have an algebraic equation which gives us a recursion relation for the anomalous dimension for each value of $k$,
\begin{align}
&\sum_{\ell=\ell'}^{\ell'+2k}\sum_{n=0}^{{(\ell'+2k-\ell)}/{2}}\sum_{m_2=0}^{k-n}\sum_{n_2=\ell'}^{\ell-m_2}\frac{(-1)^{k-n}}{(k-n)!}\left(\frac{\g_{n, \ell}}{2}\right)\mu^{(\ell)}_{m_2, n_2}\,(n-k)_{m_2}\,\chi^{(n_2)}_{\ell'}(\D_\f+k)\nn & \times\,C_{n, \ell}\,\mathcal{N}_{2\D_\f+2n+\ell, \ell}\,\G(h-2\D_\f-n-\ell-k)\, (2\D_\f+2n+\ell-1)_{\ell}\,(2h-2\D_\f-2n-\ell-1)_{\ell}\nn
&+\sum_{r=0}^{\infty} 2\,\frac{(-1)^{r}}{(k!)^2\,r!}\, \G^2(\D_\f+k+t)\,\kappa_{\ell'}(\D_\f+k)^{-1}\,Q^{2\D_\f+2k+\ell'}_{\ell', 0}(t)\nn & \times c_{\D, \ell}\,{\G(\frac{2h-\D-\ell}{2}-\D_\f-t)}{P}_{\D-h, \ell}(t+\D_\f, k)\bigg]_{t= \frac{\D-\ell}{2}-\D_\f+r}=0 \,.
\end{align}
\section{Normalisation in usual bootstrap}
%{\bf AS: shorten this section by pointing to the appendix of the long paper...give minimum number of steps}
In this section we derive the normalisation of the conformal blocks following  \cite{RGAKKSAS}. We choose the normalistion of the conformal blocks such that in the limit $u \sim 0$, $v \sim 1$ we have,
\be
c_{\D, \ell}\,G_{\D, \ell}(u, v) \sim C_{\D,\ell}u^{\frac{\D-\ell}{2}}\,(1-v)^{\ell} +\cdots\,.
\ee
We take the Mellin transform of the conformal block in the $s$ channel
%\begin{align}
%C_{\D, \ell}\,G_{\D, \ell}(u, v) & =C_{\D, \ell}\,\int ds\,dt\, u^s\,v^t\,\G(s+t)^2\,\G(-t)^2\,\G(\D_\f-s)^2\, B_{\D, \ell} (s, t)
%\end{align}
%where,
%\be
%B_{\D, \ell} (s, t)= \frac{\G(\frac{\D-\ell}{2}-s)\,\G(\frac{2h-\D-\ell}{2}-s)}{\G(\D_\f-s)^2}\,P_{\D-h, \ell}(s, t)
%\ee
and compute the residue at the physical pole $s= \frac{\D-\ell}{2}$,
\begin{align}
c_{\D, \ell}\,G_{\D, \ell}(u, v)
% & =C_{\D, \ell}\,u^{\frac{\D-\ell}{2}} \int dt\, \,v^t\,\G\left(\frac{\D-\ell}{2}+t\right)^2\,\G(-t)^2\, \G(h-\D)\,P_{\D-h, \ell}\left(\frac{\D-\ell}{2}, t\right)\nn 
& = c_{\D, \ell}\,u^{\frac{\D-\ell}{2}} \int \frac{dt}{2\pi i}\, \,v^t\,\G^2\left(\frac{\D-\ell}{2}+t\right)\,\G^2(-t)\, \G(h-\D)\,\frac{(\D-1)_{\ell}\,(2h-\D-1)_{\ell}}{4^{\ell}}\,Q^{\D}_{\ell, 0}(t)\,.
\end{align}
%where we have used the following relation to write the above in terms of the continuous Hahn polynomials,
%\be
%P_{\D-h, \ell}\left(\frac{\D-\ell}{2}, t\right)= \frac{(\D-1)_{\ell}\,(2h-\D-1)_{\ell}}{4^{\ell}}\,Q^{\D}_{\ell, 0}(t)\,.
%\ee
Now we expand $v^t$ in powers of $1-v$,
\begin{align}\label{sum}
v^t  = \sum_{m}(-1)^m\, \binom{t}{m}\,(1-v)^m
= \sum_{m}(-1)^m\, \frac{\G(t+1)}{m!\,\G(t-m+1)}\,(1-v)^m\,.
\end{align}
%{\textbf{from long paper}}
%
Note that the coefficient is a polynomial in $t$ of degree $m$. Theerefore we can write it as a sum over continuous Hahn polynomial $Q^{\D}_{\ell', 0}(t)$ for $0 \leq \ell' \leq m$. We have the following normalisation for $Q^{\D}_{\ell, 0}(t)$,
\be
Q^{\D}_{\ell, 0}(t) = 2^{\ell}\, t^{\ell} + O(t^{\ell-1})\,.
\ee
Then we have $(-1)^m\, \binom{t}{m} = \frac{(-1)^m\, 2^{-m}}{m!}Q^{\D}_{m, 0}(t) +\cdots$. We now use the orthogonality of $Q^{\D}_{\ell, 0}(t)$ polynomials \eqref{ortho} such that only $m =\ell$ survives in the expansion \eqref{sum}. Doing the $t$ integral we obtain the following,
\begin{align}
c_{\D, \ell}\,G_{\D, \ell}(u, v) & = u^{\frac{\D-\ell}{2}}\,(1-v)^{\ell}c_{\D, \ell} \bigg(\frac{(-1)^{\ell}\,2^{-\ell}\,\kappa_{\ell}(\frac{\D-\ell}{2})\,\G(h-\D)\,(\D-1)_{\ell}\,(2h-\D-1)_{\ell}}{\ell!\, 4^{\ell}}\bigg)
\end{align}
where $\kappa_{\ell}(s)$ is defined in \eqref{kappadef}. 
We thus find that $c_{\D, \ell}= \mathcal{N}_{\D, \ell}C_{\D, \ell}$,
\be\label{normu}
(\mathcal{N}_{\D, \ell})^{-1}= \frac{(-1)^{\ell}\,2^{-\ell}\,\kappa_{\ell}(\frac{\D-\ell}{2})\,\G(h-\D)\,(\D-1)_{\ell}\,(2h-\D-1)_{\ell}}{\ell!\, 4^{\ell}}\,.
\ee


\begin{thebibliography}{99}
%\bibliographystyle{unsrt}
%%%%%%%%%%%%%%introduction ref%%%%%%%%%%
%\cite{Rattazzi:2008pe}
\bibitem{Rattazzi:2008pe} 
  R.~Rattazzi, V.~S.~Rychkov, E.~Tonni and A.~Vichi,
  ``Bounding scalar operator dimensions in 4D CFT,''
  JHEP {\bf 0812}, 031 (2008)
  %doi:10.1088/1126-6708/2008/12/031
  [arXiv:0807.0004 [hep-th]].
  %%CITATION = doi:10.1088/1126-6708/2008/12/031;%%
  %358 citations counted in INSPIRE as of 09 Sep 2017

\bibitem{3dising}
S.~El-Showk, M.~F.~Paulos, D.~Poland, S.~Rychkov, D.~Simmons-Duffin and A.~Vichi,
``Solving the 3D Ising Model with the Conformal Bootstrap,''
Phys.\ Rev.\ D {\bf 86}, 025022 (2012)
%%doi:10.1103/PhysRevD.86.025022
[arXiv:1203.6064 [hep-th]].\\
S.~El-Showk, M.~F.~Paulos, D.~Poland, S.~Rychkov, D.~Simmons-Duffin and A.~Vichi,
``Solving the 3d Ising Model with the Conformal Bootstrap II. c-Minimization and Precise Critical Exponents,''
J.\ Stat.\ Phys.\  {\bf 157}, 869 (2014)
%%doi:10.1007/s10955-014-1042-7
[arXiv:1403.4545 [hep-th]].

\bibitem{mostprecise}
F.~Kos, D.~Poland, D.~Simmons-Duffin and A.~Vichi,
``Precision islands in the Ising and O(N) models,''
JHEP {\bf 1608}, 036 (2016)
%%doi:10.1007/JHEP08(2016)036
[arXiv:1603.04436 [hep-th]].

%\bibitem{dlce} 
%A.~L.~Fitzpatrick, J.~Kaplan, D.~Poland and D.~Simmons-Duffin,
%``The Analytic Bootstrap and AdS Superhorizon Locality,''
%%JHEP {\bf 1312}, 004 (2013) 
%[arXiv:1212.3616 [hep-th]]. \\
%Z.~Komargodski and A.~Zhiboedov,
%``Convexity and Liberation at Large Spin,''
%JHEP {\bf 1311}, 140 (2013)
%[arXiv:1212.4103 [hep-th]].




\bibitem{reviews}
S.~Rychkov,
``EPFL Lectures on Conformal Field Theory in $D\geq 3$  Dimensions,''
arXiv:1601.05000 [hep-th].\\
D.~Simmons-Duffin,
``TASI Lectures on the Conformal Bootstrap,''
arXiv:1602.07982 [hep-th].\\
J.~D.~Qualls,
``Lectures on Conformal Field Theory,''
arXiv:1511.04074 [hep-th].
%\cite{Penedones:2016voo}
%\bibitem{Penedones:2016voo} 
%%CITATION = doi:10.1142/9789813149441_0002;%%
%11 citations counted in INSPIRE as of 12 Sep 2017

%\bibitem{bootstrap}
%V.~S.~Rychkov and A.~Vichi,
%``Universal constraints on conformal operator dimensions,"
%Phys. Rev. \textbf{D80} (2009) 045006,
%arXiv: 0905.0211[hep-th].\\
%F.~Caracciolo and V.~S.~Rychkov,
%``Rigorous limits on the Interaction strength in Quantim field theory,"
%Phys. Rev. \textbf{D81} (2010) 085037,
%arXiv: 0912.2726[hep-th].\\
%D.~Poland and D.~Simmons-Duffin,
%``Bounds on 4D Conformal and Superconformal field theories,''
%JHEP \textbf{1105} (2011) 017,
%arXiv: 1009.2087.\\
%R.~Rattazzi, V.~S.~Rychkov and A.~Vichi,
%``Central charge bounds in 4D conformal field theory,"
%Phys. Rev \textbf{D83} (2011) 046011,
%arXiv: 1009.2725[hep-th].\\
%R.~Rattazzi, S.~Rychkov and A.~Vichi,
%``Bounds on 4D conformal field theories and global symmetry,"
%J. Phys \textbf{A44} (2011) 035402,
%arXiv:1009.5985[hep-th]\\
%D.~Poland, D.~Simmons-Duffin and A.~Vichi,
%``Carving out a space of 4D CFTs,"
%JHEP \textbf{1205} (2012) 110,
%arXiv: 1109.5176[hep-th].\\
%F.~Gliozzi,
%``More constraining conformal bootstrap,''
%Phys.\ Rev.\ Lett.\  {\bf 111}, 161602 (2013)
%% %doi:10.1103/PhysRevLett.111.161602
%[arXiv:1307.3111].\\
%F.~Gliozzi and A.~Rago,
%``Critical exponents of the 3d Ising and related models from Conformal Bootstrap,''
%JHEP {\bf 1410}, 042 (2014)
%%  %doi:10.1007/JHEP10(2014)042
%[arXiv:1403.6003 [hep-th]].\\
%Y.~Nakayama and T.~Ohtsuki,
%``Five dimensional $O(N)$-symmetric CFTs from conformal bootstrap,''
%Phys.\ Lett.\ B {\bf 734}, 193 (2014)
%% %doi:10.1016/j.physletb.2014.05.058
%[arXiv:1404.5201 [hep-th]].\\
%F.~Kos, D.~Poland, D.~Simmons-Duffin and A.~Vichi,
%``Bootstrapping the O(N) Archipelago,''
%JHEP {\bf 1511}, 106 (2015)
%% %doi:10.1007/JHEP11(2015)106
%[arXiv:1504.07997 [hep-th]].\\
%J.~D.~Qualls,
%``Universal Bounds on Operator Dimensions in General 2D Conformal Field Theories,''
%arXiv:1508.00548 [hep-th].\\
%M.~Hogervorst,
%``Dimensional Reduction for Conformal Blocks,''
%arXiv:1604.08913 [hep-th].

%\bibitem{3dising}
%S.~El-Showk, M.~F.~Paulos, D.~Poland, S.~Rychkov, D.~Simmons-Duffin and A.~Vichi,
%``Solving the 3D Ising Model with the Conformal Bootstrap,''
%Phys.\ Rev.\ D {\bf 86}, 025022 (2012)
%%%doi:10.1103/PhysRevD.86.025022
%[arXiv:1203.6064 [hep-th]].\\
%S.~El-Showk, M.~F.~Paulos, D.~Poland, S.~Rychkov, D.~Simmons-Duffin and A.~Vichi,
%``Solving the 3d Ising Model with the Conformal Bootstrap II. c-Minimization and Precise Critical Exponents,''
%J.\ Stat.\ Phys.\  {\bf 157}, 869 (2014)
%%%doi:10.1007/s10955-014-1042-7
%[arXiv:1403.4545 [hep-th]].

\bibitem{susy} 
%\cite{Beem:2013qxa}
%\bibitem{Beem:2013qxa} 
C.~Beem, L.~Rastelli and B.~C.~van Rees,
``The $\mathcal N=4$ Superconformal Bootstrap,''
Phys.\ Rev.\ Lett.\  {\bf 111}, 071601 (2013)
%doi:10.1103/PhysRevLett.111.071601
[arXiv:1304.1803 [hep-th]].\\
%%CITATION = doi:10.1103/PhysRevLett.111.071601;%%
%125 citations counted in INSPIRE as of 18 Sep 2017
M.~Lemos and P.~Liendo,
``Bootstrapping $ \mathcal{N}=2 $ chiral correlators,''
JHEP {\bf 1601}, 025 (2016)
%doi:10.1007/JHEP01(2016)025
[arXiv:1510.03866 [hep-th]].\\
%\cite{Bissi:2015qoa}
%\bibitem{Bissi:2015qoa} 
A.~Bissi and T.~Łukowski,
``Revisiting $ \mathcal{N}=4 $ superconformal blocks,''
JHEP {\bf 1602}, 115 (2016)
%doi:10.1007/JHEP02(2016)115
[arXiv:1508.02391 [hep-th]].\\
%\cite{Beem:2015aoa}
%\bibitem{Beem:2015aoa} 
C.~Beem, M.~Lemos, L.~Rastelli and B.~C.~van Rees,
``The (2, 0) superconformal bootstrap,''
Phys.\ Rev.\ D {\bf 93}, no. 2, 025016 (2016)
%doi:10.1103/PhysRevD.93.025016
[arXiv:1507.05637 [hep-th]].\\
%%CITATION = doi:10.1103/PhysRevD.93.025016;%%
%62 citations counted in INSPIRE as of 12 Sep 2017
%\cite{Kimura:2015bna}
%\bibitem{Kimura:2015bna} 
Y.~Kimura and R.~Suzuki,
``Negative anomalous dimensions in $\mathcal{N} =$ 4 SYM,''
Nucl.\ Phys.\ B {\bf 900}, 603 (2015)
%doi:10.1016/j.nuclphysb.2015.09.022
[arXiv:1503.06210 [hep-th]].
%%CITATION = doi:10.1016/j.nuclphysb.2015.09.022;%%
%4 citations counted in INSPIRE as of 12 Sep 2017


\bibitem{others}
A.~Liam Fitzpatrick, J.~Kaplan, M.~T.~Walters and J.~Wang,
``Eikonalization of conformal blocks,"
JHEP {\bf1509} (2015) 019 
arXiv:1504.01737[hep-th].\\
S.~Hellerman, D.~Orlando, S.~Reffert and M.~Watanabe,
``On the CFT operator spectrum at large global charge,"
arXiv: 1505.01537[hep-th].\\
T.~Hartman, S.~Jain and S.~Kundu,
``Causality Constraints in Conformal Field Theory,"
arXiv: 1509.00014[hep-th].\\
%\cite{Hartman:2016dxc}
%\bibitem{Hartman:2016dxc} 
%T.~Hartman, S.~Jain and S.~Kundu,
%``A New Spin on Causality Constraints,''
%JHEP {\bf 1610}, 141 (2016)
%doi:10.1007/JHEP10(2016)141
%[arXiv:1601.07904 [hep-th]].\\
%%CITATION = doi:10.1007/JHEP10(2016)141;%%
%16 citations counted in INSPIRE as of 12 Sep 2017
D.~M.~Hofman, D.~Li, D.~Meltzer, D.~Poland and F.~Rejon-Barrera,
``A Proof of the Conformal Collider Bounds,''
JHEP {\bf 1606}, 111 (2016)
% %doi:10.1007/JHEP06(2016)111
[arXiv:1603.03771 [hep-th]].\\   
%\cite{Li:2015itl}
%\bibitem{Li:2015itl} 
D.~Li, D.~Meltzer and D.~Poland,
``Conformal Collider Physics from the Lightcone Bootstrap,''
JHEP {\bf 1602}, 143 (2016)
%doi:10.1007/JHEP02(2016)143
[arXiv:1511.08025 [hep-th]].\\
%%CITATION = doi:10.1007/JHEP02(2016)143;%%
%21 citations counted in INSPIRE as of 12 Sep 2017
L.~Alvarez-Gaume, O.~Loukas, D.~Orlando and S.~Reffert,
``Compensating strong coupling with large charge,''
arXiv:1610.04495 [hep-th].

\bibitem{spins}
A.~Castedo Echeverri, E.~Elkhidir, D.~Karateev and M.~Serone,
``Deconstructing Conformal Blocks in 4D CFT,''
JHEP {\bf 1508}, 101 (2015)
% %doi:10.1007/JHEP08(2015)101
[arXiv:1505.03750 [hep-th]].\\
L.~Iliesiu, F.~Kos, D.~Poland, S.~S.~Pufu, D.~Simmons-Duffin and R.~Yacoby,
``Bootstrapping 3D Fermions,''
JHEP {\bf 1603}, 120 (2016)
%%doi:10.1007/JHEP03(2016)120
[arXiv:1508.00012 [hep-th]].\\
L.~Iliesiu, F.~Kos, D.~Poland, S.~S.~Pufu, D.~Simmons-Duffin and R.~Yacoby,
``Fermion-Scalar Conformal Blocks,''
JHEP {\bf 1604}, 074 (2016)
%%doi:10.1007/JHEP04(2016)074
[arXiv:1511.01497 [hep-th]].\\
A.~Castedo Echeverri, E.~Elkhidir, D.~Karateev and M.~Serone,
``Seed Conformal Blocks in 4D CFT,''
JHEP {\bf 1602}, 183 (2016)
%%doi:10.1007/JHEP02(2016)183
[arXiv:1601.05325 [hep-th]].\\
%L.~Iliesiu, F.~Kos, D.~Poland, S.~S.~Pufu, D.~Simmons-Duffin and %R.~Yacoby,
%``Fermion-Scalar Conformal Blocks,''
%JHEP {\bf 1604}, 074 (2016)
% %doi:10.1007/JHEP04(2016)074
%[arXiv:1511.01497 [hep-th]].\\
S.~Giombi, V.~Kirilin and E.~Skvortsov,
``Notes on Spinning Operators in Fermionic CFT,''
JHEP {\bf 1705}, 041 (2017)
%doi:10.1007/JHEP05(2017)041
[arXiv:1701.06997 [hep-th]].
%\cite{CastedoEcheverri:2016iwu}
%\bibitem{CastedoEcheverri:2016iwu}
%A.~Castedo Echeverri,
%``CFTs and the Bootstrap,''
%%CITATION = INSPIRE-1508397;%%


%\cite{Hogervorst:2017sfd}
\bibitem{hogerv} 
M.~Hogervorst and B.~C.~van Rees,
``Crossing Symmetry in Alpha Space,''
arXiv:1702.08471 [hep-th].\\
%%CITATION = ARXIV:1702.08471;%%
%11 citations counted in INSPIRE as of 11 Sep 2017
%\cite{Caron-Huot:2017vep}
%\bibitem{karateev} 
D.~Karateev, P.~Kravchuk and D.~Simmons-Duffin,
``Weight Shifting Operators and Conformal Blocks,''
arXiv:1706.07813 [hep-th].\\
%%CITATION = ARXIV:1706.07813;%%
%2 citations counted in INSPIRE as of 11 Sep 2017
%\cite{Cuomo:2017wme}
%\bibitem{cuomo} 
G.~F.~Cuomo, D.~Karateev and P.~Kravchuk,
``General Bootstrap Equations in 4D CFTs,''
arXiv:1705.05401 [hep-th].\\
%\cite{Codello:2017qek}
%\bibitem{codello} 
A.~Codello, M.~Safari, G.~P.~Vacca and O.~Zanusso,
``Leading CFT constraints on multi-critical models in $d > 2$,''
JHEP {\bf 1704}, 127 (2017)
%doi:10.1007/JHEP04(2017)127
[arXiv:1703.04830 [hep-th]].\\
%%CITATION = doi:10.1007/JHEP04(2017)127;%%
%8 citations counted in INSPIRE as of 11 Sep 2017
%\cite{Codello:2017hhh}
%\bibitem{Codello:2017hhh} 
A.~Codello, M.~Safari, G.~P.~Vacca and O.~Zanusso,
``Functional perturbative RG and CFT data in the $\epsilon$-expansion,''
arXiv:1705.05558 [hep-th].\\
%%CITATION = ARXIV:1705.05558;%%
%4 citations counted in INSPIRE as of 11 Sep 2017
%\cite{Hogervorst:2017kbj}
%\bibitem{hogerv2} 
M.~Hogervorst,
``Crossing Kernels for Boundary and Crosscap CFTs,''
arXiv:1703.08159 [hep-th].\\
%%CITATION = ARXIV:1703.08159;%%
%7 citations counted in INSPIRE as of 11 Sep 2017
%\cite{Rastelli:2017ecj}
%\bibitem{rastelli} 
L.~Rastelli and X.~Zhou,
``The Mellin Formalism for Boundary CFT$_d$,''
arXiv:1705.05362 [hep-th].\\
%%CITATION = ARXIV:1705.05362;%%
%4 citations counted in INSPIRE as of 11 Sep 2017
%%%%%%%%%%%%%%%%%%%%%%%%%%%%%%%%%%%%%%%%%%%%%%%%%%%
%\cite{Giombi:2016zwa}
%\bibitem{giombi} 
S.~Giombi, V.~Gurucharan, V.~Kirilin, S.~Prakash and E.~Skvortsov,
``On the Higher-Spin Spectrum in Large N Chern-Simons Vector Models,''
JHEP {\bf 1701}, 058 (2017)
%doi:10.1007/JHEP01(2017)058
[arXiv:1610.08472 [hep-th]].\\
%%CITATION = doi:10.1007/JHEP01(2017)058;%%
%14 citations counted in INSPIRE as of 11 Sep 2017
%\cite{Manashov:2016uam}
%\bibitem{manashov2} 
A.~N.~Manashov and E.~D.~Skvortsov,
``Higher-spin currents in the Gross-Neveu model at 1/n$^{2}$,''
JHEP {\bf 1701}, 132 (2017)
%doi:10.1007/JHEP01(2017)132
[arXiv:1610.06938 [hep-th]].\\
%%CITATION = doi:10.1007/JHEP01(2017)132;%%
%9 citations counted in INSPIRE as of 11 Sep 2017
%\cite{Lewkowycz:2016ukf}
%\bibitem{bulkloc} 
A.~Lewkowycz, G.~J.~Turiaci and H.~Verlinde,
``A CFT Perspective on Gravitational Dressing and Bulk Locality,''
JHEP {\bf 1701}, 004 (2017)
%doi:10.1007/JHEP01(2017)004
[arXiv:1608.08977 [hep-th]].\\
%%CITATION = doi:10.1007/JHEP01(2017)004;%%
%14 citations counted in INSPIRE as of 11 Sep 2017
%%CITATION = ARXIV:1705.05401;%%
%2 citations counted in INSPIRE as of 11 Sep 2017
%\bibitem{dutta} 
P.~Dutta, D.~Ghoshal and A.~Lala,
``On the Exchange Interactions in Holographic p-adic CFT,''
Phys.\ Lett.\ B {\bf 773}, 283 (2017)
%doi:10.1016/j.physletb.2017.08.042
[arXiv:1705.05678 [hep-th]].\\
%%CITATION = doi:10.1016/j.physletb.2017.08.042;%%
%1 citations counted in INSPIRE as of 11 Sep 2017
%\cite{Bagchi:2017cpu}
%\bibitem{bagchi} 
A.~Bagchi, M.~Gary and Zodinmawia,
``The nuts and bolts of the BMS Bootstrap,''
Class.\ Quant.\ Grav.\  {\bf 34}, no. 17, 174002 (2017)
%doi:10.1088/1361-6382/aa8003
[arXiv:1705.05890 [hep-th]].\\
%%CITATION = doi:10.1088/1361-6382/aa8003;%%
%4 citations counted in INSPIRE as of 11 Sep 2017
%\cite{Li:2017agi}
%\bibitem{Li} 
W.~Li,
``Inverse Bootstrapping Conformal Field Theories,''
arXiv:1706.04054 [hep-th].\\
%%CITATION = ARXIV:1706.04054;%%
%\cite{Li:2017kck}
%\bibitem{Li:2017kck} 
Z.~Li and N.~Su,
``3D CFT Archipelago from Single Correlator Bootstrap,''
arXiv:1706.06960 [hep-th].\\
%%CITATION = ARXIV:1706.06960;%%
%1 citations counted in INSPIRE as of 11 Sep 2017
%\cite{Sever:2017ylk}
%\bibitem{server} 
A.~Sever and A.~Zhiboedov,
``On Fine Structure of Strings: The Universal Correction to the Veneziano Amplitude,''
arXiv:1707.05270 [hep-th].\\
%%CITATION = ARXIV:1707.05270;%%
%3 citations counted in INSPIRE as of 11 Sep 2017
%\cite{Hikami:2017sbg}
%\bibitem{hikami} 
S.~Hikami,
``Conformal Bootstrap Analysis for Single and Branched Polymers,''
arXiv:1708.03072 [hep-th].\\
%%CITATION = ARXIV:1708.03072;%%
%\cite{DiPietro:2017kcd}
%\bibitem{di} 
L.~Di Pietro and E.~Stamou,
``Scaling dimensions in QED$_3$ from the $\epsilon$-expansion,''
arXiv:1708.03740 [hep-th].\\
%%CITATION = ARXIV:1708.03740;%%
%\cite{Parikh:2017gsf}
%\bibitem{parikh}
%\cite{Melby-Thompson:2017aip}
%\bibitem{melby} 
C.~Melby-Thompson and C.~Schmidt-Colinet,
``Double Trace Interfaces,''
arXiv:1707.03418 [hep-th].\\
%%CITATION = ARXIV:1707.03418;%%
%\cite{Isono:2017grm}
%\bibitem{isono} 
H.~Isono,
``On conformal correlators and blocks with spinors in general dimensions,''
arXiv:1706.02835 [hep-th].\\
%%CITATION = ARXIV:1706.02835;%%
%\cite{Zambelli:2016cbw}
%\bibitem{zambelli} 
L.~Zambelli and O.~Zanusso,
``Lee-Yang model from the functional renormalization group,''
Phys.\ Rev.\ D {\bf 95}, no. 8, 085001 (2017)
%doi:10.1103/PhysRevD.95.085001
[arXiv:1612.08739 [hep-th]].
%%CITATION = doi:10.1103/PhysRevD.95.085001;%%
%8 citations counted in INSPIRE as of 11 Sep 2017
%\cite{Kulaxizi:2017ixa}
\bibitem{Li2} 
F.~Kos, D.~Poland, D.~Simmons-Duffin and A.~Vichi,
``Bootstrapping the O(N) Archipelago,''
JHEP {\bf 1511}, 106 (2015)
%doi:10.1007/JHEP11(2015)106
[arXiv:1504.07997 [hep-th]].\\
D.~Li, D.~Meltzer and D.~Poland,
``Conformal Bootstrap in the Regge Limit,''
arXiv:1705.03453 [hep-th].\\
%%CITATION = ARXIV:1705.03453;%%
%7 citations counted in INSPIRE as of 12 Sep 2017
%\cite{Sleight:2017pcz}
%\cite{Giombi:2017rhm}
%\bibitem{Giombi:2017rhm} 
%S.~Giombi, V.~Kirilin and E.~Skvortsov,
%``Notes on Spinning Operators in Fermionic CFT,''
%JHEP {\bf 1705}, 041 (2017)
%%doi:10.1007/JHEP05(2017)041
%[arXiv:1701.06997 [hep-th]].\\
%%CITATION = doi:10.1007/JHEP05(2017)041;%%
%8 citations counted in INSPIRE as of 12 Sep 2017
%\cite{Hogervorst:2016hal}
%\bibitem{Hogervorst:2016hal} 
M.~Hogervorst,
``Dimensional Reduction for Conformal Blocks,''
JHEP {\bf 1609}, 017 (2016)
[arXiv:1604.08913 [hep-th]].\\
%%CITATION = doi:10.1007/JHEP09(2016)017;%%
%6 citations counted in INSPIRE as of 12 Sep 2017
%\cite{Lemos:2015awa}
%\bibitem{Lemos:2015awa} 
%M.~Lemos and P.~Liendo,
%``Bootstrapping $ \mathcal{N}=2 $ chiral correlators,''
%JHEP {\bf 1601}, 025 (2016)
%%doi:10.1007/JHEP01(2016)025
%[arXiv:1510.03866 [hep-th]].\\
%%\cite{Bissi:2015qoa}
%%\bibitem{Bissi:2015qoa} 
%A.~Bissi and T.~Łukowski,
%``Revisiting $ \mathcal{N}=4 $ superconformal blocks,''
%JHEP {\bf 1602}, 115 (2016)
%%doi:10.1007/JHEP02(2016)115
%[arXiv:1508.02391 [hep-th]].\\
%%CITATION = doi:10.1007/JHEP02(2016)115;%%
%11 citations counted in INSPIRE as of 12 Sep 2017
%%CITATION = doi:10.1007/JHEP01(2016)025;%%
%40 citations counted in INSPIRE as of 12 Sep 2017
%\cite{Fitzpatrick:2015foa}
%\bibitem{Fitzpatrick:2015foa} 
A.~L.~Fitzpatrick, J.~Kaplan, M.~T.~Walters and J.~Wang,
``Hawking from Catalan,''
JHEP {\bf 1605}, 069 (2016)
%doi:10.1007/JHEP05(2016)069
[arXiv:1510.00014 [hep-th]].\\
%%CITATION = doi:10.1007/JHEP05(2016)069;%%
%23 citations counted in INSPIRE as of 12 Sep 2017
%\cite{Rejon-Barrera:2015bpa}
%\bibitem{Rejon-Barrera:2015bpa} 
F.~Rejon-Barrera and D.~Robbins,
``Scalar-Vector Bootstrap,''
JHEP {\bf 1601}, 139 (2016)
%doi:10.1007/JHEP01(2016)139
[arXiv:1508.02676 [hep-th]].\\
%\cite{Qualls:2015bta}
%\bibitem{Qualls:2015bta} 
J.~D.~Qualls,
``Universal Bounds on Operator Dimensions in General 2D Conformal Field Theories,''
arXiv:1508.00548 [hep-th].
%%CITATION = ARXIV:1508.00548;%%
%10 citations counted in INSPIRE as of 12 Sep 2017
%%CITATION = doi:10.1007/JHEP01(2016)139;%%
%23 citations counted in INSPIRE as of 12 Sep 2017
%\cite{Basu:2015gpa}
%\bibitem{Basu:2015gpa} 
%\cite{Rychkov:2015naa}
%\bibitem{Rychkov:2015naa}
\bibitem{gliozzi2} 
F.~Gliozzi, A.~Guerrieri, A.~C.~Petkou and C.~Wen,
``Generalized Wilson-Fisher Critical Points from the Conformal Operator Product Expansion,''
Phys.\ Rev.\ Lett.\  {\bf 118}, no. 6, 061601 (2017)
%doi:10.1103/PhysRevLett.118.061601
[arXiv:1611.10344 [hep-th]].\\
%%CITATION = doi:10.1103/PhysRevLett.118.061601;%%
%19 citations counted in INSPIRE as of 11 Sep 2017
%\cite{Roumpedakis:2016qcg}
%\bibitem{roump} 
K.~Roumpedakis,
``Leading Order Anomalous Dimensions at the Wilson-Fisher Fixed Point from CFT,''
JHEP {\bf 1707}, 109 (2017)
%doi:10.1007/JHEP07(2017)109
[arXiv:1612.08115 [hep-th]]\\
%%CITATION = doi:10.1007/JHEP07(2017)109;%%
%8 citations counted in INSPIRE as of 11 Sep 2017
%\cite{Liendo:2017wsn}
%\bibitem{liendo} 
P.~Liendo,
``Revisiting the dilatation operator of the Wilson–Fisher fixed point,''
Nucl.\ Phys.\ B {\bf 920}, 368 (2017)
%doi:10.1016/j.nuclphysb.2017.04.020
[arXiv:1701.04830 [hep-th]]\\
%%CITATION = doi:10.1016/j.nuclphysb.2017.04.020;%%
%4 citations counted in INSPIRE as of 11 Sep 2017
%\cite{Gliozzi:2017hni}
%\bibitem{Gliozzi:2017hni} 
F.~Gliozzi, A.~L.~Guerrieri, A.~C.~Petkou and C.~Wen,
``The analytic structure of conformal blocks and the generalized Wilson-Fisher fixed points,''
JHEP {\bf 1704}, 056 (2017)
%doi:10.1007/JHEP04(2017)056
[arXiv:1702.03938 [hep-th]].\\
%%CITATION = doi:10.1007/JHEP04(2017)056;%%
%16 citations counted in INSPIRE as of 11 Sep 2017
%\cite{Soderberg:2017oaa}
%\bibitem{Soderberg:2017oaa} 
A.~Söderberg,
``Anomalous Dimensions in the WF O($N$) Model with a Monodromy Line Defect,''
arXiv:1706.02414 [hep-th].\\
%\cite{Behan:2017mwi}
%\bibitem{Behan:2017mwi}
  %%CITATION = ARXIV:1709.03967;%%
%\cite{Manashov:2017xtt}
%\bibitem{Manashov:2017xtt} 
A.~N.~Manashov, E.~D.~Skvortsov and M.~Strohmaier,
``Higher spin currents in the critical $O(N$) vector model at $1/N^{2}$,''
JHEP {\bf 1708}, 106 (2017)
%doi:10.1007/JHEP08(2017)106
arXiv:1706.09256 [hep-th].\\
%%CITATION = doi:10.1007/JHEP08(2017)106;%%
%1 citations counted in INSPIRE as of 24 Oct 2017
 C.~Behan,
 ``Conformal manifolds: ODEs from OPEs,''
 arXiv:1709.03967 [hep-th].
%%%%%%%%%%%%%%%%%%%%%%%%%%%%%%%%%%%%%%%%%%%%%%%%%%%%%%%%
%\cite{Alday:2015eya}
\bibitem{Alday:2015eya} 
  L.~F.~Alday, A.~Bissi and T.~Lukowski,
  ``Large spin systematics in CFT,''
  JHEP {\bf 1511}, 101 (2015)
 % doi:10.1007/JHEP11(2015)101
  [arXiv:1502.07707 [hep-th]].
  %%CITATION = doi:10.1007/JHEP11(2015)101;%%
  %44 citations counted in INSPIRE as of 09 Sep 2017

%\bibitem{largespin2AS}
%A.~Kaviraj, K.~Sen and A.~Sinha,
%``Universal anomalous dimensions at large spin and large twist,''
%JHEP {\bf 1507}, 026 (2015)
%%doi:10.1007/JHEP07(2015)026
%[arXiv:1504.00772 [hep-th]].

%\bibitem{largespin1AS}	
%A.~Kaviraj, K.~Sen and A.~Sinha,
%``Analytic bootstrap at large spin,''
%JHEP {\bf 1511}, 083 (2015)
%% doi:10.1007/JHEP11(2015)083
%[arXiv:1502.01437 [hep-th]].
%%CITATION = doi:10.1007/JHEP11(2015)083;%%
%47 citations counted in INSPIRE as of 06 Sep 2017
%\cite{Fitzpatrick:2012cg}

%\cite{Alday:2015ota}
\bibitem{Alday:2015ota} 
  L.~F.~Alday and A.~Zhiboedov,
  ``Conformal Bootstrap With Slightly Broken Higher Spin Symmetry,''
  JHEP {\bf 1606}, 091 (2016)
  %doi:10.1007/JHEP06(2016)091
  [arXiv:1506.04659 [hep-th]].
  %%CITATION = doi:10.1007/JHEP06(2016)091;%%
  %42 citations counted in INSPIRE as of 19 Aug 2017

%\cite{aldayzhiboedov}
\bibitem{aldayzhiboedov} 
  L.~F.~Alday and A.~Zhiboedov,
  ``An Algebraic Approach to the Analytic Bootstrap,''
  JHEP {\bf 1704}, 157 (2017)
  %doi:10.1007/JHEP04(2017)157
  [arXiv:1510.08091 [hep-th]].
  %%CITATION = doi:10.1007/JHEP04(2017)157;%%
  %26 citations counted in INSPIRE as of 01 Aug 2017

%\cite{Alday:2016mxe}
\bibitem{Alday:2016mxe} 
  L.~F.~Alday and A.~Bissi,
  ``Crossing symmetry and Higher spin towers,''
  arXiv:1603.05150 [hep-th].
  %%CITATION = ARXIV:1603.05150;%%
  %5 citations counted in INSPIRE as of 09 Sep 2017


%\cite{Alday:2016njk}
\bibitem{Alday:2016njk} 
  L.~F.~Alday,
  ``Large Spin Perturbation Theory,''
  arXiv:1611.01500 [hep-th].
  %%CITATION = ARXIV:1611.01500;%%
  %26 citations counted in INSPIRE as of 19 Aug 2017


%\cite{Alday:2016jfr}
\bibitem{Alday:2016jfr} 
  L.~F.~Alday,
  ``Solving CFTs with Weakly Broken Higher Spin Symmetry,''
  arXiv:1612.00696 [hep-th].
  %%CITATION = ARXIV:1612.00696;%%
  %18 citations counted in INSPIRE as of 18 Aug 2017

%\cite{Aharony:2016dwx}
\bibitem{Aharony:2016dwx} 
  O.~Aharony, L.~F.~Alday, A.~Bissi and E.~Perlmutter,
  ``Loops in AdS from Conformal Field Theory,''
  JHEP {\bf 1707}, 036 (2017)
  %doi:10.1007/JHEP07(2017)036
  [arXiv:1612.03891 [hep-th]].
  %%CITATION = doi:10.1007/JHEP07(2017)036;%%
  %18 citations counted in INSPIRE as of 18 Aug 2017



%\cite{Alday:2017gde}
\bibitem{holorecon} 
  L.~F.~Alday, A.~Bissi and E.~Perlmutter,
  ``Holographic Reconstruction of AdS Exchanges from Crossing Symmetry,''
  arXiv:1705.02318 [hep-th].
  %%CITATION = ARXIV:1705.02318;%%
  %8 citations counted in INSPIRE as of 18 Aug 2017

%\cite{Fitzpatrick:2012yx}
\bibitem{Fitzpatrick:2012yx} 
  A.~L.~Fitzpatrick, J.~Kaplan, D.~Poland and D.~Simmons-Duffin,
  ``The Analytic Bootstrap and AdS Superhorizon Locality,''
  JHEP {\bf 1312}, 004 (2013)
  %doi:10.1007/JHEP12(2013)004
  [arXiv:1212.3616 [hep-th]].
  %%CITATION = doi:10.1007/JHEP12(2013)004;%%
  %138 citations counted in INSPIRE as of 01 Aug 2017

%\cite{Komargodski:2012ek}
\bibitem{Komargodski:2012ek} 
  Z.~Komargodski and A.~Zhiboedov,
  ``Convexity and Liberation at Large Spin,''
  JHEP {\bf 1311}, 140 (2013)
 % doi:10.1007/JHEP11(2013)140
  [arXiv:1212.4103 [hep-th]].
  %%CITATION = doi:10.1007/JHEP11(2013)140;%%
  %143 citations counted in INSPIRE as of 01 Aug 2017

%\cite{KSAS}
\bibitem{dsd}
  D.~Simmons-Duffin,
  ``The Lightcone Bootstrap and the Spectrum of the 3d Ising CFT,''
  JHEP {\bf 1703}, 086 (2017)
  %doi:10.1007/JHEP03(2017)086
  [arXiv:1612.08471 [hep-th]].
  %%CITATION = doi:10.1007/JHEP03(2017)086;%%
  %26 citations counted in INSPIRE as of 13 Sep 2017


%%%%%%%%%%%%%%%%%%%%%%%%%%%%%%%%%%%%%%%%%%%%%%%


\bibitem{rychkovtan} S.~Rychkov and Z.~M.~Tan,
``The $\epsilon$-expansion from conformal field theory,''
J.\ Phys.\ A {\bf 48} (2015) no.29,  29FT01
%doi:10.1088/1751-8113/48/29/29FT01
[arXiv:1505.00963 [hep-th]].
%%CITATION = doi:10.1088/1751-8113/48/29/29FT01;%%
%45 citations counted in INSPIRE as of 12 Sep 2017
%\cite{Kos:2015mba}
%\bibitem{Kos:2015mba} 
%F.~Kos, D.~Poland, D.~Simmons-Duffin and A.~Vichi,
%``Bootstrapping the O(N) Archipelago,''
%JHEP {\bf 1511}, 106 (2015)
%%doi:10.1007/JHEP11(2015)106
%[arXiv:1504.07997 [hep-th]].\\
%%CITATION = doi:10.1007/JHEP11(2015)106;%%
%79 citations counted in INSPIRE as of 12 Sep 2017
%\cite{Vos:2014pqa}
%%CITATION = doi:10.1016/j.nuclphysb.2015.07.013;%%
%19 citations counted in INSPIRE as of 12 Sep 2017
%\cite{Kimura:2016bzo}
%\bibitem{Kimura:2016bzo}

\bibitem{KSAS} 
  K.~Sen and A.~Sinha,
  ``On critical exponents without Feynman diagrams,''
  J.\ Phys.\ A {\bf 49}, no. 44, 445401 (2016)
  %doi:10.1088/1751-8113/49/44/445401
  [arXiv:1510.07770 [hep-th]].
  %%CITATION = doi:10.1088/1751-8113/49/44/445401;%%
  %17 citations counted in INSPIRE as of 09 Sep 2017


%\cite{RGAKKSAS}
	\bibitem{RGAKKSAS} 
	R.~Gopakumar, A.~Kaviraj, K.~Sen and A.~Sinha,
	``Conformal Bootstrap in Mellin Space,''
	Phys.\ Rev.\ Lett.\  {\bf 118}, no. 8, 081601 (2017)
	%doi:10.1103/PhysRevLett.118.081601
	[arXiv:1609.00572 [hep-th]].\\
	R.~Gopakumar, A.~Kaviraj, K.~Sen and A.~Sinha,
	``A Mellin space approach to the conformal bootstrap,''
	JHEP {\bf 1705}, 027 (2017)
	%doi:10.1007/JHEP05(2017)027
	[arXiv:1611.08407 [hep-th]].
	%%CITATION = doi:10.1007/JHEP05(2017)027;%%
	%28 citations counted in INSPIRE as of 18 Jul 2017	

%\cite{PDAKAS}
\bibitem{PDAKAS} 
  P.~Dey, A.~Kaviraj and A.~Sinha,
  ``Mellin space bootstrap for global symmetry,''
  JHEP {\bf 1707}, 019 (2017)
 % doi:10.1007/JHEP07(2017)019
  [arXiv:1612.05032 [hep-th]].
  %%CITATION = doi:10.1007/JHEP07(2017)019;%%
  %12 citations counted in INSPIRE as of 09 Sep 2017

\bibitem{polya}
A.~M.~Polyakov,
``Nonhamiltonian approach to conformal quantum field theory,''
Zh.\ Eksp.\ Teor.\ Fiz.\  {\bf 66}, 23 (1974).

%\cite{Osborn:2017ucf}
\bibitem{osbornrecent} 
  H.~Osborn and A.~Stergiou,
  ``Seeking Fixed Points in Multiple Coupling Scalar Theories in the $\varepsilon$ Expansion,''
  arXiv:1707.06165 [hep-th].
  %%CITATION = ARXIV:1707.06165;%%
%charlotte,spinningads
%\cite{Bekaert:2014cea}
\bibitem{Mack:2009mi} 
G.~Mack,
``D-independent representation of Conformal Field Theories in D dimensions via transformation to auxiliary Dual Resonance Models. Scalar amplitudes,''
arXiv:0907.2407 [hep-th].
%%CITATION = ARXIV:0907.2407;%%
%93 citations counted in INSPIRE as of 07 Aug 2017
%\cite{Mellin}
%\cite{Penedones:2010ue}
\bibitem{Mellin} 
J.~Penedones,
``Writing CFT correlation functions as AdS scattering amplitudes,''
JHEP {\bf 1103}, 025 (2011)
%doi:10.1007/JHEP03(2011)025
[arXiv:1011.1485 [hep-th]].
%%CITATION = doi:10.1007/JHEP03(2011)025;%%
%133 citations counted in INSPIRE as of 11 Sep 2017
%\cite{Fitzpatrick:2011ia}
\bibitem{fitz} 
A.~L.~Fitzpatrick, J.~Kaplan, J.~Penedones, S.~Raju and B.~C.~van Rees,
``A Natural Language for AdS/CFT Correlators,''
JHEP {\bf 1111}, 095 (2011)
%doi:10.1007/JHEP11(2011)095
[arXiv:1107.1499 [hep-th]].
%%CITATION = doi:10.1007/JHEP11(2011)095;%%
%107 citations counted in INSPIRE as of 11 Sep 2017
%\cite{Paulos:2011ie}
\bibitem{pau} 
M.~F.~Paulos,
``Towards Feynman rules for Mellin amplitudes,''
JHEP {\bf 1110}, 074 (2011)
%doi:10.1007/JHEP10(2011)074
[arXiv:1107.1504 [hep-th]].
%%CITATION = doi:10.1007/JHEP10(2011)074;%%
%70 citations counted in INSPIRE as of 11 Sep 2017
%\cite{Costa:2012cb}
\bibitem{costa} 
M.~S.~Costa, V.~Goncalves and J.~Penedones,
``Conformal Regge theory,''
JHEP {\bf 1212}, 091 (2012)
%doi:10.1007/JHEP12(2012)091
[arXiv:1209.4355 [hep-th]].
%%CITATION = doi:10.1007/JHEP12(2012)091;%%
%89 citations counted in INSPIRE as of 11 Sep 2017
%%%%%%%%%%%%%%%%%%%%%%%%%%%%%%%%%%%%%%%%%%%%%%%%%%%%%%%%
\bibitem{dolanosborn} 
F.~A.~Dolan and H.~Osborn,
``Conformal partial waves and operator product expansion,"
Nucl. Phys. {\bf B678}(2004) 491-507.
arXiv: hep-th/0309180.
\bibitem{do2}
F.~A.~Dolan and H.~Osborn,
``Conformal four point functions and the operator product expansion,"
Nucl. Phys. {\bf B599}(2001) 459-496.
arXiv: hep-th/0011040.


%%%%%%%%%%%%%%%%%%9-16%%%%%%%%%%%%%%%%%%%%%%%%%%
\bibitem{Dolan:2011dv} 
F.~A.~Dolan and H.~Osborn,
``Conformal Partial Waves: Further Mathematical Results,''
arXiv:1108.6194 [hep-th].





\bibitem{fkap} 
A.~L.~Fitzpatrick and J.~Kaplan,
``AdS Field Theory from Conformal Field Theory,''
JHEP {\bf 1302}, 054 (2013)
%  doi:10.1007/JHEP02(2013)054
arXiv:1208.0337 [hep-th].

\bibitem{charlotte} 
X.~Bekaert, J.~Erdmenger, D.~Ponomarev and C.~Sleight,
``Towards holographic higher-spin interactions: Four-point functions and higher-spin exchange,''
JHEP {\bf 1503}, 170 (2015)
%doi:10.1007/JHEP03(2015)170
[arXiv:1412.0016 [hep-th]].
%%CITATION = doi:10.1007/JHEP03(2015)170;%%
%43 citations counted in INSPIRE as of 11 Sep 2017

%\cite{Sleight:2016hyl}
\bibitem{spinningads} 
  M.~S.~Costa, V.~Goncalves and J.~Penedones,
  ``Spinning AdS Propagators,''
  JHEP {\bf 1409}, 064 (2014)
  %doi:10.1007/JHEP09(2014)064
  [arXiv:1404.5625 [hep-th]].
  %%CITATION = doi:10.1007/JHEP09(2014)064;%%
  %47 citations counted in INSPIRE as of 12 Sep 2017


%\cite{Sleight:2017fpc}
\bibitem{ASnew}
R.~Gopakumar and A.~Sinha, ``Simplifying Mellin bootstrap'', to appear.

%%%%%%%%%%%%%%%%%%%%%%%%%%%%%%%%%%%%%%%%%

%%%%%%%%%%%%%%%%%%%%%%%%%Mellin ref%%%%%%%%%%%%%%%%%%%%%

%\cite{Mack:2009mi}
\bibitem{sleight} 
C.~Sleight and M.~Taronna,
``Spinning Witten Diagrams,''
JHEP {\bf 1706}, 100 (2017)
%doi:10.1007/JHEP06(2017)100
[arXiv:1702.08619 [hep-th]].\\
%%CITATION = doi:10.1007/JHEP06(2017)100;%%
%14 citations counted in INSPIRE as of 11 Sep 2017
%\cite{Giombi:2017hpr}
%\bibitem{Giombi:2017hpr} 
S.~Giombi, C.~Sleight and M.~Taronna,
``Spinning AdS Loop Diagrams: Two Point Functions,''
arXiv:1708.08404 [hep-th].\\
%%CITATION = ARXIV:1708.08404;%%
%3 citations counted in INSPIRE as of 11 Sep 2017
%\cite{Gubser:2017tsi}
%\bibitem{gubser} 
%\cite{Hijano:2015zsa}
%\bibitem{Hijano:2015zsa} 
E.~Hijano, P.~Kraus, E.~Perlmutter and R.~Snively,
``Witten Diagrams Revisited: The AdS Geometry of Conformal Blocks,''
JHEP {\bf 1601}, 146 (2016)
%doi:10.1007/JHEP01(2016)146
[arXiv:1508.00501 [hep-th]].\\
%%CITATION = doi:10.1007/JHEP01(2016)146;%%
%67 citations counted in INSPIRE as of 12 Sep 2017
S.~S.~Gubser and S.~Parikh,
``Geodesic bulk diagrams on the Bruhat-Tits tree,''
arXiv:1704.01149 [hep-th].\\
%%CITATION = ARXIV:1704.01149;%%
%5 citations counted in INSPIRE as of 11 Sep 2017
%\cite{Castro:2017hpx}
%\bibitem{Castro:2017hpx} 
A.~Castro, E.~Llabrés and F.~Rejon-Barrera,
``Geodesic Diagrams, Gravitational Interactions  OPE Structures,''
JHEP {\bf 1706}, 099 (2017)
%doi:10.1007/JHEP06(2017)099
[arXiv:1702.06128 [hep-th]].\\
%%CITATION = doi:10.1007/JHEP06(2017)099;%%
%11 citations counted in INSPIRE as of 11 Sep 2017
%\bibitem{Sleight:2017pcz} 
C.~Sleight and M.~Taronna,
``Higher spin gauge theories and bulk locality: a no-go result,''
arXiv:1704.07859 [hep-th].
%%CITATION = ARXIV:1704.07859;%%
%6 citations counted in INSPIRE as of 12 Sep 2017
%%%%%%%%%%%%%%%%%%%%%%%%%%%%%%%%%%%%%%%%%%%%%%%%%%%%%%%%%%%%%

%%%%%%%%%%%%%%%%%%%%%%%%%%others%%%%%%%%%%%%%%%%%%%%%%%%%%%%%%%
%\cite{Dutta:2017bja}
\bibitem{kss}
%\cite{Heemskerk:2009pn}
%\bibitem{Heemskerk:2009pn} 
 
  %%CITATION = doi:10.1088/1126-6708/2009/10/079;%%
  %265 citations counted in INSPIRE as of 17 Sep 2017

%L.~F.~Alday, A.~Bissi and T.~Lukowski,
%``Large spin systematics in CFT,''
%arXiv:1502.07707 [hep-th].\\
A.~Kaviraj, K.~Sen and A.~Sinha,
``Analytic bootstrap at large spin,''
JHEP {\bf 1511}, 083 (2015)
%%doi:10.1007/JHEP11(2015)083
[arXiv:1502.01437 [hep-th]].\\
A.~Kaviraj, K.~Sen and A.~Sinha,
``Universal anomalous dimensions at large spin and large twist,"
JHEP \textbf{1507} (2015) 026,
arXiv:1504.00772 [hep-th].

\bibitem{heem}
 I.~Heemskerk, J.~Penedones, J.~Polchinski and J.~Sully,
  ``Holography from Conformal Field Theory,''
  JHEP {\bf 0910}, 079 (2009)
 % doi:10.1088/1126-6708/2009/10/079
  arXiv:0907.0151 [hep-th].

\bibitem{temme} N.~M.~Temme, ``Asymptotic Methods for Integrals,'' World Scientific, 2015.

%\cite{srivastava}


%\cite{Dolan:2011dv}
\bibitem{bender} 
  C.~M.~Bender and C.~Heissenberg,
  ``Convergent and Divergent Series in Physics,''
  arXiv:1703.05164 [math-ph].

%\bibitem{caronhuot} 
  %S.~Caron-Huot,
  %``Analyticity in Spin in Conformal Theories,''
  %arXiv:1703.00278 [hep-th].
  %%CITATION = ARXIV:1703.00278;%%
  %13 citations counted in INSPIRE as of 12 Sep 2017

%%%%%%%%%%%%%%%%%%%%%%%%%%%%Discussion%%%%%%%%%%%%%%%%%%%%%
%\cite{PDAKnew}
\bibitem{caronhout} 
S.~Caron-Huot,
``Analyticity in Spin in Conformal Theories,''
arXiv:1703.00278 [hep-th].
%%CITATION = ARXIV:1703.00278;%%
%13 citations counted in INSPIRE as of 11 Sep 2017

%\cite{Karateev:2017jgd}
\bibitem{kss2}
P.~Dey, A.~Kaviraj and K.~Sen,
``More on analytic bootstrap for O(N) models,''
JHEP {\bf 1606}, 136 (2016)
%%doi:10.1007/JHEP06(2016)136
[arXiv:1602.04928 [hep-th]].\\
G.~Vos,
``Generalized Additivity in Unitary Conformal Field Theories,''
Nucl.\ Phys.\ B {\bf 899}, 91 (2015)
%%doi:10.1016/j.nuclphysb.2015.07.013
[arXiv:1411.7941 [hep-th]].\\
D.~Li, D.~Meltzer and D.~Poland,
``Non-Abelian Binding Energies from the Lightcone Bootstrap,''
arXiv:1510.07044 [hep-th].
%%CITATION = ARXIV:1510.07044;%%
%2 citations counted in INSPIRE as of 03 Feb 2016

\bibitem{PDAKnew}
P.~Dey and A.~Kaviraj, to appear.
%\cite{Dymarsky:2017yzx}
\bibitem{zhiboedov2} 
M.~Kulaxizi, A.~Parnachev and A.~Zhiboedov,
``Bulk Phase Shift, CFT Regge Limit and Einstein Gravity,''
arXiv:1705.02934 [hep-th].
%%CITATION = ARXIV:1705.02934;%%
%6 citations counted in INSPIRE as of 12 Sep 2017
%\bibitem{kos}
%F.~Kos, D.~Poland, D.~Simmons-Duffin and A.~Vichi,
%``Bootstrapping the O(N) Archipelago,''
%JHEP {\bf 1511}, 106 (2015)
%%doi:10.1007/JHEP11(2015)106
%[arXiv:1504.07997 [hep-th]].

%%%%%%%%%%%%%%%%%%%%%%%%cite%%%%%%%%%%%%%%%%
%\cite{Li:2017lmh}
\bibitem{polandnew} 
A.~Dymarsky, F.~Kos, P.~Kravchuk, D.~Poland and D.~Simmons-Duffin,
``The 3d Stress-Tensor Bootstrap,''
arXiv:1708.05718 [hep-th].
%%CITATION = ARXIV:1708.05718;%%

%\bibitem{sasha1} ...already cited zhiboedov2..ref 34
%  M.~Kulaxizi, A.~Parnachev and A.~Zhiboedov,
%  ``Bulk Phase Shift, CFT Regge Limit and Einstein Gravity,''
%  arXiv:1705.02934 [hep-th].
%  %%CITATION = ARXIV:1705.02934;%%

%\cite{Qiao:2017xif}
\bibitem{tauberian} 
J.~Qiao and S.~Rychkov,
``A tauberian theorem for the conformal bootstrap,''
arXiv:1709.00008 [hep-th].
%%CITATION = ARXIV:1709.00008;%%

%%%%%%%%%%%%%%%%Appendix A%%%%%%%%%%%%%%%%%%%%%%%%%%%%%

\bibitem{AAR}
 G.~E.~Andrews, R.~Askey and R~Roy, ``Special functions,'' Cambridge University Press, 1999.
%\cite{Wilson}
\bibitem{nist-hbk}
http://dlmf.nist.gov/18.26
\bibitem{Wilson} 
J.~A.~Wilson, ``Asymptotics for the $_4F_3$ Polynomials,''
Journal of Approximation Theory {\bf 66}, 58-71 (1991).

\bibitem{Fields1964}
J.~L.~Fields, ``A note on the asymptotic expansion of a ratio of gamma functions,'' Proc. Edinburgh Math. Soc. {\bf 15}, 43-45.
\bibitem{srivastava}
H.~M.~Srivastava and P.~G.~Todorov, ``An explicit formula for the generalized Bernoulli polynomials,'' Journal of Mathematical Analysis and Applications, 1988, Volume 130, Issue 2, 509-513. 

\end{thebibliography}
\end{document}